\documentclass[letterpaper,titlepage,11pt]{article}
\usepackage{hyperref}
\usepackage{amssymb,amsmath,amsfonts}
\usepackage{epsfig}

\def\clock{{\count0=\time
           \divide\count0 60
           \ifnum\count0<10 0\fi\the\count0
           \multiply\count0 -60 \advance\count0 \time
           :\ifnum\count0<10 0\fi \the\count0
         }}
\newcommand{\timestamp}{{\small\vbox{\hbox{\tt\jobname.tex}
\hbox{\the\day/\the\month/\the\year, \clock}}}}

\newcommand{\be}{\begin{equation}} \newcommand{\ee}{\end{equation}}
\newcommand{\bea}{\begin{eqnarray}} \newcommand{\eea}{\end{eqnarray}}

\newcommand{\CO}{\mathcal{O}}

\newcommand{\CT}{\mathcal{T}}

\newcommand{\CM}{\mathcal{M}}

\newcommand{\CR}{\mathcal{R}}

\newcommand{\id}{\hbox{1\kern-.27em l}}
\newcommand{\sid}{\hbox{\scriptsize1\kern-.27em l}}

\newcommand{\we}{\kern-.1em\wedge\kern-.1em}
\newcommand{\scal}{\kern-.13em\cdot\kern-.13em}

\newcommand{\II}{I\kern-.09em I}

\newcommand{\al}{\alpha}

\newcommand{\Z}{\mathbb{Z}} 
\newcommand{\R}{\mathbb{R}}

\newcommand{\spa}{\ , \ \ }

\newcommand{\refeq}[1]{(\ref{#1})}

\newcommand{\rom}[1]{\mathrm{#1}}

\newcommand{\beastar}{\begin{eqnarray*}}
\newcommand{\eeastar}{\end{eqnarray*}}

\numberwithin{equation}{section}

\begin{document}

\begin{titlepage}

\rightline{\vbox{\small\hbox{\tt hep-th/0407050} }} \vskip 3cm

\centerline{\Large \bf Sequences of Bubbles and Holes:} \vskip
0.2cm \centerline{\Large \bf New Phases of Kaluza-Klein Black
Holes}

\vskip 1.6cm \centerline{\bf Henriette Elvang$\,^{1}$, Troels
Harmark$\,^{2}$, Niels A. Obers$\,^{2}$} \vskip 0.5cm
\begin{center}
\sl $^1$ Department of Physics, UCSB \\
\sl Santa Barbara, CA 93106, USA \\
\vskip 0.5cm
\sl $^2$ The Niels Bohr Institute \\
\sl Blegdamsvej 17, 2100 Copenhagen \O , Denmark
\end{center}

\vskip 0.5cm

\centerline{\small\tt elvang@physics.ucsb.edu, harmark@nbi.dk,
obers@nbi.dk}

\vskip 1.2cm \noindent We construct and analyze a large class of
exact five- and six-dimensional regular and static solutions of
the vacuum Einstein equations. These solutions describe sequences
of Kaluza-Klein bubbles and black holes, placed alternately so
that the black holes are held apart by the bubbles. Asymptotically
the solutions are Minkowski-space times a circle,
i.e.~Kaluza-Klein space, so they are part of the $(\mu,n)$ phase
diagram introduced in \mbox{hep-th/0309116}. In particular, they
occupy a hitherto unexplored region of the phase diagram, since
their relative tension exceeds that of the uniform black string.
The solutions contain bubbles and black holes of various
topologies, including six-dimensional black holes with ring
topology $S^3 \times S^1$ and tuboid topology $S^2 \times S^1
\times S^1$.
The bubbles support the $S^1$'s of the horizons against
gravitational collapse. We
find two maps between solutions, one that relates five- and
six-dimensional solutions, and another that relates solutions in
the same dimension by interchanging bubbles and black holes. To
illustrate the richness of the phase structure and the
non-uniqueness in the $(\mu,n)$ phase diagram, we consider in
detail particular examples of the general class of solutions.


\end{titlepage}

\pagestyle{empty}

\tableofcontents

\newpage

\pagestyle{plain} \setcounter{page}{1}

\section{Introduction}

In four-dimensional vacuum gravity, a black hole in an
asymptotically flat space-time is uniquely specified by the ADM
mass and angular momentum measured at infinity
\cite{Israel:1967wq,Carter:1971,Hawking:1972vc,Robinson:1975}.
Uniqueness theorems \cite{Gibbons:2002bh,Gibbons:2002av} for
$D$-dimensional ($D > 4$) asymptotically flat space-times state
that the only static black holes in pure gravity are given by the
Schwarzschild-Tangherlini black hole solutions
\cite{Tangherlini:1963}. However, in pure gravity there are no
uniqueness theorems for non-static black holes with $D > 4$, or
for black holes in space-times with non-flat asymptotics. On the
contrary, there are known cases of non-uniqueness. An explicit
example of this occurs in five dimensions for stationary solutions
in an asymptotically flat space-time: for a certain range of mass
and angular momentum there exist both a rotating black hole with
$S^3$ horizon \cite{Myers:1986un} and rotating black rings with
$S^2 \times S^1$ horizons \cite{Emparan:2001wn}.

The topic of this paper is static black hole space-times that are
asymptotically Minkowski space $\CM^{d}$ times a circle $S^1$, in
other words, we study static black
holes%
\footnote{We use ``black hole'' to denote any black object, no
matter its horizon topology.} in Kaluza-Klein theory. For brevity,
we generally refer to these solutions as \emph{Kaluza-Klein black
holes}. Changing the boundary conditions from asymptotically flat
space $\CM^{d+1}$ to asymptotically $\CM^{d} \times S^1$ opens up
for a rich spectrum of black holes. As we shall see, the
non-uniqueness of Kaluza-Klein black holes goes even further than
for black holes in asymptotically flat space $\CM^{d+1}$.

Much numerical
\cite{Gubser:2001ac,Wiseman:2002zc,Wiseman:2002ti,Choptuik:2003qd,Sorkin:2003ka,Kudoh:2003ki,Sorkin:2004qq}
and analytical \cite{Harmark:2002tr,Harmark:2003fz,Harmark:2003dg,Kol:2003if,Harmark:2003eg,%
Harmark:2003yz,Gorbonos:2004uc} work has been done to investigate
the ``phase space'' of black hole solutions in Kaluza-Klein
theory.  Part of the motivation has been by the wish to reveal the
endpoint for the classical evolution of the unstable uniform black
string \cite{Gregory:1993vy}.

Recently, Refs.~\cite{Harmark:2003dg,Harmark:2003eg} proposed a
\emph{phase diagram} as part of a program for classifying all
black hole solutions of Kaluza-Klein theory. The input for the
phase diagram consist of two physical parameters that are measured
asymptotically: the dimensionless mass $\mu \propto M/L^{d-2}$,
where $M$ is the ADM mass and $L$ is the proper length of the
Kaluza-Klein circle at infinity, and the relative tension $n$
\cite{Harmark:2003dg,Harmark:2003eg,Harmark:2004ch}. This is the
tension per unit mass of a string winding the Kaluza-Klein circle.
The $(\mu,n)$ phase diagram makes it possible to illustrate the
different branches of solutions and exhibit their possible
relationships. The main purpose of this paper is to construct and
analyze a large class of exact five- and six-dimensional
Kaluza-Klein black hole solutions occupying an hitherto unexplored
region of the $(\mu,n)$ phase diagram.

In Kaluza-Klein space-times, it is well-known that there exist
both uniform and non-uniform black strings with the same mass
$\mu$
\cite{Horowitz:2001cz,Gubser:2001ac,Wiseman:2002zc,Gregory:1988nb,%
Choptuik:2003qd,Sorkin:2004qq} (see also
\cite{Horowitz:2002dc,Kol:2002xz,Kol:2003ja}). There is also a
family of topologically spherical black holes localized on the
Kaluza-Klein circle
\cite{Harmark:2002tr,Sorkin:2003ka,Kudoh:2003ki,Harmark:2003yz,Gorbonos:2004uc}.
These solutions, however, can be told apart at infinity, because
they exist for different values of the relative tension $n$. We
show in this paper that there is non-uniqueness of black holes in
Kaluza-Klein theory, and we argue that for a certain open set of
values of $\mu$ and $n$ there is even infinite non-uniqueness of
Kaluza-Klein black holes.
Infinite non-uniqueness has been seen before in \cite{Emparan:2004wy}
for black rings with dipole charges in asymptotically flat
five-dimensional space.
The solutions we present here are, on the contrary, solutions
of pure gravity and the non-uniqueness
involves regular space-times with multiple
black holes. While some configurations have black holes whose
horizons are topologically spheres, we also encounter black rings
with horizon topologies $S^{d-2} \times S^1$ (for $d=4,5$), and in
six dimensions a black tuboid with $S^{2}\times S^1 \times S^1$
horizon.

A crucial feature of the black hole space-times studied in this
paper is that they all involve Kaluza-Klein ``bubbles of
nothing''. Expanding Kaluza-Klein bubbles were first studied by
Witten in \cite{Witten:1982gj} as the endstate of the
semi-classical decay of the Kaluza-Klein vacuum $\CM^4 \times
S^1$. The bubble is the minimal area surface that arises as the
asymptotic $S^1$ smoothly shrinks to zero at a non-zero radius.
The expanding Witten bubbles are non-static space-times with zero
ADM mass. The Kaluza-Klein bubbles appearing in the solutions of
this paper are on the other hand static with positive ADM mass.
Kaluza-Klein bubbles will be reviewed early in the present paper.

The first solution combining a black hole and a Kaluza-Klein
bubble was found by Emparan and Reall \cite{Emparan:2001wk} as an
example of an axisymmetric static space-time in the class of
generalized Weyl solutions. Later Ref.~\cite{Elvang:2002br}
studied space-times with two black holes held apart by a
Kaluza-Klein bubble and argued that the bubble balances the
gravitational attraction between the two black holes, thus keeping
the configuration in static equilibrium.

One natural question to ask is what the role of these solutions is
in the phase diagram of Kaluza-Klein black holes. To address this
issue we recall that one useful property of the $(\mu,n)$ phase
diagram is that physical solutions lie in the region
\cite{Harmark:2003dg}
\begin{equation}
\label{boundtot} \mu \geq 0 \spa 0 \leq n \leq d-2 \, .
\end{equation}
These  bounds were derived using various energy theorems
\cite{Traschen:2003jm,Shiromizu:2003gc,Harmark:2003dg,Harmark:2004ch}.
 However, so far only solutions in the lower region,
\begin{equation}
\label{boundbs} 0 \leq n \leq \frac{1}{d-2} \, ,
\end{equation}
have been discussed in connection to the phase diagram. This
region includes the following three known branches:
\begin{itemize}
\item The uniform black string branch which has relative tension
$n=1/(d-2)$ and hence bounds the region \eqref{boundbs} from
above. The uniform black string is classically unstable
\cite{Gregory:1993vy} for $\mu < \mu_\rom{GL}$, where
$\mu_\rom{GL}$ is the Gregory-Laflamme mass, and it is believed to
be stable for $\mu > \mu_\rom{GL}$. \item The non-uniform string
branch \cite{Gubser:2001ac,Wiseman:2002zc,Sorkin:2004qq} which
emanates from the uniform string branch at $\mu = \mu_{\rm GL}$
and has decreasing $n$ and increasing (decreasing) $\mu$ for $4
\leq d \leq 12$ ($d > 12$). \item The black hole on cylinder
branch
\cite{Harmark:2002tr,Harmark:2003yz,Kol:2003if,Harmark:2003dg,Harmark:2003eg,%
Sorkin:2003ka,Kudoh:2003ki,Gorbonos:2004uc} which starts in the
point $(\mu,n)=(0,0)$ and has increasing $n$ as $\mu$ increases.
This is the branch of topologically spherical black holes
localized on the Kaluza-Klein circle.
\end{itemize}
An obvious question is thus whether there are Kaluza-Klein black
hole solutions occupying the upper region
\begin{equation}
\label{boundbub}
 \frac{1}{d-2} < n \leq d-2 \, .
\end{equation}
 We will find in this paper that it is in fact the
solutions involving Kaluza-Klein bubbles that occupy this region.
A special point in the phase diagram is the static Kaluza-Klein
bubble which corresponds to the single point $(\mu,n)=(\mu_{\rm
b},d-2)$ in the phase diagram, where $\mu_{\rm b}$ is the
dimensionless mass of the static Kaluza-Klein bubble.

More generally, we construct exact metrics for bubble-black hole
configurations with $p$ bubbles and $q=p,p\pm 1$ black holes in
$D=5,6$ dimensions. These are regular and static solutions of the
vacuum Einstein equations, describing sequences of Kaluza-Klein
bubbles and black holes placed alternately, e.g.~for $(p,q)=(2,3)$
we have the sequence: \beastar
  \rom{black~hole}~-~\rom{bubble}~-~\rom{black~hole}
  ~-~\rom{bubble}~-~\rom{black~hole} \, .
\eeastar We will call this class of solutions \emph{bubble-black
hole sequences} and refer to particular elements of this class as
$(p,q)$ solutions. This large class of solutions, which was
anticipated in Ref.~\cite{Emparan:2001wk}, includes as particular
cases the $(1,1)$, $(1,2)$ and $(2,1)$ solutions obtained and
analyzed in \cite{Emparan:2001wk,Elvang:2002br}. All of these
solutions have $1/(d-2) < n < d-2$.

Besides their explicit construction, we present a comprehensive
analysis of various aspects of these bubble-black hole sequences.
This includes the regularity and topology of the Kaluza-Klein
bubbles in the sequences, the topology of the event horizons, and
general thermodynamical properties. An important feature is that
the  $(p,q)$ solutions are subject to constraints enforcing
regularity, but this leaves $q$ independent dimensionless
parameters allowing for instance the relative sizes of the black
holes to vary.  The existence of $q$ independent parameters in
each $(p,q)$ solution is the reason for the large degree of
non-uniqueness in the $(\mu,n)$ phase diagram, when considering
bubble-black hole sequences.

The Kaluza-Klein bubbles play a key role in keeping these
configurations in static equilibrium: not only do they balance the
mutual attraction between the black holes, they also balance the
gravitational self-attraction of black holes with non-trivial
horizon topologies such as black rings. A different example of
five-dimensional multi-black hole space-times based on the
generalized Weyl ansatz was studied in Ref.~\cite{Tan:2003jz}.
Those solutions differ from ours in that they are asymptotically
flat, and instead of bubbles, the black holes are
held in static equilibrium by struts due to conical singularities.%
\footnote{ In four dimensional asymptotically flat space, the
analogue of the configuration in \cite{Tan:2003jz} is the
Israel-Kahn multi-black hole solution, where the gravitational
attraction of the black holes is balanced by struts between the
black holes (or cosmic strings extending out to infinity). In
Kaluza-Klein theory the black holes are balanced by the bubbles
and the metrics are regular and free of conical singularities
\cite{Emparan:2001wk,Elvang:2002br}. We stress that when
discussing non-uniqueness we always restrict ourselves to
solutions that are regular everywhere on and outside the
horizon(s); thus we do not consider solutions with singular
horizons or solutions with conical singularities. All bubble-black
hole solutions discussed in this paper are regular.}

For the simplest cases, we will plot the corresponding solution
branches in the $(\mu,n)$ phase diagram, where they are seen to
lie in the upper region \eqref{boundbub}. Moreover, these examples
illustrate the richness of the phase structure and the
non-uniqueness in the phase diagram.

The structure and main results of the paper are as follows. We
introduce in Section \ref{s:phase} the $(\mu,n)$ phase diagram and
explain how $\mu$ and $n$ are easily computed from the asymptotic
behavior of the metric. We also briefly review the three known
solution branches that occupy the region \eqref{boundbs}, i.e. the
uniform and non-uniform black strings and the localized spherical
black holes.

Section \ref{bubble} provides a review of the static Kaluza-Klein
bubble. In particular we review the argument that the static
bubbles are classically unstable, and decay by either expanding or
collapsing. We find a critical dimension $D=10$ below which the
mass of the static bubble is smaller than the Gregory-Laflamme
mass for the uniform black string. Hence, for $D\le 10$ the
endstate of the static bubble decay can be expected to be the
endstate of the uniform black string, rather than the black string
itself.

The bubble-black hole sequences are constructed using the general
Weyl ansatz of \cite{Emparan:2001wk}. We review this method in
section \ref{s:gweyl}, where we also write down metrics for the
simplest Kaluza-Klein space-times and explain how to read of the
asymptotic quantities using Weyl coordinates.

In Section \ref{s:5dKK} we construct the solution for the general
bubble-black hole sequence in five dimensions. We analyze the
constraints of regularity, the structure of the Kaluza-Klein
bubbles, and the event horizons and their topology. We also
compute the physical quantities relevant for the phase diagram and
the thermodynamics. Section \ref{s:6dKK} provides a parallel
construction and analysis for the six-dimensional bubble-black
hole sequences.

It is shown that the five- and six-dimensional solutions are quite
similar in structure and are in fact related by an explicit map.
In particular, we find a map that relates the physical quantities,
so that we can use it to obtain the phase diagram for the
six-dimensional solutions from the five-dimensional one. This map
is derived in Subsection \ref{secmap56}.

For static space-times with more than one black hole horizon we
can associate a temperature to each black hole by analytically
continuing the solution to Euclidean space and performing the
proper identifications needed to make the Euclidean solution
regular where the horizon was located in the Lorentzian solution.
The temperatures of the black holes need not be equal, and we
derive a generalized Smarr formula that involves the temperature
of each black hole. The Euclidean solution is regular everywhere
only when all the temperatures are equal. It is always possible to
choose the $q$ free parameters of the $(p,q)$ solution to give a
one-parameter family of regular equal temperature solutions, which
we shall denote by $(p,q)_{\mathfrak{t}}$.

We show in Section \ref{s:eqt} that the equal temperature
$(p,q)_{\mathfrak{t}}$ solutions  are of special interest for two
reasons: First, the two solutions, $(p,q)_{\mathfrak{t}}$ and
$(q,p)_{\mathfrak{t}}$, are directly related by a double Wick
rotation which effectively interchanges the time coordinate and
the coordinate parameterizing the Kaluza-Klein circle. This
provides a duality map under which bubbles and black holes are
interchanged. The duality also implies an explicit map between the
physical quantities of the solutions, in particular between the
curves in the $(\mu,n)$ phase diagram.

Secondly, we show that for a given family of $(p,q)$ solutions,
the equal temperature solution extremizes the entropy for fixed
mass $\mu$ and fixed size of the Kaluza-Klein circle at infinity.
For all explicit cases considered we find that the entropy is
minimized for equal temperatures. This is a feature that is
particular to black holes, independently of the presence of
bubbles. As an analog, consider two Schwarzschild black holes very
far apart. It is straightforward to see that for fixed total mass,
the entropy of such a configuration is minimized when the black
holes have the same radius (hence same temperature), while the
maximal entropy configuration is the one where all the mass is
located in a single black hole.

In Section \ref{s:prop} we consider in detail particular examples
of the general five-  and six-dimensional bubble-black hole
sequences obtained in Sections \ref{s:5dKK}-\ref{s:6dKK}. For
these examples, we plot the various solution branches in the
$(\mu,n)$ phase diagram and discuss the total entropy of the
sequence as a function of the mass.

We find that the entropy of the $(1,1)$ solution is always lower
than the entropy of the uniform black string of the same mass
$\mu$. We expect that all other bubble-black hole sequences
$(p,q)$ have entropy lower than the $(1,1)$ solution; we confirm
this for all explicitly studied examples in Section \ref{s:prop}.
The physical reason to expect that all bubble-black hole sequences
have lower entropy than a uniform string of same mass, is that
some of the mass has gone into the bubble rather than the black
holes, giving a smaller horizon area for the same mass.

\begin{figure}[t]
\centerline{\epsfig{file=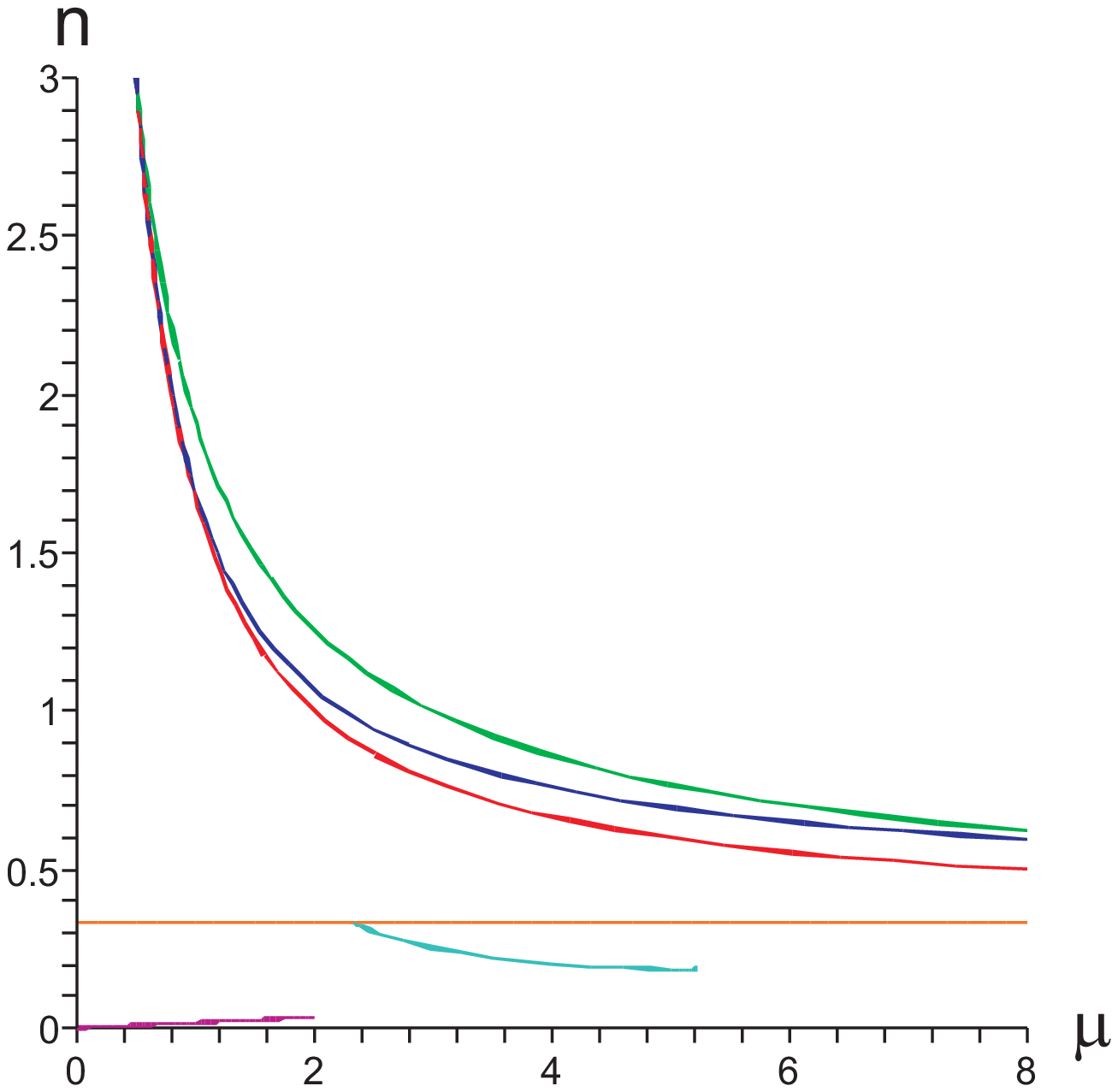,width=12cm,height=9cm}}
\caption{The $(\mu,n)$ phase diagram for six dimensions.}
\label{introfig}
\begin{picture}(0,0)(0,0)
\put(100,95){\footnotesize uniform black string}
\put(261,78){\footnotesize non-uniform black string}
\put(150,70){\footnotesize black hole} \put(85,122){\footnotesize
black ring} \put(85,112){\footnotesize on bubble}
\put(120,130){\vector(2,1){15}} \put(133,170){\footnotesize black
tuboid} \put(205,145){\footnotesize two equal-size black rings}
\put(205,135){\footnotesize on bubble}
\put(200,145){\vector(-1,-1){16}}
\end{picture}
\end{figure}

As an appetizer, and to give a representative taste of the type of
results we find  for the phase diagram, we show in Figure
\ref{introfig} the phase diagram for six dimensions (the phase
diagram for five dimensions is similar). Here we briefly summarize
what is shown in the plot. The horizontal line $n=1/3$ is the
uniform black string branch. This branch separates the diagram
into two regions: $0\le n \le 1/3$ and $1/3<n \le 3$.
The region $0 \le n \le 1/3$ has been the focus of many recent
studies. The branch coming out of the uniform black string branch
at $\mu_\rom{GL} \approx 2.3$ (the Gregory-Laflamme mass) is the
non-uniform black string branch, which is reproduced here using
numerical data courtesy Wiseman \cite{Wiseman:2002zc}. The curve
starting at $(\mu,n)=(0,0)$ is the branch of spherical black holes
localized on the Kaluza-Klein circle. Here we plot the slope of
the first part of the branch using the analytical results for
small black holes \cite{Harmark:2003yz}.

It is the region $1/3 < n \le 3$ that contains the bubble-black
hole sequences. The top point of the curves in this region is the
static bubble solution located at $(\mu,n)=(1/2,3)$. All $(p,q)$
solutions
 start out at this point and approach the uniform black string
branch at $n=1/3$ as $\mu \to \infty$. The lowest lying of these
curves is the $(p,q)=(1,1)$ solution with a single black ring
(with horizon topology $S^3 \times S^1$) supported by a single
Kaluza-Klein bubble. There are also solutions lying ``inside'' the
wedge bounded by the $(1,1)$ solution and the $(1,2)_\mathfrak{t}$
solution with one bubble balancing two equal size black rings with
equal temperatures. The solutions in this wedge are all $(1,2)$
solutions where the two black rings are at different temperatures.
We note that for any given $\mu > 1/2$, the solutions in the wedge
provide a continuous set of bubble-black hole sequences with the
same mass. These solutions can be told apart since they have
different values of $n$. But even specifying both $\mu$ and $n$
does not give a unique solution. Though not visible in the figure,
the top curve crosses into the wedge of $(1,2)$ solutions at the
value $\mu = 117/10$. This curve is the one parameter family
$(p,q)=(2,1)$ describing two equal-size Kaluza-Klein bubbles
supporting a black tuboid, a black hole with horizon topology $S^2
\times S^1 \times S^1$.

We conclude in Section \ref{s:concl} with a discussion of our
results and outlook for future developments. An appendix treats
details of the analysis for the $(p,q)=(2,3)$ solution.

\subsubsection*{Notation}
Throughout the paper we use $d$ to denote the space-time dimension
of the Minkowski part of the metric. The space-times are
asymptotically $\CM^d \times S^1$, and we use $D=d+1$ for the
dimension of the full space-time.

\section{Review of the $(\mu,n)$ phase diagram}
\label{s:phase}

In \cite{Harmark:2003dg,Harmark:2003eg} a program was set forth to
categorize --- in higher-dimensional General Relativity --- all
static solutions of the vacuum Einstein equations that asymptote
to $\CM^d \times S^1$ for $d \geq 4$, with $\CM^d$ being the
$d$-dimensional Minkowski space-time. When an event horizon is
present we call these solutions \emph{static neutral Kaluza-Klein
black holes}, since $\CM^d \times S^1$ is a Kaluza-Klein type
space-time. In this section we review the ideas and results of
\cite{Harmark:2003dg,Harmark:2003eg} that are important for this
paper.

The general idea is to define a ``phase diagram'' and plot in it
the branches of different types of static Kaluza-Klein black
holes. The physical parameters used in defining such a phase
diagram should be measurable at asymptotic infinity. In
\cite{Harmark:2003dg,Kol:2003if,Harmark:2003eg} it was suggested
that besides the proper length $L$ of the $S^1$ at infinity, the
relevant physical parameters are the mass $M$ and the tension
$\CT$, which can be defined for any solution asymptoting to $\CM^d
\times S^1$. The tension $\CT$ was defined in
\cite{Traschen:2001pb,Townsend:2001rg,Harmark:2003dg,Kol:2003if,Harmark:2004ch}.

Let the $S^1$ be parameterized by the coordinate $\phi$, which we
take to have period $L$. Define $(t,x^1,...,x^{d-1})$ to be
Cartesian coordinates for $\CM^d$, so that the radial coordinate
is $\rho = \sqrt{(x^1)^2 + \cdots + (x^{d-1})^2}$. We consider
black holes localized in $\R^{d-1}$, so the asymptotic region is
defined by $\rho \rightarrow \infty$, and we write the asymptotic
behavior of the metric components $g_{tt}$ and $g_{\phi\phi}$ as
\begin{equation}
\label{gttzz} g_{tt} \simeq -1 + \frac{c_t}{\rho^{d-3}} \spa ~~~~
g_{\phi\phi} \simeq 1 + \frac{c_\phi}{\rho^{d-3}}
\end{equation}
for $\rho \rightarrow \infty$. Note that we have chosen $\phi$
such that the period $L$ of $\phi$ is the proper length of the
$S^1$ at infinity. It was shown in
\cite{Harmark:2003dg,Kol:2003if,Harmark:2004ch} that the ADM mass
$M$ and the tension $\CT$ along the $\phi$-direction
can be computed from the asymptotic metric as%
\begin{equation}
\label{MT} M = \frac{\Omega_{d-2} L}{16 \pi G_{\rm N}} \left[
(d-2) c_t - c_\phi \right] \spa ~~~~ \CT = \frac{\Omega_{d-2}}{16
\pi G_{\rm N}} \left[ c_t - (d-2) c_\phi \right] \, ,
\end{equation}
where $\Omega_m = 2 \pi^{\frac{m+1}{2}} / \Gamma (\frac{m+1}{2})$
is the surface volume of the $m$-sphere.

Since for given $L$ we wish to compare solutions with the same
mass, it is natural to normalize the mass with respect to $L$. We
then work with the rescaled mass $\mu$ and the  relative tension
$n$, which are dimensionless quantities defined as
\begin{equation}
\label{mun} \mu = \frac{16 \pi G_{\rm N}}{L^{d-2}} M =
\frac{\Omega_{d-2}}{L^{d-3}} \left[ (d-2) c_t - c_\phi \right]
\spa~~ n = \frac{\CT L}{M} = \frac{c_t - (d-2) c_\phi}{(d-2)c_t -
c_\phi} \, .
\end{equation}

Not all values of $\mu$ and $n$ correspond to physically
reasonable solutions. We have $\mu \geq 0$ from the Weak Energy
Condition, and $n$ must satisfy the bounds \cite{Harmark:2003dg}
\begin{equation}
\label{nbound} 0 \leq n \leq d-2 \, .
\end{equation}
The lower bound comes from positivity of the tension
\cite{Traschen:2003jm,Shiromizu:2003gc}. The upper bound is due to
the Strong Energy Condition. There is a more physical way to
understand the upper bound: for a solution with $n < d-2$ the
gravitational force on a test particle at infinity is attractive,
while it would be repulsive if $n > d-2$.

The aim of the work initiated in
\cite{Harmark:2003dg,Harmark:2003eg} is to plot all static vacuum
solutions that asymptote to $\CM^d \times S^1$ in the $(\mu,n)$
phase diagram. In other words, we categorize all these solutions
according to their physical parameters measured at asymptotic
infinity. In this way one can get an overview of the possible
solutions and one can for example see for a given mass $\mu$ what
possible branches of solutions are available.

Previously only solutions with $0 \leq n \leq 1/(d-2)$ have been
considered for the $(\mu,n)$ phase diagram. We focus on that part
of the phase diagram in the remainder of this section; the rest of
the paper will discuss solutions in the $1/(d-2) < n \leq d-2$
region of the phase diagram.

According to our present knowledge, the solutions with $0 \leq n
\leq 1/(d-2)$ all have a local $SO(d-1)$ symmetry and two possible
topologies for the event horizons: 1) $S^{d-1}$, which we call
black holes on cylinders, and 2) $S^{d-2} \times S^1$, which we
call black strings.

There are three known branches of solutions:
\begin{itemize}
\item {\bf Uniform black string branch}. The metric for the
uniform black string is constructed as the $d$-dimensional
Schwarzschild metric times a circle:
\begin{equation}
\label{unblstr} ds^2 = - \left( 1 -
\frac{\rho_0^{d-3}}{\rho^{d-3}} \right) dt^2 + \left( 1 -
\frac{\rho_0^{d-3}}{\rho^{d-3}}\right)^{-1} d\rho^2 + \rho^2
d\Omega_{d-2}^2 + d\phi^2 \ .
\end{equation}
We note that $c_\phi=0$, so by \refeq{mun} a uniform black string
has $n=1/(d-2)$. Gregory and Laflamme
\cite{Gregory:1993vy,Gregory:1994bj} discovered that the uniform
string is classically unstable for $\mu < \mu_{\rm GL}$, and the
critical mass $\mu_{\rm GL}$ can be obtained numerically for each
dimension $d$. In Table \ref{tabmuc} we list the explicit values
of $\mu_{\rm GL}$ for $d=4,\dots,14$. The uniform black strings
are believed to be classically stable for $\mu > \mu_{\rm GL}$.
\item {\bf The non-uniform black string branch}. This branch was
discovered in \cite{Gregory:1988nb,Gubser:2001ac}. For $d=4$ the
beginning of the branch was studied in \cite{Gubser:2001ac}, and
for $d=5$ a large piece of the branch was found numerically by
Wiseman \cite{Wiseman:2002zc}. Recently, Sorkin
\cite{Sorkin:2004qq} studied the non-uniform strings for general
dimensions $d$. The non-uniform string branch starts at $\mu =
\mu_{\rm GL}$ with $n=1/(d-2)$ in the uniform string branch and
then it has decreasing $n$ and increasing $\mu$ for $d \leq 12$.
Sorkin \cite{Sorkin:2004qq} found that for $d > 12$ it has instead
decreasing $n$ and $\mu$, which means that we have a critical
dimension at $d=12$ where the physics of the non-uniform string
branch changes. For $d \le 12$ the non-uniform black string has
lower entropy than the uniform black string with the same mass
$\mu$, while for $d > 12$ the non-uniform string has the higher
entropy \cite{Sorkin:2004qq}.
\item {\bf The black hole on cylinder branch}. This branch has
been studied analytically in
\cite{Harmark:2002tr,Harmark:2003yz,Kol:2003if,Harmark:2003eg,%
Gorbonos:2004uc} (see also \cite{Harmark:2003dg}) and numerically
for $d=4$ in \cite{Sorkin:2003ka} and for $d=5$ in
\cite{Kudoh:2003ki}. The branch starts in $(\mu,n)=(0,0)$ and has
increasing $n$ and $\mu$. The first part of the branch is known
analytically \cite{Harmark:2003yz}.
\end{itemize}

All three branches mentioned above can be described with the same
ansatz for the metric. This ansatz was proposed in
\cite{Harmark:2002tr,Harmark:2003fz}, and proven
in \cite{Wiseman:2002ti,Harmark:2003eg}.

\begin{figure}[ht]
\centerline{\epsfig{file=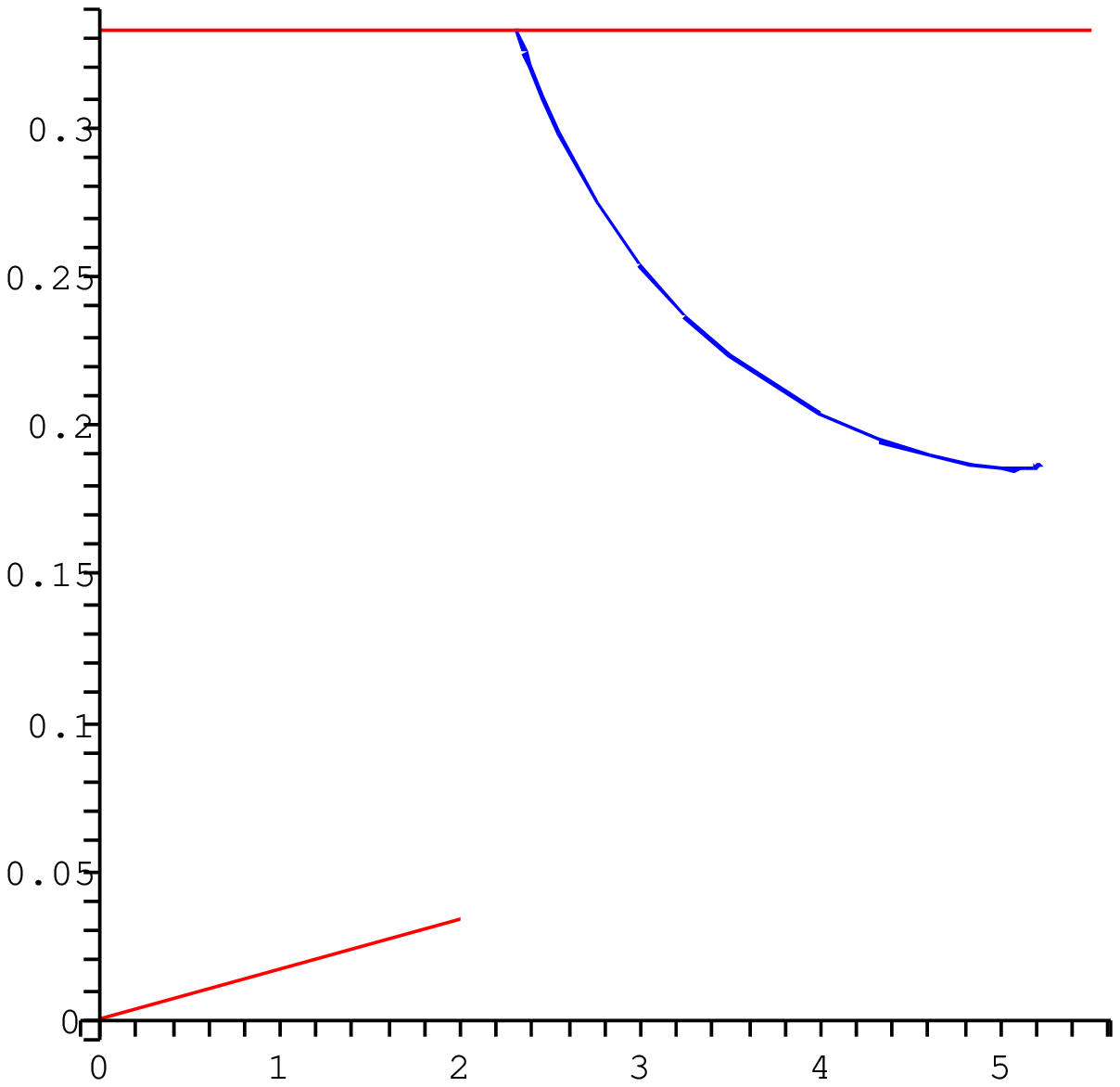,width=7 cm,height=6cm} \hskip
.5cm \epsfig{file=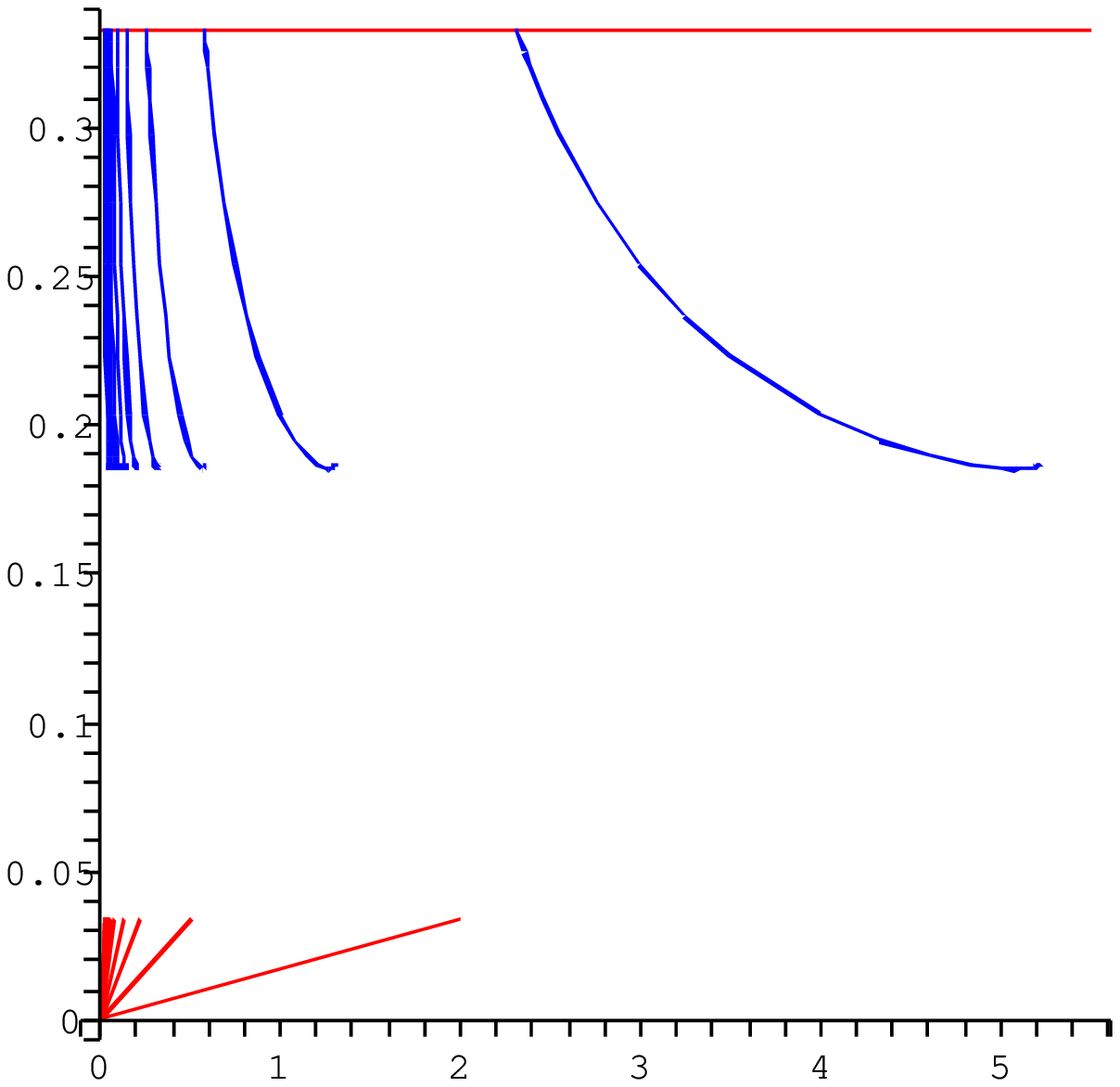,width=7 cm,height=6cm}}
 \caption{Phase diagram for $d=5$ with $n \leq 1/3$.
To the left we have the diagram without copies while we included
the copies to the right.} \label{fig_neut}
\begin{picture}(0,0)(0,0)
\put(180,60){\Large $\mu$} \put(398,60){\Large $\mu$}
\put(5,205){\Large $n$} \put(220,205){\Large $n$}
\put(60,215){\footnotesize uniform black string}
\put(275,215){\footnotesize uniform black string}
\put(75,155){\footnotesize non-uniform}
\put(75,145){\footnotesize black string}
\put(325,140){\footnotesize non-uniform}
\put(325,130){\footnotesize black string}
\put(90,90){\footnotesize black hole}
\put(305,90){\footnotesize black hole}
\put(250,140){\footnotesize copies}
\put(250,98){\footnotesize copies}
\end{picture}
\end{figure}

In Figure \ref{fig_neut} we display for $d=5$ the known solutions
with $n \leq 1/3$ in the $(\mu,n)$ phase diagram. The non-uniform
black string branch was drawn in \cite{Harmark:2003dg} using the
data of \cite{Wiseman:2002zc}. We included in the right part of
Figure \ref{fig_neut} the  copies of the black hole on cylinder
branch and the non-uniform string branch
\cite{Horowitz:2002dc,Harmark:2003eg}.%
\footnote{ We find the $k$'th copy of that solution by
``repeating'' the solution $k$ times on the circle
\cite{Horowitz:2002dc,Harmark:2003eg}. Furthermore, if the
original solution is in the point $(\mu,n)$ of the phase diagram,
then the $k$'th copy will be in the point $(\mu',n') =
(\mu/k^{d-3},n)$~\cite{Harmark:2003eg}.}

For the temperature and entropy we use the rescaled temperature
$\mathfrak{t}$ and entropy $\mathfrak{s}$ defined by
\begin{equation}
\label{tsdef} \mathfrak{t} = L T \spa ~~~~ \mathfrak{s} = \frac{16
\pi G_{\rm N}}{L^{d-1}} S \, .
\end{equation}
The dimensionless quantities
 $\mu$, $n$, $\mathfrak{t}$, and $\mathfrak{s}$ are connected
through the Smarr formula \cite{Harmark:2003dg,Kol:2003if}
\begin{equation}
\label{smarr1} (d-1) \mathfrak{t} \, \mathfrak{s} = (d-2-n) \mu \,
.
\end{equation}
We also have the first law of thermodynamics
\cite{Harmark:2003dg,Kol:2003if}
\begin{equation}
\label{first1} \delta \mu =  \mathfrak{t} \, \delta \mathfrak{s}
\, .
\end{equation}
Using the two relations \eqref{smarr1} and \eqref{first1} one can
pick a curve in the $(\mu,n)$ phase diagram and integrate the
thermodynamics just from the points in the $(\mu,n)$ diagram alone
\cite{Harmark:2003dg}. In this way, the phase diagram contains the
essential information about the thermodynamics of each branch.
Note that for this argument we assumed that there was only one
black hole present, so that there was only one temperature in the
system. Later in this paper we present solutions with several
disconnected event horizons, which can have different
temperatures.

\section{The static Kaluza-Klein bubble}
\label{bubble}

Static Kaluza-Klein bubbles belong to the class of the solutions
we wish to categorize in Kaluza-Klein theory, and they turn out to
play a crucial role for the solutions in the $1/(d-2) < n \le d-2$
region of the $(\mu,n)$ phase diagram. In this section, we
introduce Kaluza-Klein bubbles and describe in detail the
properties of static Kaluza-Klein bubbles.

Kaluza-Klein bubbles were discovered by Witten in
\cite{Witten:1982gj}, where it was explained that (in the absence
of fundamental fermions) the Kaluza-Klein vacuum $\CM^4 \times
S^1$ is semi-classically unstable to creation of expanding
Kaluza-Klein bubbles. The expanding Kaluza-Klein bubble solutions
can be obtained by a double Wick rotation of the 5D Schwarzschild
solution and the resulting space-time is asymptotically $\CM^4
\times S^1$. The Kaluza-Klein bubble is located at the place where
the $S^1$ direction smoothly closes off: this defines a minimal
two-sphere in the space-time, i.e. a ``bubble of nothing''. In
this expanding Kaluza-Klein bubble solution the minimal two-sphere
expands until all of the space-time is gone. Since it is not a
static solution, the expanding Kaluza-Klein bubble is not in the
class of solutions we consider for the $(\mu,n)$ phase diagram. We
comment further on the expanding Kaluza-Klein bubble below.

However, there exist also static Kaluza-Klein bubbles. A
$(d+1)$-dimensional static Kaluza-Klein bubble solution can be
constructed by adding a trivial time-direction to the Euclidean
section of a $d$-dimensional Schwarzschild black hole. This gives
the metric
\begin{equation}
\label{statbub} ds^2 = - dt^2 + \left( 1 -
\frac{R^{d-3}}{\rho^{d-3}} \right) d\phi^2 + \left( 1 -
\frac{R^{d-3}}{\rho^{d-3}}\right)^{-1} d\rho^2 + \rho^2
d\Omega_{d-2}^2 \  .
\end{equation}
Clearly, this solution is static. It has a minimal $(d-2)$-sphere
of radius $R$ located at $\rho=R$. To avoid a conical singularity
at $\rho=R$, we need $\phi$ to be periodic with period
\begin{equation}
\label{LandR} L = \frac{4\pi R}{d-3} \  .
\end{equation}
With this choice of $L$ the Kaluza-Klein circle $\phi$ shrinks
smoothly to zero as $\rho \to R$, and the location $\rho=R$
becomes a point with respect to the $\rho$ and $\phi$ directions.
Around the location $\rho = R$ the space-time is locally of the
form $\R \times \R^2 \times S^{d-2}$, where $\R$ is the time, the
two-plane is parameterized by $\rho$ and $\phi$, and the $S^{d-2}$
becomes the minimal $(d-2)$-sphere at $\rho=R$. Clearly, there are
no boundaries at $\rho=R$ and no space-time for $\rho < R$, which
is the reason for calling the Kaluza-Klein bubble a ``bubble of
nothing''. Note that the bubble space-time is a regular manifold
with topology $\R \times \R^2 \times S^{d-2}$, where the $S^{d-2}$
is non-contractible.

Due to the periodicity of $\phi$ we see that the static bubble
\eqref{statbub} asymptotically (i.e. for $\rho \rightarrow
\infty$) goes to $\CM^d \times S^1$. Thus, it belongs to the class
of solutions we are interested in, i.e. static pure gravity
solutions that asymptote to $\CM^d \times S^1$, and we can plot
the solution in the $(\mu,n)$ phase diagram. Using \eqref{gttzz}
we read off $c_t = 0$ and $c_\phi = - R^{d-3}$. {}From \eqref{mun}
and \eqref{LandR} we then find the dimensionless mass $\mu$ and
the relative binding energy $n$ to be
\begin{equation}
\label{munbub} \mu = \mu_{\rm b} \equiv \Omega_{d-2} \left(
\frac{d-3}{4\pi} \right)^{d-3} \spa ~~~~~~~ n = d-2 \, ,
\end{equation}
where we named this special value of $\mu$ as $\mu_{\rm b}$. Note
that the static bubble only exists at one point in the $(\mu,n)$
diagram. This is because the relation between $L$ and $R$ in
\eqref{LandR} means that once $L$ is chosen then all parameters
are fixed. We note that since the static Kaluza-Klein bubble has
$n=d-2$, it precisely saturates the upper bound in \eqref{nbound}.
Indeed, the Newtonian gravitational force on a test particle at
infinity is zero.

The static Kaluza-Klein bubble is known to be classically
unstable. This can be seen from the fact that the static bubble is
the Euclidean section of the Schwarzschild black hole times a
trivial time direction. The Euclidean flat space $\R^3\times S^1$
(hot flat space) is semi-classically unstable to nucleation of
Schwarzschild black holes. This was shown by Gross, Perry, and
Yaffe \cite{Gross:1982cv}, who found that the Euclidean
Lichnerowicz operator $\Delta_\rom{E}$ for the Euclidean section
of the four dimensional Schwarzschild solution with mass $M$ has a
negative eigenvalue: $\Delta_\rom{E} u_{ab} = \lambda u_{ab}$ with
$\lambda=-0.19 (GM)^{-2}$. The Lichnerowicz equation for the
perturbations of the Lorentzian static bubble space-time is
$\Delta_\rom{L} h_{ab}=0$ (in the transverse traceless gauge), so
taking the ansatz $h_{ab} = u_{ab} \, e^{i\Omega t}$, the
Lichnerowicz equation requires $\Omega^2=\lambda$, ie.\
$\Omega=\pm i \sqrt{-\lambda}$. So this is an instability mode of
the static bubble, and the perturbation causes the bubble to
either expand or collapse exponentially fast.

That the static Kaluza-Klein bubble is classically unstable poses
the question: what does it decay to? We discuss this in the
following.

Note first that the static bubble is massive, while the expanding
Witten bubble obtained as the double Wick rotation of the
Schwarzschild black hole is massless. Therefore the instability of
the static bubble does not connect it directly to the Witten
bubble. Furthermore, if $L$ is the size of the circle at infinity,
the minimal radius of the Witten bubble is $\CR =
\frac{(d-2)}{4\pi} L$, and the radius of the static bubble is $R =
\frac{d-3}{4\pi} L$. Therefore we have \bea R = \frac{d-3}{d-2}
\CR \, , \eea and this means that for given size of the
Kaluza-Klein circle at infinity, the radius $R$ of the static
Kaluza-Klein bubble is smaller than the minimal radius $\CR$ of
the expanding Kaluza-Klein bubble.

In the following, we discuss initial data describing massive
bubbles that are initially expanding or collapsing, and we shall
see that for given size of the Kaluza-Klein circle the initially
expanding bubbles have larger radii than the static bubble, and
collapsing bubbles have smaller radii. Presumably the collapse of
a massive bubble results in a black hole or black string, so we
shall also compare the mass $\mu_{\rm b}$ of the Kaluza-Klein
bubble to the Gregory-Laflamme mass $\mu_\rom{GL}$.

\subsubsection*{Initial data for massive expanding and collapsing bubbles}

We now consider a subclass of a more general family of initial
data for five-dimensional vacuum bubbles found by Brill and
Horowitz \cite{Brill:1991qe}. The metric on the initial surface is
\begin{equation}
 \label{bubdata}
 ds^2 = U(\rho) d\phi^2 + U(\rho)^{-1} d\rho^2 + \rho^2 d\Omega_2^2
\end{equation}
with $U(\rho) = 1- 2m/\rho - b/\rho^2$ for arbitrary parameters
$m$ and $b$ satisfying $-m^2<b<\infty$. The bubble is located at
the positive zero $\rho_+$ of $U$, $\rho_+ = m + \sqrt{m^2+b}$,
and by fixing the period of $\phi$ to be $L= 2\pi
\rho_+^2/(\rho_+-m)$ we avoid a conical singularity at
$\rho=\rho_+$. The ADM mass is $G_{\rm N} M=\frac{1}{2} m L$, so
\bea
  \mu = \frac{4 m \sqrt{m^2 + b}}{(m+\sqrt{m^2 +b})^2} \, .
\eea  In the following we assume that $m>0$. In this case we find
$0 < \mu \le 1$. Since this initial data exists for masses $\mu$
less than or equal to static bubble mass $\mu_{\rm b}=1$, it is not
inconceivable that the evolution of this initial data can guide us
about the possible endstates of the decay of the static
Kaluza-Klein bubble.

For the initial data \refeq{bubdata}, Corley and Jacobson
\cite{Corley:1994mc} showed that depending on the mass of the
bubble and its size (or alternatively, the size of the $S^1$ at
infinity), the bubbles are initially going to expand or collapse
according to the initial acceleration of the bubble area which is
given by \bea
  \label{Addot}
  \ddot{A} = 8 \pi \left( 1- \frac{2m}{\rho_+}\right)
\eea (the derivatives are with respect to proper time). The
relation \refeq{Addot} shows that for $b>0$ the bubble is
initially expanding, but for $-m^2<b<0$ it is initially
collapsing. We note that for $b=0$, we have $\mu=1$ and
$\rho_+=2m$, so in this case the initial data describes an
initially non-accelerating bubble. It is clear that
\refeq{bubdata} with $b=0$ is initial data for the static
Kaluza-Klein bubble.

It may appear surprising that there are expanding as well as
collapsing bubbles with the same mass $\mu$. The difference relies
not on the mass, but in the radius of the bubble compared to the
size of the circle at infinity. Consider for the initial data
\refeq{bubdata} the size of the bubble relative to the size $L$ of
the Kaluza-Klein circle:
\begin{equation}
  \label{bubsize}
  2\pi \frac{\rho_+}{L} = 1-\frac{m}{\rho_+} \, .
\end{equation}
For the static bubble ($b=0$), we have $\rho_+ = 2m$. Initially
expanding bubbles ($b>0$) have $\rho_+>2m$ and initially
collapsing bubbles ($-m^2<b<0$) have $\rho_+<2m$. So by
\refeq{bubsize} we see that bubbles that are bigger than the
static bubble are going to expand and bubbles that are smaller are
going to collapse.

In an interesting paper \cite{Sarbach:2003dz}, Lehner and Sarbach
recently studied numerically the evolution of the initial data
\refeq{bubdata} (as well as more general bubbles). They found that
initially expanding bubbles continue to expand, and the smaller
the mass of the bubble, the more rapidly the area of the bubble
grows as a function of proper time. By studying such massive
expanding bubbles, one can hope to learn about the endstate of the
instability of perturbations of the static bubble.

The instability can also cause the static Kaluza-Klein bubble to
collapse and thereby decay to another static solution, presumably
one
with an event horizon.%
\footnote{For vacuum solutions, the initial data for an initially
collapsing bubble requires a positive mass. With strong gauge
fields, negative mass bubbles can also initially collapse, but
after a while the collapse is halted and the bubbles bounce and
begin to expand again \cite{Sarbach:2003dz}.} The numerical
results \cite{Sarbach:2003dz} show that for an initially
collapsing massive bubble, the collapse continues until at some
point an apparent horizon forms. A comparison of curvature
invariants suggests that the resulting black hole is a uniform
black string \cite{Sarbach:2003dz}.

Is the endstate of the instability of the static Kaluza-Klein
bubble then a black string? One would expect the endpoint of a
classical evolution to be a classically stable configuration, so
since the uniform black string is classically unstable for
$\mu<\mu_\rom{GL}$ we must compare the mass of bubble $\mu_{\rm
b}$ to the Gregory-Laflamme mass $\mu_\rom{GL}$.

\subsubsection*{Comparison of Kaluza-Klein bubble mass
and Gregory-Laflamme mass}

The mass $\mu_{\rm b}$ of the static Kaluza-Klein bubble is given
in \refeq{munbub}, and for $4 \leq d \leq 9$ we can use the
Gregory-Laflamme masses $\mu_\rom{GL}$ computed in
\cite{Gregory:1993vy,Gregory:1994bj}. In Table \ref{tabmuc} we
list the approximate values of $\mu_{\rm GL}$ and $\mu_{\rm b}$.
We see that for $4 \leq d \leq 9$ the static bubble mass $\mu_{\rm
b}$ is always lower than the Gregory-Laflamme mass $\mu_{\rm GL}$,
so in the microcanonical ensemble the static Kaluza-Klein bubble
cannot decay to the uniform black string branch but must decay to
another branch of solutions, possibly the black hole on cylinder
branch.

\begin{table}[h]
\begin{center}
\begin{tabular}{|c||c|c|c|c|c|c|c|c|c|c|c|}
\hline $d$ & $4$ & $5$ & $6$ & $7$ & $8$ & $9$ & $10$ & $11$ &
$12$ & $13$ & $14$
\\ \hline
$\mu_{\rm GL}$ & $3.52$ & $2.31$ & $1.74$ & $1.19$ & $0.79$ &
$0.55$ &
$0.37$ & $0.26$ & $0.18$ & $0.12$ & $0.08$ \\
\hline $\mu_{\rm b}$ & $1$ & $0.5$ & $0.36$ & $0.32$ & $0.33$ &
$0.38$ & $0.49$ &
$0.69$ & $1.03 $ & $1.63$ & $2.74$ \\
\hline
\end{tabular}
\caption{The mass $\mu_{\rm b}$ of the static Kaluza-Klein bubble
compared to the critical masses $\mu_{\rm GL}$ for the
Gregory-Laflamme instability. For $d\le 9$, we use $\mu_{\rm GL}$
from \cite{Gregory:1993vy,Gregory:1994bj}, and for $d>9$ we use
the result of \cite{Sorkin:2004qq}.
 \label{tabmuc}}
\end{center}
\end{table}

\begin{figure}[ht]
\centerline{\epsfig{file=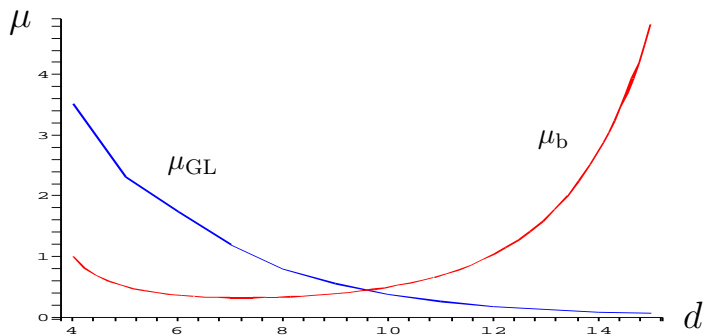,width=10cm,height=5cm}}
\caption{Plot of the static Kaluza-Klein bubble mass $\mu_{\rm b}$
and the Gregory-Laflamme mass $\mu_{\rm GL}$ versus $d$.}
\label{fig_mub}
\begin{picture}(0,0)(0,0)
\put(335,70){\Large $d$}
\put(80,184){\Large $\mu$}
\put(140,130){$\mu_{\rm GL}$}
\put(280,140){$\mu_{\rm b}$}
\end{picture}
\end{figure}

For higher $d$, we can use the results of \cite{Sorkin:2004qq}.
Here it is found that the Gregory-Laflamme mass is $\mu_{\rm GL} =
16.2 \cdot 0.686^d$ for all $d$ to a good approximation. We have
used this to plot in Figure \ref{fig_mub} the static Kaluza-Klein
bubble mass $\mu_{\rm b}$ versus the Gregory-Laflamme mass
$\mu_{\rm GL}$, and also listed the approximate values for the
dimensionless masses in Table \ref{tabmuc}. We see from Table
\ref{tabmuc} and Figure \ref{fig_mub} that $\mu_{\rm b}$ is
greater than $\mu_{\rm GL}$ for $d > 9$. Therefore, for $d>9$ it
is possible that the endpoint of the classical decay of the static
Kaluza-Klein bubble is a uniform black string.

The fact that we have a critical dimension at $d=9$ ($D=10$) is
interesting in view of the recent results of \cite{Sorkin:2004qq}
showing that the non-uniform black string branch starts having
decreasing $\mu$ as $n$ decreases for $d > 12$, ie.\ with a
critical dimension $d=12$ ($D=13$). Moreover, in Ref.~\cite{Kol:2002xz}
the critical dimension $d=9$ ($D=10$) appeared in studying the
stability of the cone metric, as a model for the black hole-black string
transition. It would therefore be interesting to examine whether there is
any relation between these critical dimensions.

As we have seen, the classical instability of the static
Kaluza-Klein bubble causes the bubble to either expand or
collapse. For five-dimensional Kaluza-Klein space-times, there
exists initial data \cite{Brill:1991qe} for massive bubbles that
are initially expanding or collapsing \cite{Corley:1994mc}, and
numerical studies \cite{Sarbach:2003dz} shows us that there exist
massive expanding bubbles and furthermore the numerical analysis
indicates that contracting massive bubbles collapse to a black
hole with an event horizon.

If the unstable mode of the static bubble preserves the
translational invariance around the $S^1$-direction, the endstate
of a collapsing bubble would be expected to be a uniform black
string. Consider then more general perturbations causing the decay
of the static bubble. For $d>9$ we found $\mu_{\rm b}>
\mu_\rom{GL}$, so here it is possible that the static bubble
decays to a stable uniform black string. Moreover, for $d>12$ the
non-uniform strings have higher entropy than the uniform strings
for a given $\mu$ \cite{Sorkin:2004qq}, so here it is possible
that a non-uniform string can be the endpoint of the bubble decay.

However, for $4\le d \le 9$, we have seen that the mass of the
bubble lies in the range for which the uniform black string is
classically unstable, and therefore we do not expect the uniform
black string to be the endstate of the decay of the static bubble.
It seems therefore plausable that for $4\le d \le 9$ the
instability of the bubble develops inhomogeneities in the
$S^1$-direction so that the likely endstate of the bubble decay is
a black hole localized on the Kaluza-Klein circle (ie.\ a solution
on the black hole on cylinder branch).

\subsubsection*{Summary}
The static Kaluza-Klein bubble is massive and exists at a single
point $(\mu_{\rm b},d-2)$ in the $(\mu,n)$ phase diagram. It is
classically unstable and will either expand or collapse. In the
latter case, the endstate is presumably an object with an event
horizon. For $d\le 9$ we have argued that it cannot be the uniform
black string, but should be whatever is the endstate of the
uniform black string.

It is important to emphasize that the static Kaluza-Klein bubble
does not have any event horizon. This also means that it does not
have entropy or temperature (i.e. the temperature is zero).
However, in the following sections we discuss solutions in five and
six dimensions with both
Kaluza-Klein bubbles and event horizons present. In sections
\ref{s:5dKK}-\ref{s:6dKK}, we shall see that in the region
$1/(d-2)<n<d-2$ of the $(\mu,n)$ phase diagram, all known
solutions describe combinations of Kaluza-Klein bubbles and black
hole event horizons, and for each value of $n$ in the range
$1/(d-2)<n<d-2$ there exist continuous families of such solutions.

\section{Generalized Weyl solutions}
\label{s:gweyl}

In this section we review the generalized Weyl solutions. We use
this method in sections \ref{s:5dKK} and \ref{s:6dKK} to find
exact solutions describing sequences of Kaluza-Klein bubbles and
black holes. We examine the asymptotics of  the generalized Weyl
solutions and show how to read off the physical quantities from
the asymptotic metric. Furthermore, as a warm-up to the following
sections, we discuss the uniform black string and the static
Kaluza-Klein bubble metrics in Weyl coordinates.

\subsection{Review of generalized Weyl solutions}
\label{revweyl}

Emparan and Reall showed in \cite{Emparan:2001wk} that for any
$D$-dimensional static space-time with  $D-1$ additional commuting
orthogonal Killing vectors, i.e. with a total of $D-2$ commuting
orthogonal Killing vectors, the metric can be written in the form
\begin{equation}
  ds^2 = - e^{2U_1} dt^2
         + \sum_{a=2}^{D-2} e^{2U_a} d\phi_a^2
         + e^{2\nu}(dr^2 + dz^2) \ .
  \label{genweyl}
\end{equation}
where $U_a = U_a (r,z)$ and $\nu = \nu(r,z)$. For the metric
\refeq{genweyl} to be a solution of pure gravity, i.e. of the
Einstein equations without matter, the potentials $U_a$,
$a=1,\dots,D-2$, must obey
\begin{equation}
\label{lapl} \left( \frac{\partial^2}{\partial r^2} + \frac{1}{r}
\frac{\partial}{\partial r} + \frac{\partial^2}{\partial z^2}
\right) U_a = 0 \ ,
\end{equation}
and are therefore axisymmetric solutions of Laplace's equation in
a three-dimensional flat Euclidean space with metric
\begin{equation}
  dr^2 + r^2 d\phi^2 + dz^2 \ .
  \label{flat}
\end{equation}
The $U_a$ potentials are furthermore required to obey the
constraint
\begin{equation}
\label{constr} \sum_{a=1}^{D-2} U_a = \log r \ .
\end{equation}
Given the potentials $U_a$, $a=1,...,D-2$, the function
$\nu=\nu(r,z)$ is determined, up to a constant, by the integrable
system of differential equations
\begin{equation}
\label{nueqs}
\partial_r \nu = - \frac{1}{2r} + \frac{r}{2}
\sum_{a=1}^{D-2} \left[ (\partial_r U_a)^2 - (\partial_z U_a)^2
\right] \spa
\partial_z \nu = r \sum_{a=1}^{D-2} \partial_r U_a \partial_z U_a \ .
\end{equation}
Therefore, we can find solutions to the Einstein equations by
first solving the Laplace equations \eqref{lapl} for the
potentials $U_a$, $a=1,...,D-2$, subject to the constraint
\eqref{constr}, and subsequently solve \eqref{nueqs} to find
$\nu$.

In four dimensions this method of finding static axisymmetric
solutions was pioneered by Weyl in \cite{Weyl:1917}. Emparan and
Reall then generalized Weyl's results to higher dimensions in
\cite{Emparan:2001wk}. We refer therefore to solutions of the kind
described above as {\sl generalized Weyl solutions}.

In general, sources for the $U_a$ potentials at $r > 0$ lead to
naked singularities, so we consider only sources at $r=0$. The
location $r=0$ corresponds to a straight line in the unphysical
three-dimensional space with metric \eqref{flat} mentioned above.
The constraint \eqref{constr} then means that the total sum of the
potentials is equivalent to the potential of an infinitely long
rod of zero thickness lying along the $z$-axis at $r=0$. In the
unphysical three-dimensional space \eqref{flat}, this infinite rod
has mass $1/2$ per unit length, with Newton's constant in this
space set to one. We demand furthermore that for a given value of
$z$ there is only one rod, except in isolated points. Thus, we
build solutions by combining rods of mass $1/2$ per unit length
for the different $U_a$ potentials under the restrictions that the
rods do not overlap and that they add up to the infinite rod.

We use here and in the following the notation that $[z_1,z_2]$
denotes a rod from $z=z_1$ to $z=z_2$.

We now review which rod sources to use for the potentials $U_a$ to
write familiar static axisymmetric solutions in the generalized
Weyl form:
\begin{figure}[t]
   \centerline{\epsfig{file=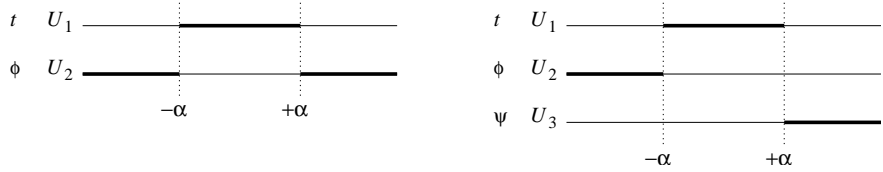,height=2.2cm}}
   \caption{Rod configurations for the four-dimensional (left) and
   five-dimensional (right) Schwarzschild black holes. The ends of the
   figures are supposed to represent $z=\pm\infty$, so that the
   semi-infinite rods for $U_2$ and $U_3$ extend all the way out to
   infinity.}
   \label{rodpic0}
\end{figure}
\begin{itemize}
\item $D=4$ Minkowski space. No source for the $U_1$-potential of
the $t$-direction, but an infinite rod $[-\infty,\infty]$ for the
potential $U_2$ of the $\phi_2$-direction.%
\footnote{Minkowski space can also be constructed from other rod
configurations, see \cite{Emparan:2001wk}.}

\item A $D=4$ Schwarzschild black hole. A finite rod
$[-\alpha,\alpha]$ for the potential $U_1$ of the $t$-direction,
and two semi-infinite rods $[-\infty,-\alpha]$ and
$[\alpha,\infty]$ for the potential $U_2$ of the
$\phi_2$-direction. This is illustrated in the left part of Figure
\ref{rodpic0}. Here $\alpha = G_{\rm N} M$ with $M$ being the mass
of the black hole and $G_{\rm N}$ the four-dimensional Newton's
constant.

\item $D=5$ Minkowski space. Two semi-infinite rods, one rod
$[-\infty,0]$ for the potential $U_2$ in the $\phi_2$-direction,
and the other rod $[0,\infty]$ for the potential $U_3$ for the
$\phi_3$-direction. No sources for the potential $U_1$.

\item A $D=5$ Schwarzschild black hole. The rod configuration is
illustrated in the right part of Figure \ref{rodpic0}. It consists
of a finite rod $[-\alpha,\alpha]$ for the potential $U_1$ in the
$t$-direction, a semi-infinite rod $[-\infty,-\alpha]$ for $U_2$
of $\phi_2$-direction, and another semi-infinite rod
$[\alpha,\infty]$ for $U_3$ of the $\phi_3$-direction. Here
$\alpha = \frac{2}{3\pi} G_{\rm N} M$ with $M$ being the mass of
the black hole and $G_{\rm N}$ the five-dimensional Newton's
constant.
\end{itemize}
Furthermore, we review in detail particular solutions that are
crucial to this paper in Section \ref{revkal}.

As will become clearer in the following we have, for our purposes
at least, the rule of thumb that a semi-infinite or infinite rod
gives rise to a rotational axis (so that the corresponding
coordinate becomes an angle in the metric), a finite rod in the
time-direction $t$ gives rise to an event horizon, while a finite
rod in the spatial directions results in a static Kaluza-Klein
bubble.

\subsection{Kaluza-Klein space-times as generalized Weyl solutions}
\label{revkal} In this section we show how we can write the five-
and six-dimensional Kaluza-Klein space-times $\CM^4 \times S^1$
and $\CM^5 \times S^1$ as generalized Weyl solutions, and we
explain how to read off the physical quantities for generalized
Weyl solutions asymptoting to these space-times. We furthermore
describe the uniform black string branch and the static
Kaluza-Klein bubbles in five and six dimensions, since this
clarifies our use of the generalized Weyl solution technique, and
also since these solutions will be the building-blocks of the
solutions presented below. We begin in five dimensions and then
move on to six dimensions.

\subsubsection*{Five-dimensional Kaluza-Klein space-time $\CM^4 \times S^1$
and asymptoting solutions}

For $D=5$ we have the generalized Weyl ansatz Eq.~\eqref{genweyl}
which we write as
\begin{equation}
\label{weylansD5}
  ds^2 = - e^{2U_1} dt^2 + e^{2U_2} d\phi^2 + e^{2U_3} d\psi^2
         + e^{2\nu}(dr^2 + dz^2) \ .
\end{equation}
Notice that we have renamed $\phi_2= \phi$ and $\phi_3 = \psi$.

We begin by describing the $D=5$ Kaluza-Klein space-time $\CM^4
\times S^1$ as a generalized Weyl solution. This corresponds to
the potentials
\begin{equation}
\label{flat5D} e^{2U_1} = 1 \spa e^{2U_2} = 1 \spa e^{2U_3} = r^2
\ .
\end{equation}
We see that this is an infinitely long rod $[-\infty,\infty]$ for
the $U_3$ potential, i.e.~in the $\psi$-direction. Note that
$e^{2\nu} = 1$ as can be checked from \eqref{nueqs}. Making the
coordinate transformation
\begin{equation}
r = \rho \sin \theta \spa z = \rho \cos \theta \ ,
\end{equation}
we get four-dimensional Minkowski-space in spherical coordinates
times a circle
\begin{equation}
ds^2 = - dt^2 + d\rho^2 + \rho^2 \left( d\theta^2 + \sin^2 \theta
d\psi^2  \right) + d\phi^2 \ .
\end{equation}
We see that $\phi$ is the circle direction which we take to be
periodic with period $L$.

We now consider generalized Weyl solutions that asymptote to the
five-dimensional Kaluza-Klein space-time $\CM^4 \times S^1$. The
asymptotic region of a generalized Weyl solution is the region
$\sqrt{r^2 + z^2} \rightarrow \infty$.

In Section \ref{s:phase} we explained how to read off the rescaled
mass $\mu$ and the relative tension $n$ for static solutions
asymptoting to the five-dimensional Kaluza-Klein space-times
$\CM^4 \times S^1$. From \eqref{mun} we see that we need to read
off $c_t$, $c_\phi$, and $L$ in order to find $\mu$ and $n$. While
the circumference $L$ is clearly the period of $\phi$ we can read
off $c_t$ and $c_\phi$ from the potentials $U_1$ and $U_2$ as
\begin{equation}
\label{asyD51} e^{2U_1} = 1 - \frac{c_t}{\sqrt{r^2+z^2}} + \CO
\left( \frac{1}{r^2+z^2} \right) \spa e^{2U_2} = 1 +
\frac{c_\phi}{\sqrt{r^2+z^2}} + \CO \left( \frac{1}{r^2+z^2}
\right)  \ ,
\end{equation}
for $\sqrt{r^2 + z^2} \rightarrow \infty$. Using this in
\eqref{mun} with $d=4$ then gives $\mu$ and $n$.

\subsubsection*{Uniform black string and static Kaluza-Klein
bubble in $\CM^4 \times S^1$}

\begin{figure}[t]
   \centerline{\epsfig{file=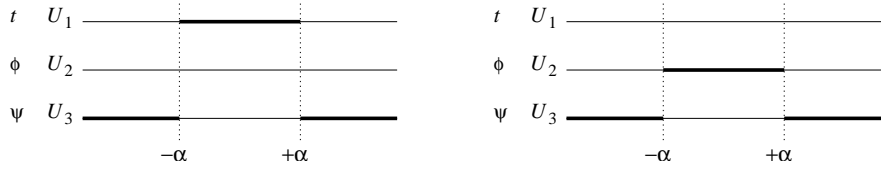,height=2.2cm}}
   \caption{Rod configurations for the uniform black string (left) and
   the static Kaluza-Klein bubble (right) in five dimensions ($d=4$).}
   \label{rodpic1}
\end{figure}

In Section \ref{s:phase} we reviewed the uniform black string in
$\CM^d \times S^1$, in particular the metric was given in
\refeq{unblstr}. The five-dimensional uniform black string metric
($d=4$) can be written in Weyl coordinates by choosing the
potentials
\begin{equation}
\label{bspot} e^{2U_1} = \frac{R_+ - \zeta_+}{R_- - \zeta_-} \spa
e^{2U_2} = 1 \spa e^{2U_3} = ( R_+ + \zeta_+ ) ( R_- - \zeta_- ) \
,
\end{equation}
with
\begin{equation}
\label{zRpm} \zeta_\pm = z \pm \alpha \spa R_\pm = \sqrt{r^2 +
(\zeta_\pm)^2} \ .
\end{equation}
This means that we have a finite rod $[-\alpha,\alpha]$ for the
potential $U_1$ of the $t$-direction, no rod sources for the $U_2$
potential of the $\phi$-direction, and two semi-infinite rods
$[-\infty,-\alpha]$ and $[\alpha,\infty]$ for the potential $U_3$
of the $\psi$-direction. We have depicted this rod configuration
in the left part of Figure \ref{rodpic1}. Using \eqref{asyD51} we
see that $c_t = 2 \alpha$ and $c_\phi = 0$. We have
\begin{equation}
\label{nuD5} e^{2\nu} = \frac{R_+ R_- + \zeta_+ \zeta_- + r^2}{2
R_+ R_-} \frac{R_- - \zeta_-}{R_+ - \zeta_+} \ ,
\end{equation}
as can be checked using \eqref{nueqs}. If we make the coordinate
transformation
\begin{equation}
\label{rztorho} r = \rho \sqrt{1 - \frac{2\alpha}{\rho}}  \sin
\theta \spa z = ( \rho - \alpha ) \cos \theta \ ,
\end{equation}
and set $\rho_0=2\alpha$ we get back the metric \eqref{unblstr}
for the uniform black string with $d=4$, which explicitly exhibits
the $SO(3)$ spherical symmetry.

To get instead a static Kaluza-Klein bubble as a generalized Weyl
solution we can make a double Wick rotation of the $t$ and $\phi$
directions. This gives the potentials
\begin{equation}
\label{bubD5} e^{2U_1} = 1 \spa e^{2U_2} = \frac{R_+ -
\zeta_+}{R_- - \zeta_-} \spa e^{2U_3} = ( R_+ + \zeta_+ ) ( R_- -
\zeta_- ) \ ,
\end{equation}
with $\zeta_\pm$ and $R_\pm$ given by \eqref{zRpm}. This means
that we have two semi-infinite rods $[-\infty,-\alpha]$ and
$[\alpha,\infty]$ in the $\psi$-direction and a finite rod
$[-\alpha,\alpha]$ in the $\phi$-direction. We have depicted this
rod-configuration in the right half of Figure \ref{rodpic1}. The
function $\nu(r,z)$ is again given by \eqref{nuD5}. To avoid a
conical singularity at $r \to 0$ for $|z|<\alpha$, where the orbit
of $\partial_\phi$ shrinks to zero, we need to fix the period of
$\phi$ to be $L = 8 \pi \alpha$. Using \eqref{asyD51} we see that
$c_t = 0$ and $c_\phi = -2\alpha$, so $\mu=1$ and $n=2$ by Eq.
\eqref{mun}. If we make the coordinate transformation
\eqref{rztorho} we get the explicitly spherically symmetric metric
\eqref{statbub} for the static Kaluza-Klein bubble metric with
$d=4$.

\subsubsection*{Six-dimensional Kaluza-Klein space-time $\CM^5 \times S^1$
and asymptoting solutions}

For $D=6$ we have the generalized Weyl ansatz Eq.~\eqref{genweyl}
which we write as
\begin{equation}
  ds^2 = - e^{2U_1} dt^2
         + e^{2U_2} d\phi^2
         + e^{2U_3} d\psi^2
         + e^{2U_4} d\chi^2
         + e^{2\nu}(dr^2 + dz^2) \ .
\end{equation}
Notice that we have renamed $\phi_2= \phi$, $\phi_3 = \psi$ and
$\phi_4 = \chi$.

Written as a generalized Weyl solution the $D=6$ Kaluza-Klein
space-time $\CM^5 \times S^1$ corresponds to the potentials
\begin{equation}
\label{flatD6} e^{2U_1} = 1 \spa e^{2U_2} = 1 \spa e^{2U_3} =
\sqrt{r^2+z^2} + z \spa e^{2U_4} = \sqrt{r^2+z^2} - z \ .
\end{equation}
We see that this is a semi-infinite rod $[-\infty,0]$ sourcing the
potential for the $\psi$-direction and a semi-infinite rod
$[0,\infty]$ for the potential for the $\chi$-direction. We have
\begin{equation}
\label{flatD6nu} e^{2\nu} = \frac{1}{2\sqrt{r^2 + z^2}} \ ,
\end{equation}
as can be verified using \eqref{nueqs}. Making the coordinate
transformation
\begin{equation}
\label{6dtransf} r = \frac{1}{2} \rho^2 \sin 2\theta \spa z =
\frac{1}{2} \rho^2 \cos 2\theta  \ ,
\end{equation}
where $\theta \in [0,\frac{\pi}{2}]$, we see that we get
five-dimensional Minkowski-space in spheroidal coordinates times a
circle
\begin{equation}
ds^2 = - dt^2 + d\rho^2 + \rho^2 \left( d\theta^2 + \cos^2 \theta
d\psi^2 + \sin^2 \theta d\chi^2 \right) + d\phi^2 \ .
\end{equation}

We consider now six-dimensional generalized Weyl solutions that
asymptote to $\CM^5 \times S^1$. We can read off $c_t$ and
$c_\phi$ from the asymptotics of the potentials since we have
\begin{equation}
\label{asyD61} e^{2U_1} = 1 - \frac{\frac{1}{2}
c_t}{\sqrt{r^2+z^2}} + \CO \left( \frac{1}{r^2+z^2} \right) \spa
e^{2U_2} = 1 + \frac{\frac{1}{2} c_\phi}{\sqrt{r^2+z^2}} + \CO
\left( \frac{1}{r^2+z^2} \right)  \ ,
\end{equation}
for $\sqrt{r^2 + z^2} \rightarrow \infty$. Using this with
\eqref{mun} for $d=5$ we get $\mu$ and $n$.

\subsubsection*{Uniform black string and static Kaluza-Klein
bubble in $\CM^5 \times S^1$}

\begin{figure}[t]
   \centerline{\epsfig{file=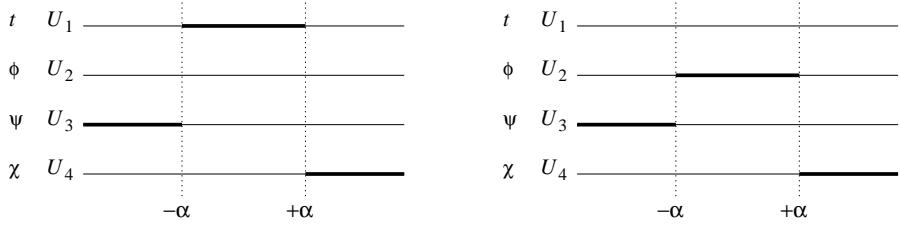,height=2.9cm}}
   \caption{Rod configurations for the uniform black string (left) and
   the static Kaluza-Klein bubble (right) in six dimensions ($d=5$).}
   \label{rodpic2}
\end{figure}

The $D=6$ uniform black string as a generalized Weyl solution has
the potentials
\begin{equation}
e^{2U_1} = \frac{R_+ - \zeta_+}{R_- - \zeta_-} \spa e^{2U_2} = 1
\spa e^{2U_3} = R_+ + \zeta_+ \spa e^{2U_4} = R_- - \zeta_- \ ,
\end{equation}
with
\begin{equation}
\zeta_\pm = z \pm \alpha \spa R_\pm = \sqrt{r^2 + (\zeta_\pm)^2} \
.
\end{equation}
The source configuration for the potentials $U_a$ are therefore a
finite rod $[-\alpha,\alpha]$ for $U_1$, no source the potential
$U_2$, a semi-infinite rod $[-\infty,-\alpha]$ for $U_3$, and a
semi-infinite rod $[\alpha,\infty]$ for $U_4$. We have depicted
this rod configuration in the left half of Figure \ref{rodpic2}.
Using \eqref{asyD61} we see that $c_t = \alpha$ and $c_\phi = 0$.
The function $\nu$ is given by
\begin{equation}
\label{nuD6} e^{2\nu} = \frac{\sqrt{R_+ R_- + \zeta_+ \zeta_- +
r^2}}{2 \sqrt{2} R_+ R_-} \sqrt{\frac{R_- - \zeta_-}{R_+ -
\zeta_+}} \ ,
\end{equation}
as can be checked using \eqref{nueqs}. Notice Eq.~\eqref{nuD6}
reduces to Eq.~\eqref{flatD6nu} for $\alpha = 0$. If we set
$\rho_0 = 2\sqrt{\alpha}$ and make the coordinate transformation
\begin{equation}
\label{rztorho2} r = \frac{1}{2} \rho^2
   \sqrt{ 1 - \frac{4\alpha}{\rho^2}} \sin 2 \theta
\spa z = \frac{1}{2} \rho^2
  \left( 1 - \frac{2\alpha}{\rho^2} \right) \cos 2 \theta \ ,
\end{equation}
with $\theta \in [0,\frac{\pi}{2}]$, we get the metric
\eqref{unblstr} for the $d=5$ uniform black string metric, which
explicitly exhibits the $SO(4)$ spherical symmetry.

If we make a double Wick rotation of the $t$ and $\phi$-directions
we get the $D=6$ static Kaluza-Klein bubble. This corresponds to
the potentials
\begin{equation}
e^{2U_1} = 1 \spa e^{2U_2} = \frac{R_+ - \zeta_+}{R_- - \zeta_-}
\spa e^{2U_3} = R_+ + \zeta_+ \spa e^{2U_4} = R_- - \zeta_- \ ,
\end{equation}
This means we have no rod sources for $U_1$, a finite rod
$[-\alpha,\alpha]$ for $U_2$, a semi-infinite rod
$[-\infty,-\alpha]$ for $U_3$, and a semi-infinite rod
$[\alpha,\infty]$ for $U_4$. We have depicted this
rod-configuration in the right part of Figure \ref{rodpic2}. The
function $\nu(r,z)$ is again given by \eqref{nuD6}. We need $L =
2\pi \sqrt{\alpha}$ to avoid a conical singularity at $\rho\to 0$
for $|z|<\alpha$. Using \eqref{asyD61} we see that $c_t = 0$ and
$c_\phi = - \alpha$, so $\mu=1/2$ and $n=3$ by Eq. \eqref{mun}. If
we make the coordinate transformation \eqref{rztorho2} we get the
explicitly spherically symmetric metric \eqref{statbub} for the
static Kaluza-Klein bubble metric with $d=5$.

%
%

\section{Five-dimensional bubble-black hole sequences}
\label{s:5dKK}

In this section we derive the metrics for bubble-black hole
sequences in five dimensions, using the generalized Weyl
construction reviewed above. We discuss some general aspects, such
as regularity, topology of the Kaluza-Klein bubbles and event
horizons, and the asymptotics of the solution. Specific cases as
well as further general physical properties of these
five-dimensional  solutions are presented in Section \ref{s:prop}.

\subsection{Five dimensional $(p,q)$ solutions}
\label{s:5dsol}

In this section we construct five-dimensional solutions with $p$
static Kaluza-Klein bubbles and $q$ black holes. We use method of
the generalized Weyl solutions reviewed in Section \ref{s:gweyl}
to construct the solutions. This means we use the ansatz
\begin{equation}
  ds^2 = - e^{2U_1} dt^2 + e^{2U_2} d\phi^2 + e^{2U_3} d\psi^2
         + e^{2\nu}(dr^2 + dz^2) \ .
\end{equation}
for the metric. The black holes and bubbles are placed alternately
along the $z$-axis in the Weyl coordinates, like pearls on a
string, for instance, \beastar
  \rom{black~hole}~-~\rom{bubble}~-~\rom{black~hole}
  ~-~\rom{bubble}~- \cdots -~\rom{black~hole} \, .
\eeastar Two black holes (or two bubbles) cannot sit next to each
other, so we have that $|p-q| \leq 1$.

To generate the black holes we place $q$ finite rods sourcing the
potential $U_1$ for the $t$-direction. The static Kaluza-Klein
bubbles are generated by placing $p$ finite rods sourcing the
potential $U_2$ for the $\phi$-direction. The potential $U_3$ is
then determined from the constraint \eqref{constr}.

For each Kaluza-Klein bubble in the solution we have a possible
conical singularity which is absent only if the periodicity of
$\phi$ is chosen appropriately. This gives rise to constraints
which we examine in Section \ref{s:regu5d}.

The period of $\psi$ is $2\pi$, since all the solutions asymptote
to $D=5$ Kaluza-Klein space $\CM^4 \times S^1$ as described in
Section \ref{revkal}. We study the asymptotics of the solutions in
Section \ref{s:asym5d}.

We introduce $N=p+q+1$ along with the set of numbers
$a_1<a_2<\dots<a_N$, where $a_i$ denote the endpoints of the rods.
In order to write the solution in a compact way, we follow
\cite{Emparan:2001wk} and introduce the following notation
\begin{equation}
\label{defzRY}
  \zeta_i = z - a_i
  \spa
  R_i = \sqrt{r^2 + \zeta_i^2}
  \spa
  Y_{ij} = R_i R_j + \zeta_i \zeta_j + r^2
\end{equation}
for $i,j=1,\dots,N$. We use this below to write down the
solutions.

We now have three different cases: $p=q-1$, $p=q+1$ and $p=q$. We
give the solutions for each of these cases in the following.

\subsubsection*{The case $p=q-1$}

\begin{figure}[t]
   \centerline{\epsfig{file=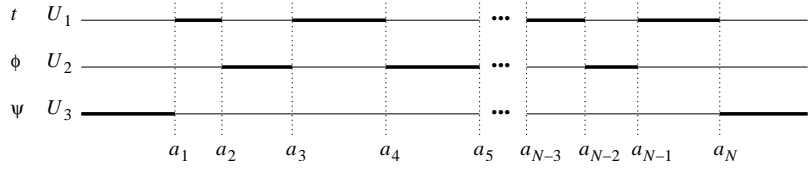,height=2.2cm}}
   \caption{Rod configurations for $p=q-1$ in five dimensions ($d=4$): The sequence begins and ends with a black hole.}
   \label{fig5D1}
\end{figure}

In this case the black hole-bubble sequence begins and ends with a
black hole: \beastar
  \rom{black~hole}~-~\rom{bubble}~-~\rom{black~hole}
  ~-~\rom{bubble}~- \cdots -~\rom{black~hole} \, .
\eeastar Note that $N=2p+2=2q$ so $N$ is even. The potentials are
given by
\begin{equation}\label{metric5d}
  \begin{array}{rccclcl}
  e^{2U_1}
  &=& \prod_{i=1}^N (R_i- \zeta_i)^{(-1)^{i+1}}
  &=& \displaystyle
  \frac{R_1 - \zeta_1}{R_2 - \zeta_2}
  \frac{R_3 - \zeta_3}{R_4 - \zeta_4} \cdots
  \frac{R_{N-1} - \zeta_{N-1}}{R_N - \zeta_N} \, ,\\[4mm]
  e^{2U_2}
  &=& \prod_{i=2}^{N-1}(R_i- \zeta_i)^{(-1)^{i}}
  &=& \displaystyle
  \frac{R_2 - \zeta_2}{R_3 - \zeta_3}
  \frac{R_4 - \zeta_4}{R_5 - \zeta_5} \cdots
  \frac{R_{N-2} - \zeta_{N-2}}{R_{N-1} - \zeta_{N-1}} \, ,\\[4mm]
  e^{2U_3} &=& (R_1+\zeta_1)(R_N-\zeta_N) \, .
\end{array}
\end{equation}
The corresponding rod configuration has $q$ finite rods
$[a_1,a_2]$, $[a_3,a_4]$,...,$[a_{N-1},a_N]$ sourcing the
potential $U_1$ (giving the $q$ black holes), and $p$ finite rods
$[a_2,a_3]$, $[a_4,a_5]$,...,$[a_{N-2},a_{N-1}]$ sourcing $U_2$
(giving the $p$ Kaluza-Klein bubbles).%
\footnote{As mentioned earlier, we use here and in the following
the notation that $[z_1,z_2]$ denotes a rod from $z=z_1$ to
$z=z_2$.} The potential $U_3$ is sourced by two semi-infinite rods
$[-\infty,a_1]$ and $[a_N,\infty]$. We have depicted this
rod-configuration in Figure \ref{fig5D1}.

One can now use \eqref{nueqs} for the potentials in
Eq.~\eqref{metric5d}.
This gives%
\footnote{Note that one can use the integrals given in
\cite{Emparan:2001wk} to derive Eq.~\eqref{gennu5D}.}
\begin{equation}
e^{2\nu} = \frac{Y_{1 N}}{2^{N/2}} \left(
  \prod_{i=1}^N \frac{1}{R_i}
\right) \left(
  \prod_{2\le i < j \le N-1} Y_{ij}^{(-1)^{i+j+1}}
\right) \sqrt{
  \prod_{i=2}^{N-1}
  \left(
    \frac{Y_{1i}}{Y_{iN}}
  \right)^{(-1)^{i}} } \,
\frac{R_N-\zeta_N}{R_1-\zeta_1} \, . \label{gennu5D}
\end{equation}

For $(p,q)=(0,1)$ we see that we recover the uniform black string
solution Eqs.~\eqref{bspot}-\eqref{nuD5} (setting $a_1 = - \alpha$
and $a_2 = \alpha$). For $(p,q)=(1,2)$ we get instead the solution
with two black holes on a Kaluza-Klein bubble --- this
configuration was previously studied in \cite{Elvang:2002br}. The
$(p,q)=(1,2)$ and $(2,3)$ solutions will be discussed in detail in
Sections \ref{sec12} and \ref{sec23} respectively.

\subsubsection*{The case $p=q+1$}

\begin{figure}[t]
   \centerline{\epsfig{file=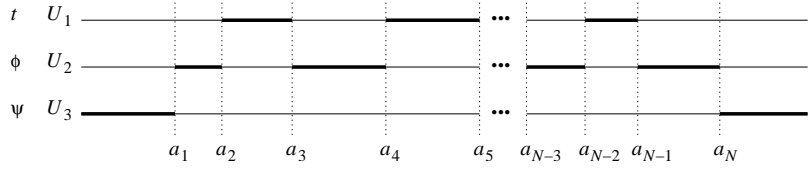,height=2.2cm}}
   \caption{Rod configurations for $p=q+1$ in five dimensions ($d=4$):
   The sequence begins and ends with a Kaluza-Klein bubble.}
   \label{fig5D2}
\end{figure}

In this case the sequence begins and ends with a Kaluza-Klein
bubble: \beastar
  \rom{bubble}~-~\rom{black~hole}~-~\rom{bubble}~-~\rom{black~hole}-
\cdots -~\rom{bubble} \, . \eeastar Note that $N=2q$ so $N$ is
even. This case can be obtained from the previous case --- where
the solution started and ended with black holes --- by a double
Wick rotation of the $t$- and $\phi$-directions since the double
Wick rotation interchanges the black holes and the bubbles. The
potentials are given by
\begin{equation}\label{met5D2}
  \begin{array}{rccclcl}
  e^{2U_1}
  &=& \prod_{i=2}^{N-1}(R_i- \zeta_i)^{(-1)^{i}}
  &=& \displaystyle
  \frac{R_2 - \zeta_2}{R_3 - \zeta_3}
  \frac{R_4 - \zeta_4}{R_5 - \zeta_5} \cdots
  \frac{R_{N-2} - \zeta_{N-2}}{R_{N-1} - \zeta_{N-1}} \, ,\\[4mm]
  e^{2U_2}
  &=& \prod_{i=1}^N (R_i- \zeta_i)^{(-1)^{i+1}}
  &=& \displaystyle
  \frac{R_1 - \zeta_1}{R_2 - \zeta_2}
  \frac{R_3 - \zeta_3}{R_4 - \zeta_4} \cdots
  \frac{R_{N-1} - \zeta_{N-1}}{R_N - \zeta_N} \, ,\\[4mm]
  e^{2U_3} &=& (R_1+\zeta_1)(R_N-\zeta_N) \, .
\end{array}
\end{equation}
The configuration thus has $q$ finite rods $[a_2,a_3]$,
$[a_4,a_5]$,...,$[a_{N-2},a_{N-1}]$ sourcing the potential $U_1$
(giving the $q$ black holes), and $p$ finite rods $[a_1,a_2]$,
$[a_3,a_4]$,...,$[a_{N-1},a_N]$ sourcing the potential $U_2$
(giving the $p$ Kaluza-Klein bubbles). The potential $U_3$ is
sourced by two semi-infinite rods $[-\infty,a_1]$ and
$[a_N,\infty]$. We have depicted this rod-configuration in Figure
\ref{fig5D2}. The function $\nu(r,z)$ is again given by
\eqref{gennu5D}. Note that one obtains Figure \ref{fig5D2} from
Figure \ref{fig5D1} by interchanging the $t$- and
$\phi$-directions. This is exactly the effect of the double Wick
rotation mentioned above.

For $(p,q)=(1,0)$ we regain the static Kaluza-Klein bubble
solution given in Eqs.~\eqref{bubD5}, \eqref{nuD5} and
\eqref{zRpm} (putting $a_1 = - \alpha$ and $a_2 = \alpha$). The
$(p,q)=(2,1)$ solution will be considered in detail in Section
\ref{sec12} and we also comment on the $(3,2)$ solution in Section
\ref{sec23}.

\subsubsection*{The case $p=q$}

In this case we start with a black hole
and end with a Kaluza-Klein bubble:%
\footnote{We could also consider the case where we start with a
Kaluza-Klein bubble and end with a black hole. This we can get
either by a double Wick rotation as above, or by the
transformation $z \rightarrow -z$. However, since these solutions
clearly have equivalent physics to the ones we consider here we do
not regard them as a separate class of solutions.} \beastar
  \rom{black~hole}~-~\rom{bubble}~-~\rom{black~hole}-
\cdots -~\rom{bubble} \, . \eeastar

Note that $N=2p+1$ so $N$ is odd. We have the potentials
\begin{equation}
  \begin{array}{rccclcl}
  e^{2U_1}
  &=& \prod_{i=1}^{N-1} (R_i- \zeta_i)^{(-1)^{i+1}}
  &=& \displaystyle
  \frac{R_1 - \zeta_1}{R_2 - \zeta_2}
  \frac{R_3 - \zeta_3}{R_4 - \zeta_4} \cdots
  \frac{R_{N-2} - \zeta_{N-2}}{R_{N-1} - \zeta_{N-1}} \, , \\[4mm]
  e^{2U_2}
  &=& \prod_{i=2}^{N}(R_i- \zeta_i)^{(-1)^{i}}
  &=& \displaystyle
  \frac{R_2 - \zeta_2}{R_3 - \zeta_3}
  \frac{R_4 - \zeta_4}{R_5 - \zeta_5} \cdots
  \frac{R_{N-1} - \zeta_{N-1}}{R_{N} - \zeta_{N}} \, ,\\[4mm]
  e^{2U_3} &=& (R_1+\zeta_1)(R_N-\zeta_N) \, .
\end{array}
\label{pqcaseU}
\end{equation}
\begin{figure}[t]
   \centerline{\epsfig{file=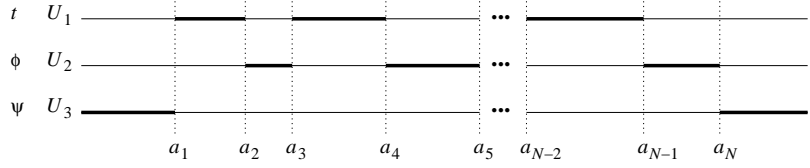,height=2.2cm}}
   \caption{Rod configurations for $p=q$ in five dimensions ($d=4$):
   This sequence begins with a black hole and ends with a
   Kaluza-Klein bubble.}
   \label{fig5D3}
\end{figure}%
The configuration thus has $q$ finite rods $[a_1,a_2]$,
$[a_3,a_4]$,...,$[a_{N-2},a_{N-1}]$ sourcing the potential $U_1$
(giving the $q$ black holes), and $p$ finite rods $[a_2,a_3]$,
$[a_4,a_5]$,...,$[a_{N-1},a_{N}]$ sourcing $U_2$ (giving the $p$
Kaluza-Klein bubbles). We have depicted this rod-configuration in
Figure \ref{fig5D3}. From \eqref{nueqs} one can find
\begin{equation}
e^{2\nu} = \frac{\sqrt{Y_{1 N}}}{2^{N/2}} \left(
  \prod_{i=1}^N \frac{1}{R_i}
\right) \left(
  \prod_{2\le i < j \le N-1} Y_{ij}^{(-1)^{i+j+1}}
\right) \sqrt{
  \prod_{i=2}^{N-1}
  \left(
    Y_{1i} Y_{iN}
  \right)^{(-1)^{i}} } \,
  \frac{R_N-\zeta_N}{R_1-\zeta_1} \, .
\label{pqcasenu}
\end{equation}
Note that the solution \eqref{pqcaseU}-\eqref{pqcasenu} for the
$p=q$ case that we consider here can formally be obtained from the
$p=q-1$ case above. This is done by considering a configuration
with $p+1$ black holes and $p$ bubbles and then setting $a_{N} =
a_{N-1}$ where $N=2p+2$, and finally substituting $N-1 \rightarrow
N$.

For $(p,q)=(1,1)$ we get the solution with one black hole and one
static Kaluza-Klein bubble studied previously in
\cite{Emparan:2001wk}. The $(p,q)=(1,1)$ solution will be
considered in detail in Section \ref{sec11}, and we also comment
on the $(2,2)$ solution in Section \ref{sec23}.

\subsection{Regularity and topology of the Kaluza-Klein bubbles}
\label{s:regu5d}

We examine in this section the behavior of the solutions near the
Kaluza-Klein bubbles. As stated above, for each of the $p$
Kaluza-Klein bubbles we have a possible conical singularity which
is absent only if the periodicity of $\phi$ is chosen
appropriately. Thus, in order to have a regular solution we have
$p$ constraints that need to be obeyed. Below we write down these
constraints for the solutions. We also examine the topology of the
Kaluza-Klein bubbles by considering the metric on the bubbles.

We can count the number of free parameters characterizing a
$(p,q)$ solution as follows. If we, as above, let $L$ be the
circumference of the Kaluza-Klein circle parameterized by $\phi$,
we see that for a given $L$ we have $p$ constraints on the
$N=p+q+1$ parameters in order to have a regular solution.
Therefore, we have $q+1$ parameters left by demanding regularity
of the solution. By translational invariance we can disregard one
of these, leaving now $q$ dimensionful parameters describing our
solution (for a given circumference $L$). This means that we need
$q$ dimensionless parameters to describe the space of regular
$(p,q)$ solutions.

Before considering specific solutions we examine here the general
features of how a solution behaves near a Kaluza-Klein bubble.
Consider a rod $[z_1,z_2]$ sourcing $U_2$, i.e. the potential for
the $\phi$-direction. For $r \rightarrow 0$ with $z_1 < z < z_2$
we have in general
\begin{equation}
\label{nearr0} ds^2 = - g(z) dt^2 + f(z) d\psi^2 +
\frac{1}{f(z)g(z)} \left[ r^2 d\phi^2 + c^2 ( dr^2 + dz^2 )
\right] \, ,
\end{equation}
where $f(z)$ and $g(z)$ are functions and $c$ is a number.%
\footnote{For the specific solutions the explicit form of $f(z)$
and $g(z)$ and value of $c$ depend on which interval is
considered. Note also that $f(z)$, $g(z)$ and $c$ will in general
be different for Eqs.~\eqref{timerod}, \eqref{bubmet6D} and
\eqref{BHmet6D}.} Clearly, this metric has a conical singularity
at $r=0$, unless we take $\phi$ to have period $2\pi c$. Thus, we
need that $L = 2\pi c$. More generally, it is useful to write the
regularity condition as
\begin{equation}
\label{delphi} L = \Delta \phi = 2\pi \lim_{r\rightarrow 0} \sqrt{
\frac{r^2 g_{rr}}{g_{\phi\phi}}} \, .
\end{equation}
To see that the $(p,q)$ solutions reduce to the form
\eqref{nearr0} for $r \rightarrow 0$ near a bubble, one can use
that for $r \rightarrow 0$ we have
\begin{equation}
\label{Rzero}
  \frac{R_i - \zeta_i}{R_{j} - \zeta_{j}}
  \simeq \left\{
  \begin{array}{ll}
    \frac{|\zeta_i|}{|\zeta_j|}
    & \mbox{for}~z < a_i < a_j \\[2mm]
    \frac{r^2}{4|\zeta_i\zeta_j|}
    & \mbox{for}~a_i < z < a_{j} \\[2mm]
    \frac{|\zeta_j|}{|\zeta_i|}
    & \mbox{for}~a_{i} < a_j < z \, ,
  \end{array}
  \right.
\end{equation}
and
\begin{equation}
\label{Yzero} Y_{ij} \simeq
  \left\{
    \begin{array}{ll}
       2|\zeta_i \zeta_j| &
       \mbox{ for } z < a_i , a_j  \mbox{ or } z > a_i , a_j \\[1mm]
       \frac{r^2 (a_j - a_i)^2}{2|\zeta_i \zeta_j|} &
       \mbox{ for } a_i < z < a_j
    \end{array}
  \right.
\end{equation}
for the definitions \eqref{defzRY}. Using these formulas, one can
obtain explicit expressions for $f(z)$, $g(z)$ and $c$ for each of
the bubbles.

Note that we can read off the topology of the Kaluza-Klein bubbles
from the metric \eqref{nearr0}. At $r=0$ and for fixed time $t$,
we have the metric
\begin{equation}
\label{bubbles} (ds_{2})^2 = f(z) d\psi^2 + \frac{c^2}{f(z)g(z)}
dz^2 \, .
\end{equation}
The topology of the Kaluza-Klein bubble as a two-dimensional
surface is now determined from the behavior of $f(z)$ and $g(z)$
for $z \rightarrow z_1,z_2$.

Since the bubble-rod is next to a rod either in the $U_1$
potential, corresponding to an event horizon, or the $U_3$
potential, if the bubble is at either end of the bubble-black hole
sequence, we have that either $f(z)$ or $g(z)$ goes to zero for $z
\rightarrow z_1$ (or for $z \rightarrow z_2$), but never both of
them. We have therefore three different cases:
\begin{itemize}
\item
The pure bubble space-time: there are no sources for the $U_1$
potential. This corresponds to the case $(p,q)=(1,0)$, which is
the static bubble without any event horizon. In this case $f(z)$
goes to zero in both endpoints $z_1$ and $z_2$, so the bubble has
the topology of a two-sphere $S^2$.
\item The bubble sits between two event horizons. In this case
$f(z)$ vanishes nowhere in the interval $[z_1,z_2]$, so $\psi$
parameterizes a circle whose circumference varies with $z$ but
never shrinks to zero size. The function $g(z)$ goes to zero at
both endpoints $z_1$ and $z_2$. Since these are simple poles in
the metric component $g_{zz}$ the $z$-coordinate parameterizes a
finite interval $I$ of length \bea
  s = c \int_{z_1}^{z_2} \big[ f(z)g(z) \big]^{-1/2} dz \, ,
\eea which we may think of as the proper distance between the two
event horizons sitting on either side of the bubble. We conclude
from the above that a bubble sitting between two event horizons is
topologically a finite cylinder $S^1 \times I$.
\item The bubble is at either end of the bubble-black hole
sequence. In terms of the rod configuration, there is a
semi-infinite rod for the $U_3$ potential on one side of the
bubble, say at $z=z_1$, and on the other side $z=z_2$ there is a
finite rod for the $U_1$ potential (this gives rise to the event
horizon sitting next to the bubble). In this case $f(z)$ goes to
zero only at $z=z_1$, and $g(z)$ goes to zero only at $z=z_2$.
Thus the $\psi$-circle shrinks smoothly to a point as $z \to z_1$,
and the topology is therefore a disk $D$.
\end{itemize}

It should be noted that the bubble topologies determined above are
inferred from the coordinate patch described by the metric
\refeq{bubbles}. In the cases where an event horizon is present,
the coordinates can be continued in a way analogous to that of the
maximally extended Schwarzschild black hole. If the space-time has
more than one event horizon, this extension is not uniquely given,
but for the simplest bubble-black hole solutions we shall comment
on this point when we consider examples of specific solutions in
Section \ref{s:prop}.

We now consider the constraints imposed on parameters of the
solutions by the requirement of regularity. Again we consider the
three cases $p=q-1$, $p=q+1$ and $p=q$ separately.

\subsubsection*{The case $p=q-1$}

The sequence begins and ends with a black hole. From our above
considerations, this means that each of the Kaluza-Klein bubbles
has topology as a cylinder $S^1 \times I$.

Using \eqref{delphi} together with \eqref{Rzero}-\eqref{Yzero}, we
see that the bubble corresponding to the $[a_{2k},a_{2k+1}]$ rod
(sourcing $U_2$) requires $\phi$ to have the period
\begin{equation}
\label{genL}
 (\Delta \phi)_k =
4\pi  (a_N-a_1) \prod_{i=2}^{2k} \prod_{j=2k+1}^{N-1}
(a_j-a_i)^{(-1)^{i+j+1}} \prod_{i=2}^{2k}
[\sqrt{a_N-a_i}]^{(-1)^{i+1}} \prod_{i=2k+1}^{N-1}
[\sqrt{a_i-a_1}]^{(-1)^{i}}
\end{equation}
with $k=1,...,p$, in order to avoid a conical singularity on the
bubble. Given the circumference $L$ of the Kaluza-Klein circle
parameterized by $\phi$, we see then that we must require the $p$
constraints
\bea
 L =
  (\Delta \phi)_k \, , ~~~\rom{for~all}~k=1, \ldots , p
  \label{Deltaphi}
\eea
in order for the solution to be regular.

\subsubsection*{The case $p=q+1$}

In this case we have Kaluza-Klein bubbles in both ends of the
bubble-black hole sequence. If $p \geq 2$, this means that each of
the two bubbles in the ends is topologically a disk $D$. Any other
bubble in the solution has topology $S^1 \times I$. If $p=1$ we
have just one bubble with topology $S^2$.

Regularity of the $k$'th bubble, corresponding to the rod
$[a_{2k-1},a_{2k}]$ sourcing $U_2$, requires $\phi$ to have period
\begin{equation}
\label{genL2}
 (\Delta \phi)_k =
4\pi  (a_N-a_1) \prod_{i=2}^{2k-1} \prod_{j=2k}^{N-1}
(a_j-a_i)^{(-1)^{i+j+1}} \prod_{i=2}^{2k-1}
[\sqrt{a_N-a_i}]^{(-1)^{i+1}} \prod_{i=2k}^{N-1}
[\sqrt{a_i-a_1}]^{(-1)^{i}}
\end{equation}
with $k=1,...,p$. Given $L$, the $p$ constraints are then that $L
= (\Delta \phi)_k$ for $k=1,...,p$.

\subsubsection*{The case $p=q$}

In the left end of the sequence we have a black hole, while in the
right end we have a bubble with topology as a disk $D$. Any other
bubble in the solution has topology $S^1 \times I$.

Regularity of the $k$'th bubble, corresponding to the rod
$[a_{2k},a_{2k+1}]$ sourcing $U_2$, requires $\phi$ to have period
\begin{equation}
\label{genL3}
 (\Delta \phi)_k =
 4\pi  (a_N-a_1) \prod_{i=2}^{2k} \prod_{j=2k+1}^{N}
(a_j-a_i)^{(-1)^{i+j+1}} \prod_{i=2}^{2k}
[\sqrt{a_N-a_i}]^{(-1)^{i+1}} \prod_{i=2k+1}^{N}
[\sqrt{a_i-a_1}]^{(-1)^{i}}
\end{equation}
with $k=1,...,p$. Again, given $L$, the $p$ constraints are then
that $L = (\Delta \phi)_k$ for $k=1,...,p$.

\subsection{Event horizons, topology, thermodynamics and balance}
\label{s:eh5D}

In the above solutions we have $q$ finite rods sourcing $U_1$,
where $U_1$ is the potential associated with the time-direction
$t$. This gives
$q$ event horizons. In the following we examine these event
horizons, discuss their topologies and the associated temperatures
and entropies.

We first consider the general features of how a solution behaves
near an event horizon. Consider a rod $[z_1,z_2]$ sourcing $U_1$,
i.e.~the potential for the time direction $t$. For $r \rightarrow
0$ with $z_1 < z < z_2$ we have in general
\begin{equation}
\label{timerod} ds^2 = g(z) d\phi^2 + f(z) d\psi^2 +
\frac{1}{f(z)g(z)} \left[ - r^2 dt^2 + c^2 ( dr^2 + dz^2 ) \right]
\, ,
\end{equation}
where $f(z)$ and $g(z)$ are functions and $c$ is a number. We see
that we have an event horizon at $r=0$ since $g_{tt}$ goes to zero
there, and the metric is otherwise regular at $r=0$. It is easy to
see using \eqref{Rzero}-\eqref{Yzero} that the $(p,q)$ solutions
reduce to the form \eqref{timerod} for $r\rightarrow 0$.

By Wick rotating the time-coordinate $t$ to the coordinate
$\omega=it$, we can see that in the Euclidean section of the
solution $\omega$ should have period $2\pi c$ in order to avoid a
conical singularity at $r \to 0$. This means that the horizon has
inverse temperature $2\pi c$. More generally, we can write this as
\begin{equation}
\label{tempD5} \beta = \frac{1}{T} = 2\pi \lim_{r\rightarrow 0}
\sqrt{ \frac{r^2 g_{rr}}{-g_{tt}}} \, .
\end{equation}
We can also read off the topology of the event horizon. For $r=0$
and fixed $t$, we have the metric
\begin{equation}
\label{ehD5} (ds_3)^2 = g(z) d\phi^2 + f(z) d\psi^2 +
\frac{c^2}{f(z)g(z)} dz^2 \, .
\end{equation}
This is the metric on the event horizon. Thus, we can find the
topology of the event horizon as a three-dimensional surface by
considering the behavior of the functions $f(z)$ and $g(z)$ for $z
\rightarrow z_1,z_2$. The analysis is similar to that of the
bubble topology.
There are three different cases:

\begin{itemize}
\item No bubbles present: this corresponds to the case
$(p,q)=(0,1)$. Since there are no bubbles, $g(z)$ stays non-zero,
and the $\phi$-coordinate parameterizes an $S^1$ which is just the
Kaluza-Klein circle. The function $f(z)$ goes to zero at both
endpoints $z_1$ and $z_2$, so $(z,\psi)$ parameterizes a
two-sphere $S^2$. The topology of the event horizon is therefore
$S^2 \times S^1$, as expected since the solution $(p,q)=(0,1)$ is
the uniform black string wrapping around the Kaluza-Klein circle.
\item The event horizon has a bubble on both sides. In this case
$g(z)$ goes to zero at both endpoints $z_1$ and $z_2$, so
$(z,\phi)$ parameterizes an $S^2$. Since $f(z)$ stays non-zero we
see that the $\psi$-direction corresponds to an $S^1$. The
topology of the event horizon is therefore that of a black ring
$S^2 \times S^1$. Note that we call a black hole with horizon
topology $S^2 \times S^1$ a black ring if the $S^1$ is not
topologically supported, i.e.~if the $S^1$ direction is a
contractible circle in the space-time and not the Kaluza-Klein
circle parameterized by $\phi$. The $S^1$ of the black ring is
supported by the Kaluza-Klein bubble against its gravitational
self-attraction.
\item The event horizon is at either end of the bubble-black hole
sequence. In this case $f(z)$ goes to zero at one endpoint and
$g(z)$ goes to zero at the other endpoint. It is not hard to see
that the topology of the event horizon is a three-sphere $S^3$.
\end{itemize}
Note that none of the black holes are localized on the
Kaluza-Klein circle.

We can read off the entropy of the event horizon by computing the
area using the metric \eqref{ehD5}. Since the square-root of the
determinant of the metric \eqref{ehD5} is equal to $c$, we find
the entropy to be
\begin{equation}
\label{entrD5} S = \frac{2\pi L (z_2-z_1) c}{4 G_{\rm N}} =
\frac{2\pi L (z_2-z_1) }{4 G_{\rm N}} \lim_{r\rightarrow 0} \sqrt{
\frac{r^2 g_{rr}}{-g_{tt}}} \, .
\end{equation}
Note that combining \eqref{tempD5} and \eqref{entrD5}, we get
\begin{equation}
\label{TSD5} TS = \frac{L (z_2-z_1)}{4 G_{\rm N}} \, .
\end{equation}
This will be useful below.

\subsubsection*{The case $p=q-1$}

The sequence  begins and ends with a black hole. From the above
results on the topology of the event horizons and bubbles, we see
that this class of solutions for $q \geq 2$ looks as follows:
\[
  \begin{array}{ccccccccc}
  \rom{black~hole}&-&\rom{bubble} &-& \rom{black~ring}
  &\ \cdots \ &\rom{bubble} &-& \rom{black~hole} \\
  S^3 &&  S^1 \times I  && S^2 \times S^1 &\ \cdots \ &   S^1 \times I   && S^3
  \end{array}
\]
For $q=1$, we have the black string with topology $S^2 \times
S^1$.

Using \eqref{tempD5}  we find the inverse temperature for the
$k$'th event horizon corresponding to the rod $[a_{2k-1},a_{2k}]$
sourcing $U_1$ to be
\begin{equation}
\label{betak1} \beta_k = \frac{1}{T_k} = 4\pi  (a_N-a_1)
\prod_{i=2}^{2k-1} \prod_{j=2k}^{N-1} (a_j-a_i)^{(-1)^{i+j+1}}
\prod_{i=2}^{2k-1} [\sqrt{a_N-a_i}]^{(-1)^{i+1}}
\prod_{i=2k}^{N-1} [\sqrt{a_i-a_1}]^{(-1)^{i}}
\end{equation}
with $k=1,...,p$. {} From \eqref{TSD5} we see that the entropy of
the $k$'th black hole can be computed from $S_k = \beta_k L
(a_{2k}-a_{2k-1}) / (4 G_{\rm N})$.

\subsubsection*{The case $p=q+1$}

In this case, we have Kaluza-Klein bubbles in both ends of the
sequence. This class of solutions has the following general
structure (for $p \geq 2$):
\[
  \begin{array}{ccccccccc}
  \rom{bubble}&-&\rom{black~ring} &-& \rom{bubble}
  &\ \cdots \ &\rom{black~ring} &-& \rom{bubble} \\
  D &&  S^2 \times S^1  && S^1 \times I &\ \cdots \ & S^2 \times S^1 && D
  \end{array}
\]
For $p=1$, we have the Kaluza-Klein bubble with topology $S^2$.

Using \eqref{tempD5} we find the inverse temperature for the
$k$'th event horizon corresponding to the rod $[a_{2k},a_{2k+1}]$
sourcing $U_1$ to be
\begin{equation}
\label{betak2} \beta_k = \frac{1}{T_k} = 4\pi  (a_N-a_1)
\prod_{i=2}^{2k} \prod_{j=2k+1}^{N-1} (a_j-a_i)^{(-1)^{i+j+1}}
\prod_{i=2}^{2k} [\sqrt{a_N-a_i}]^{(-1)^{i+1}}
\prod_{i=2k+1}^{N-1} [\sqrt{a_i-a_1}]^{(-1)^{i}}
\end{equation}
with $k=1,...,p$. {}From \eqref{TSD5} we see that the entropy can
be computed from $S_k = \beta_k L (a_{2k+1}-a_{2k}) / (4 G_{\rm
N})$.

\subsubsection*{The case $p=q$}

In this case, the sequence starts with a black hole and ends with
a Kaluza-Klein bubble. This class of solutions has the following
structure:
\[
  \begin{array}{ccccccc}
  \rom{black~hole}&-&\rom{bubble} &-& \rom{black~ring}
  &\ \cdots \ &\rom{bubble}  \\
  S^3 &&  S^1 \times I  && S^2 \times S^1 &\ \cdots \ &  D
  \end{array}
\]
Using \eqref{tempD5} we find the inverse temperature for the
$k$'th event horizon corresponding to the rod $[a_{2k-1},a_{2k}]$
sourcing $U_1$ to be
\begin{equation}
\label{betak3} \beta_k = \frac{1}{T_k} = 4\pi  (a_N-a_1)
\prod_{i=2}^{2k-1} \prod_{j=2k}^{N} (a_j-a_i)^{(-1)^{i+j+1}}
\prod_{i=2}^{2k-1} [\sqrt{a_N-a_i}]^{(-1)^{i+1}} \prod_{i=2k}^{N}
[\sqrt{a_i-a_1}]^{(-1)^{i}}
\end{equation}
with $k=1,...,p$. From \eqref{TSD5} we see that the entropy can be
computed from $S_k = \beta_k L (a_{2k}-a_{2k-1}) / (4 G_{\rm N})$.

\subsubsection*{Balance}

We now address the physical reason why the static Kaluza-Klein bubbles
keep the black holes in a static equilibrium. For this we
consider the configuration with two black holes on a bubble. The black holes
attract each other, but nonetheless the configuration is held in static
equilibrium by the bubble. This balance can be examined by using the bubble
initial data discussed in Section \ref{bubble}.
Combining Eqs.~\eqref{Addot}-\eqref{bubsize}, the initial acceleration of a
Kaluza-Klein bubble is
\begin{equation}
  8\pi \left( \frac{4\pi \rho_+}{L}-1 \right) \, ,
  \end{equation}
where $\rho_+$ is the size of the bubble and $L$ is the length of the
Kaluza-Klein circle at infinity. Keeping the asymptotics fixed, the initial
acceleration grows linearly with the size of the bubble.

Now for a static bubble, the initial acceleration vanishes. In
\cite{Elvang:2002br} it was found that adding two small black holes to the
static bubble increases its size and hence the bubble wants to expand. The
static equilibrium can then be understood as the balance between the attraction
of the black holes and the acceleration of the bubble. Furthermore the bubble
(as we have seen) can accommodate black holes of unequal size.
Even the attraction of a single black hole can prevent the bubble from
expanding and this configuration is described by the static solution of one
black hole on a bubble.

The balance is closely related to the regularity conditions. If the regularity
constraints are not satisfied, there will be conical singularities on the
bubbles. It can then be shown \cite{Elvang:2002br} that the combined push of the
bubble and an excess angle of the conical singularity can then balance bigger
black holes. Or if the conical singularity is associated with a deficit angle
(providing a pull), the black holes of the static black hole-bubble configuration
will be smaller compared to the case of the regular solutions. The study of the
conically singular solutions provide insight into the balance of the solutions,
however, in this paper we shall focus only on regular solutions.

\subsection{Asymptotics}
\label{s:asym5d}

The asymptotic region of the $(p,q)$ solutions is at $\sqrt{r^2 +
z^2} \rightarrow \infty$. It is easily seen from the explicit
expressions for the $(p,q)$ solutions that they asymptote to the
$\CM^4 \times S^1$ solution given by \eqref{flat5D}
for $\sqrt{r^2+z^2} \rightarrow \infty$.%
\footnote{Note that this means $\psi$ has period $2\pi$, which in
fact one also gets by considering the two semi-infinite rods
$[-\infty,a_1]$ and $[a_N,\infty]$ which both requires $\psi$ to
have period $2\pi$ in order to avoid a conical singularity.}

Using the identity \bea
  \label{5dasymp1}
  \frac{R_i - \zeta_i}{R_j - \zeta_j} &=& 1 + \frac{a_i - a_j}{\sqrt{r^2 + z^2}} + \CO ( (r^2+z^2)^{-1} )
\eea for $\sqrt{r^2+z^2} \rightarrow \infty$, we can furthermore
see that the $U_1$ and $U_2$ potentials for $(p,q)$ solutions
become of the form \eqref{asyD51} for $\sqrt{r^2 + z^2}
\rightarrow \infty$. Using \eqref{asyD51} we can then read off
$c_t$ and $c_\phi$ for the $(p,q)$ solutions. For $p=q-1$ and
$p=q$, we find
\begin{equation}
\label{ct1_5D} c_t = \sum_{k=1}^{q} (a_{2k} - a_{2k-1}) \spa
c_\phi = -\sum_{k=1}^{p} (a_{2k+1}- a_{2k}) \, ,
\end{equation}
while for $p=q+1$ we find
\begin{equation}
\label{ct2_5D} c_t = \sum_{k=1}^{q} (a_{2k+1} - a_{2k}) \spa
c_\phi = -\sum_{k=1}^{p} (a_{2k}- a_{2k-1}) \, .
\end{equation}
{}From the above we see that $c_t$ is the sum of the lengths of
the rods giving the event horizons, while $-c_\phi$ is the sum of
the lengths of the rods giving the bubbles. Clearly, this means
that $c_t - c_\phi = a_N - a_1$. Using \refeq{mun}, we can now
determine the dimensionless mass $\mu$ and the relative binding
energy $n$ from
\begin{equation}
\label{mu5} \mu = \frac{4 \pi}{L} [ 2 c_t - c_\phi] \spa n =
\frac{c_t - 2 c_\phi}{2c_t - c_\phi } \, .
\end{equation}
Note that since $a_1 < a_2 < \cdots < a_N$, we have $c_t > 0$ and
$c_\phi < 0$. This means that $1/2 < n < 2$ (for $p,q \geq 1$).

Since $c_t$ is the sum of the lengths of the rods giving the event
horizons, we see from \eqref{TSD5} that $\sum_{k=1}^q T_k S_k = L
c_t / (4 G_{\rm N})$. {}From this we get, in terms of the
dimensionless entropy $\mathfrak{s}_k$ and temperature
$\mathfrak{t}_k$ defined in \eqref{tsdef}, the generalized Smarr
formula \bea \label{gendsmarr}
  \sum_{k=1}^{q}\mathfrak{t}_k \mathfrak{s}_k
  = \frac{2-n}{3} \mu \, .
\eea This is a new realization of the generalized Smarr formula
\eqref{smarr1}, involving the temperature and entropy of each
individual black hole. More generally, this relation can be
derived along the lines of
Refs.~\cite{Harmark:2003dg,Kol:2003if,Townsend:2001rg} by assuming
the space-time to contain several disconnected horizons.

The generalized First Law of thermodynamics takes the form \bea
 \label{Gen1stLaw}
 \delta \mu  = \sum_{k=1}^{q} \mathfrak{t}_k \delta \mathfrak{s}_k \ .
\eea We have verified this for the examples in Section
\ref{s:prop}.

\section{Six-dimensional bubble-black hole sequences}
\label{s:6dKK}
%
%

In this section we derive the metrics for bubble-black hole sequences
in six dimensions. We discuss general aspects such as
regularity, topology of the black holes and bubbles, and the
asymptotics of the solutions. The analysis is shown to be related
to that of the five-dimensional bubble-black hole sequences
of Section \ref{s:5dKK}, a fact that we use extensively. A more
detailed analysis of the physical properties of specific solutions
is given in Section \ref{s:prop}.

\subsection{Six dimensional $(p,q)$ solutions \label{s:6dsol}}

In this section we construct six-dimensional solutions with $p$ static Kaluza-Klein bubbles and $q$ black holes. We
use the method of generalized Weyl solutions reviewed in Section
\ref{s:gweyl} to construct the solutions. The ansatz for the
metric is
\bea
  ds^2 = - e^{2U_1} dt^2
         + e^{2U_2} d\phi^2
         + e^{2U_3} d\psi^2
         + e^{2U_4} d\chi^2
         + e^{2\nu}(dr^2 + dz^2) \, .
  \label{6dweyl}
\eea Like in five dimensions, the black holes and bubbles are
placed alternately along the $z$-axis in the Weyl
coordinates, like pearls on a string. Two black holes (or two
bubbles) cannot sit next to each other, so $|p-q| \leq 1$.

To generate the black holes we place $q$ finite rods sourcing the
potential $U_1$ for the $t$-direction. The static Kaluza-Klein
bubbles are generated by placing $p$ finite rods sourcing the
potential $U_2$ for the $\phi$-direction. The $U_3$ and $U_4$
potential each have a semi-infinite rod, such that the constraint
\eqref{constr} is obeyed.

Note that the periods of the $\psi$ and $\chi$ directions are
$2\pi$, as can be seen from the fact that the solutions asymptote
to $\CM^5 \times S^1$ (which we described in Section
\ref{revkal}). We examine the asymptotics of our solutions in
Section \ref{s:asym6d}.

\begin{figure}[t]
   \centerline{\epsfig{file=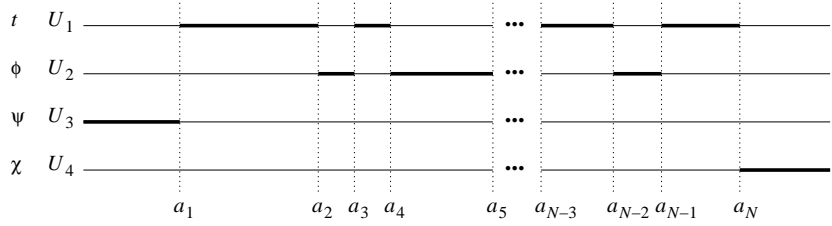,height=2.9cm}}
   \caption{Rod configurations for $p=q-1$ configurations
   in six dimensions ($d=5$):
   This sequence begins and ends with a black hole.}
   \label{fig6Da}
\end{figure}

\begin{figure}[t]
   \centerline{\epsfig{file=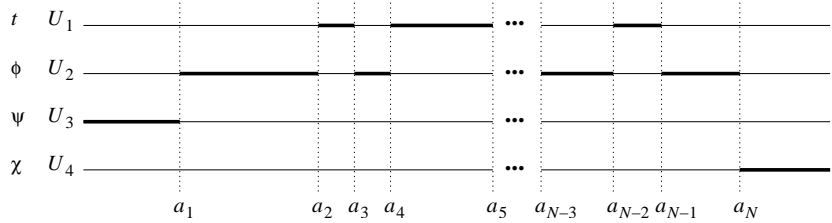,height=2.9cm}}
   \caption{Rod configurations for $p=q+1$ configurations
   in six dimensions ($d=5$):
   This sequence begins and ends with Kaluza-Klein bubbles.}
   \label{fig6Db}
\end{figure}

\begin{figure}[t]
   \centerline{\epsfig{file=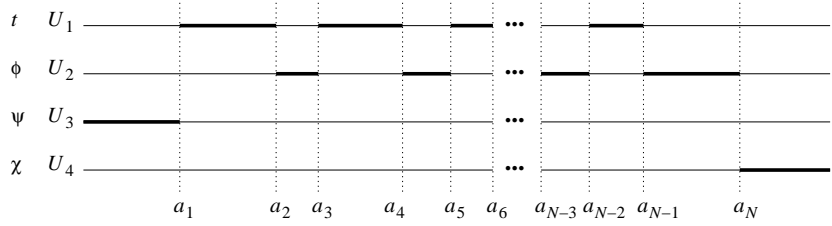,height=2.9cm}}
   \caption{Rod configurations for $p=q$ configurations
   in six dimensions ($d=5$):
   This sequence begins with a black hole and ends with a
   Kaluza-Klein bubble.}
   \label{fig6Dc}
\end{figure}

We introduce $N=p+q+1$ along with the set of numbers
$a_1<a_2<\dots<a_N$, where $a_i$ denote the endpoints of the rods.
We use the notation defined in Eqs.~\eqref{defzRY} of $R_i$
and $Y_{ij}$ introduced in Section \ref{s:5dKK}. We again have
three different cases: $p=q-1$, $p=q+1$ and $p=q$. The rod
configurations for these cases are:
\begin{itemize}
\item The case $p=q-1$. The bubble-black hole sequence begins and ends
with a black hole. We have $q$ finite rods $[a_1,a_2]$,
$[a_3,a_4]$,...,$[a_{N-1},a_{N}]$ sourcing the potential $U_1$,
giving the $q$ event horizons. We have $p$ finite rods
$[a_2,a_3]$, $[a_4,a_5]$,...,$[a_{N-2},a_{N-1}]$ sourcing $U_2$,
giving the $p$ Kaluza-Klein bubbles. Finally, we have a
semi-infinite rod $[-\infty,a_1]$ sourcing $U_3$, and a
semi-infinite rod $[a_N,\infty]$ sourcing $U_4$. The rod configuration
is illustrated in Figure \ref{fig6Da}.
\item The case $p=q+1$. The sequence starts and ends with a
Kaluza-Klein bubble. We have $q$ finite rods $[a_2,a_3]$,
$[a_4,a_5]$,...,$[a_{N-2},a_{N-1}]$ sourcing the potential $U_1$,
giving the $q$ event horizons. We have $p$ finite rods
$[a_1,a_2]$, $[a_3,a_4]$,...,$[a_{N-1},a_{N}]$ sourcing $U_2$,
giving the $p$ Kaluza-Klein bubbles. Finally, we have a
semi-infinite rod $[-\infty,a_1]$ sourcing $U_3$, and a
semi-infinite rod $[a_N,\infty]$ sourcing $U_4$. This is depicted in
Figure \ref{fig6Db}.
\item The case $p=q$. The sequence begins with a black hole and ends
with a Kaluza-Klein bubble.
In this case we have $q$ finite rods
$[a_1,a_2]$, $[a_3,a_4]$,...,$[a_{N-2},a_{N-1}]$ sourcing the
potential $U_1$, giving the $q$ event horizons. We have $p$ finite
rods $[a_2,a_3]$, $[a_4,a_5]$,...,$[a_{N-1},a_{N}]$ sourcing
$U_2$, giving the $p$ Kaluza-Klein bubbles. Finally, we have a
semi-infinite rod $[-\infty,a_1]$ sourcing $U_3$, and a
semi-infinite rod $[a_N,\infty]$ sourcing $U_4$. The rod configuration
is given in Figure \ref{fig6Dc}.
\end{itemize}
We can now use the solutions constructed in Section \ref{s:5dKK}
to write down the six-dimensional solutions. Given the parameters
$a_1,...,a_N$, the six-dimensional $(p,q)$ solution is given by
\begin{equation}
\label{potD6} e^{2U_1} = e^{2U_1^{(5D)}} \spa e^{2U_2} =
e^{2U_2^{(5D)}} \spa e^{2U_3} = R_1 + \zeta_1 \spa e^{2U_4} = R_N
- \zeta_N
\end{equation}
where we write $U_1^{(5D)}$ and $U_2^{(5D)}$ for the potentials in
the five-dimensional $(p,q)$ solutions listed in Section
\ref{s:5dsol}. Furthermore, the function $\nu$ for the six dimensional
solution can be written
\bea \label{nu5nu6} e^{2\nu} = e^{2\nu^{(5D)}}
\sqrt{\frac{1}{2Y_{1N}}\frac{R_1-\zeta_1}{R_N-\zeta_N}} \, ,
\eea
where $\nu^{(5D)}$ is from the five-dimensional solutions given
in Section \ref{s:5dsol}. Specifically, for $p=q-1$ we should use
$U_1^{(5D)}$, $U_2^{(5D)}$ and $\nu^{(5D)}$ as given in
\eqref{metric5d} and \eqref{gennu5D}, for $p=q+1$ we should use
\eqref{met5D2} and \eqref{gennu5D}, and for $p=q$ we should use
\eqref{pqcaseU} and \eqref{pqcasenu}.

We note that there is a subtlety in relating the five- and
six-dimensional solutions: In five dimensions, the parameters
$a_1,...,a_N$ are of dimension length, while in six dimensions,
the parameters $a_1,...,a_N$ are of dimension length squared.
Therefore, it is important to remark that
Eqs.~\eqref{potD6}-\eqref{nu5nu6}, and similar formulas below,
should be understood in the sense that we formally use the same
formulas as for the five-dimensional solutions, but with the
parameters $a_1,...,a_N$ of dimension length squared formally
inserted into the expression obtained for the five-dimensional
$(p,q)$ solutions. For the physical quantities one should use then
the six-dimensional Newton's constant.

\subsection{Regularity and topology of Kaluza-Klein bubbles \label{s:regu6d}}

We examine in this section the behavior of the $(p,q)$ solutions
near the Kaluza-Klein bubbles. For each of the $p$ Kaluza-Klein
bubbles we have a possible conical singularity which is
absent only if the periodicity of $\phi$ is chosen appropriately. For a
given period $L$ of the circle parameterized by $\phi$, this gives $p$
constraints on the parameters of our solution.

If we consider a Kaluza-Klein bubble corresponding to a finite rod
$[z_1,z_2]$ sourcing $U_2$ in a six-dimensional $(p,q)$ solution,
we get for $z_1 < z < z_2$ in the limit $r\rightarrow 0$ the
metric
\begin{equation}
\label{bubmet6D} ds^2 = - g(z) dt^2 + f_1(z) d\psi^2 + f_2(z)
d\chi^2 + \frac{1}{f_1(z)f_2(z)g(z)} \left[ r^2 d\phi^2 + c^2 (
dr^2 + dz^2 ) \right]
\end{equation}
with $f_1(z)$, $f_2(z)$ and $g(z)$ being functions and $c$ a
number. This can be seen using Eqs.~\eqref{Rzero}-\eqref{Yzero}.
If we take $\phi$ to have period $L$, we see now that we need $L =
2\pi c$ in order to avoid a conical singularity.

Using the results for the five-dimensional case, we can easily
find the explicit condition for the period of $\phi$ imposed by the
requirement of regularity of the $k$'th bubble in a
six-dimensional $(p,q)$ solution. From
\eqref{potD6}-\eqref{nu5nu6} we see that regularity
on the $k$'th bubble requires $\phi$ to have period
\begin{equation}
\label{delphik} (\Delta \phi)_k = \frac{1}{\sqrt{2(a_N-a_1)}}
(\Delta \phi^{(5D)})_k
\end{equation}
for $k=1,...,p$, where $(\Delta \phi^{(5D)})_k$ is given by
\eqref{genL}, \eqref{genL2}, or \eqref{genL3}, depending on whether
we consider the case $p=q-1$, $p=q+1$ or $p=q$. For a given period
$L$ of $\phi$, the $p$ constraints on the parameters $a_1,...,a_N$
are then $L = (\Delta \phi)_k$, for $k=1,...,p$.

We can read off the topology of the Kaluza-Klein bubble from the
metric \eqref{bubmet6D}. For $r=0$ and fixed time $t$, the metric
on the bubble is
\begin{equation}
(ds_3)^2 = f_1(z) d\psi^2 + f_2(z) d\chi^2 +
\frac{c^2}{f_1(z)f_2(z)g(z)} dz^2 \ .
\end{equation}
For the $(p,q)$ solutions we know that if $z_1=a_1$ we have that
$f_1(z_1) = 0$, and that if $z_2=a_N$ we have that $f_2(z_2)=0$,
and otherwise $f_1(z)$ and $f_2(z)$ are non-zero. On the other
hand, $g(z)$ goes to zero in one of the endpoints $z_1$ or $z_2$
if the bubble is connected to an event horizon in that endpoint.
{}From this we have three possible cases:
\begin{itemize}
\item The pure bubble space-time.
This corresponds to the
case $(p,q)= (1,0)$. In this case $g(z)$ is non-zero, while
$f_i(z) \rightarrow 0$ for $z\rightarrow z_i$ for $i=1,2$. This
gives that the topology of the bubble is a three-sphere $S^3$.
\item The bubble sits between two event horizons. In this case
$g(z)$ goes to zero for $z \rightarrow z_1,z_2$, while $f_1(z)$
and $f_2(z)$ are non-zero. Thus $\psi$ and $\chi$ each parameterize a
circle, and $z$ parameterizes an interval $I$. The proper distance
between the two event horizons is
\bea
  s = c \int_{z_1}^{z_2} \big[ f_1(z) f_2(z) g(z)\big]^{-1/2} dz \, .
\eea
The topology of the bubble is a torus times an interval,
$T^2 \times I$, where $T^2 = S^1 \times S^1$ is a rectangular torus.
\item The bubble is located at either end of the bubble-black hole
sequence, so that it
has an event horizon only on one side. Assume
without loss of generality that $z_1=a_1$, i.e.~that the bubble
sits at the left end and has an event horizon to the
right of it. Then $f_1(z)$ goes to zero for $z\rightarrow
z_1$, $f_2(z)$ is non-zero, and $g(z)$ goes to zero for $z
\rightarrow z_2$. This gives the topology of a disk times a circle,
$D \times S^1$, for the bubble.
\end{itemize}

\subsection{Event horizons, topology and thermodynamics \label{s:eh6D}}

We discuss here the $q$ event horizons that are present in a
$(p,q)$ solution. As stated above, each event horizon corresponds
to a finite rod sourcing the potential $U_1$.

Consider an event horizon corresponding to a rod $[z_1,z_2]$
sourcing $U_1$ in a six-dimensional $(p,q)$ solution. For
$r\rightarrow 0$ and with $z_1 < z < z_2$ we have
\begin{equation}
\label{BHmet6D} ds^2 = g(z) d\phi^2 + f_1(z) d\psi^2 + f_2(z)
d\chi^2 + \frac{1}{f_1(z)f_2(z)g(z)} \left[ - r^2 dt^2 + c^2 (
dr^2 + dz^2 ) \right]
\end{equation}
with $f_1(z)$, $f_2(z)$ and $g(z)$ being functions and $c$ a
number. That the six-dimensional $(p,q)$ solution becomes of the
form Eq.~\eqref{BHmet6D} can be seen using
Eqs.~\eqref{Rzero}-\eqref{Yzero}.

By Wick rotating the time-coordinate we see that the period of the
Wick-rotated time $it$ should be $2\pi c$ in order to avoid a
conical singularity. This means that the horizon has an inverse
temperature $\beta = 1/T = 2\pi c$.

We can use the results for the five-dimensional $(p,q)$ solutions
of Section \ref{s:eh5D} to find the temperatures for the event
horizons of six-dimensional $(p,q)$ solutions. Using the
\refeq{potD6}-\refeq{nu5nu6}, we get that the inverse temperature of
the $k$'th event horizon of the six-dimensional $(p,q)$ solution is
\begin{equation}
\label{invtemp6D} \beta_k = \frac{1}{T_k} =
\frac{1}{\sqrt{2(a_N-a_1)}} \beta^{(5D)}_k
\end{equation}
with $k=1,...,q$, where $\beta^{(5D)}_k$ is given by
\eqref{betak1}, \eqref{betak2}, or \eqref{betak3} depending on
whether we are considering the case $p=q-1$, $p=q+1$, or $p=q$.

If we consider the metric \eqref{BHmet6D} for $r=0$ and fixed $t$,
we get the metric for the event horizon
\begin{equation}
\label{ds42} (ds_4)^2 = g(z) d\phi^2 + f_1(z) d\psi^2 + f_2(z)
d\chi^2 + \frac{c^2 }{f_1(z)f_2(z)g(z)} dz^2 \ .
\end{equation}
We can now find the topology of the event horizon from this metric
by considering the behavior of $f_1(z)$, $f_2(z)$ and $g(z)$ for
$z\rightarrow z_1,z_2$. We have three cases:
\begin{itemize}
\item No bubbles present. This is the case $(p,q)= (0,1)$
corresponding to the uniform black string.
The function $g(z)$ is non-zero, while
$f_i(z) \rightarrow 0$ for $z\rightarrow z_i$ for $i=1,2$. This
gives that the topology of the event horizon is $S^3 \times S^1$,
where the $S^1$ is parameterized by the $\phi$-coordinate of the
Kaluza-Klein circle.
\item The event horizon has a bubble on both sides. In this case
$g(z)$ goes to zero in both endpoints $z_1$ and $z_2$, while
$f_1(z)$ and $f_2(z)$ are non-zero. Hence $(\phi,z)$ parameterizes a
two-sphere $S^2$, and $\psi$ and $\chi$ each parameterize an
$S^1$. Note that the size of these $S^1$'s depends on $z$, but they
never close off to zero size.
The topology of the event horizon is $S^2 \times T^2$ with $T^2 = S^1
\times S^1$. Note that neither of the $S^1$'s are topologically
supported (i.e.~are not wrapping the Kaluza-Klein direction). We call
black holes with this topology \emph{black tuboids}. We will discuss
features of this new horizon topology in Section \ref{sec12}.
\item The event horizon is at either end of the bubble-black hole
sequence. Assume without loss of generality that $z_1=a_1$,
i.e.~that the event horizon is at the beginning of the sequence.
Then $f_1(z)$ goes to zero for $z\rightarrow z_1$, $f_2(z)$ is
non-zero, and $g(z)$ goes to zero for $z \rightarrow z_2$. Hence
$(\phi,\psi,z)$ parameterizes a three-sphere $S^3$, while $\chi$
parameterizes an $S^1$. The horizon topology is $S^3 \times S^1$,
and since the $S^1$ is not topologically supported, the black hole
is a black ring.
\end{itemize}

In the last two cases we have black holes, tuboids and rings, for
which the $S^1$'s of the horizons are not wrapped on the
Kaluza-Klein circle. These $S^1$'s are instead supported by the
Kaluza-Klein bubbles which keep the configurations in static
equilibrium. We discuss examples in Section \ref{s:prop}.

Using the above analysis of bubbles and black holes, we see that the
structure of the $(p,q)$ solutions is as follows. For $p=q-1$ with $q
\geq 2$, the configuration looks like:
\[
  \begin{array}{ccccccccc}
  \rom{black~ring}&-&\rom{bubble} &-& \rom{black~tuboid}
  &\ \cdots \ &\rom{bubble} &-& \rom{black~ring} \\
  S^3 \times S^1 &&  T^2 \times I  && S^2 \times T^2 &\ \cdots \ &   T^2 \times I   && S^3 \times S^1
  \end{array}
\]
For $p=q+1$, with $p \geq 2$, we have instead:
\[
  \begin{array}{ccccccccc}
  \rom{bubble}&-&\rom{black~tuboid} &-& \rom{bubble}
  &\ \cdots \ &\rom{black~tuboid} &-& \rom{bubble} \\
  D \times S^1 &&  S^2 \times T^2  && T^2 \times I &\ \cdots \ & S^2 \times T^2 && D \times S^1
  \end{array}
\]
Finally, for $p=q$, we have:
\[
  \begin{array}{ccccccc}
  \rom{black~ring}&-&\rom{bubble} &-& \rom{black~tuboid}
  &\ \cdots \ &\rom{bubble}  \\
  S^3 \times S^1 &&  T^2 \times I  && S^2 \times T^2 &\ \cdots \ &  D \times S^1
  \end{array}
\]

We can also find the entropy associated with the event horizon
from the metric \eqref{ds42} by computing the area. Since the
square-root of the determinant of the metric \eqref{ds42} is equal
to $c$, we get the entropy
\begin{equation}
S = \frac{(2\pi)^2 L (z_2-z_1) c}{4 G_{\rm N}} \, .
\end{equation}
Using that the temperature $T = 1/(2\pi c)$ we get moreover
\begin{equation}
\label{TS6D} TS = \frac{2\pi L (z_2-z_1)}{4 G_{\rm N}} \, .
\end{equation}
We can use this to find the entropy for the $k$'th event horizon
in a $(p,q)$ solution. If $p=q-1$ or $p=q$, the $k$'th event
horizon has entropy
\begin{equation}
\label{Sk16D} S_k = \beta_k \frac{2\pi L (a_{2k}-a_{2k-1})}{4
G_{\rm N}} \ ,
\end{equation}
with $k=1,...,q$ and where $\beta_k$ is the inverse temperature of
the $k$'th event horizon given in Eq.~\eqref{invtemp6D}. If
$p=q+1$ we get instead
\begin{equation}
\label{Sk26D} S_k = \beta_k \frac{2\pi L (a_{2k+1}-a_{2k})}{4
G_{\rm N}} \ ,
\end{equation}
with $k=1,...,q$ and $\beta_k$ given by Eq.~\eqref{invtemp6D}.

\subsection{Asymptotics \label{s:asym6d}}

The asymptotic region of the six-dimensional $(p,q)$ solutions is
$\sqrt{r^2 + z^2 } \rightarrow \infty$. From the $(p,q)$ solutions
given by Eqs.~\eqref{potD6}-\eqref{nu5nu6} it is easy to see that
they asymptote to the $\CM^5 \times S^1$ solution
\eqref{flatD6}-\eqref{flatD6nu} for
$\sqrt{r^2 + z^2 } \rightarrow \infty$.%
\footnote{Note that this means $\psi$ and $\chi$ both have periods
$2\pi$, which one can also get by considering the two
semi-infinite rods $[-\infty,a_1]$ sourcing $U_3$, and
$[a_N,\infty]$ sourcing $U_4$ and demanding regularity of
the solution.}

Using the identity \eqref{5dasymp1} we see that the $U_1$ and
$U_2$ potentials for $(p,q)$ solutions become of the form
\eqref{asyD61} for $\sqrt{r^2 + z^2 } \rightarrow \infty$ and we
can read off $c_t$ and $c_\phi$. For $p=q-1$ and $p=q$, we find
\begin{equation}
\label{ct1_6D} c_t = 2 \sum_{k=1}^{q} (a_{2k} - a_{2k-1}) \spa
c_\phi = - 2\sum_{k=1}^{p} (a_{2k+1}- a_{2k}) \ ,
\end{equation}
while for $p=q+1$ we find
\begin{equation}
\label{ct2_6D} c_t = 2 \sum_{k=1}^{q} (a_{2k+1} - a_{2k}) \spa
c_\phi = - 2\sum_{k=1}^{p} (a_{2k}- a_{2k-1}) \ .
\end{equation}
{}From the above we see that $c_t$ is two times the sum of the
lengths of the rods giving the event horizons, while $-c_\phi$ is
two times the sum of the lengths of the rods giving the bubbles.
It follows from this that $c_t - c_\phi = 2 (a_N - a_1 )$. Using
\refeq{mun} we can now determine the dimensionless mass $\mu$ and
the relative binding energy $n$ from
\begin{equation}
\label{mu6} \mu = \frac{2 \pi^2}{L^2} [ 3 c_t - c_\phi] \spa n =
\frac{c_t - 3 c_\phi}{3c_t - c_\phi } \ .
\end{equation}
Note that we have $c_t > 0$ and $c_\phi < 0$. This means that $1/3
< n < 3$ (for $p,q \geq 1$). Finally, we can use
\eqref{Sk16D}-\eqref{Sk26D} together with
\eqref{ct1_6D}-\eqref{ct2_6D} to get the generalized Smarr formula
for six-dimensional $(p,q)$ solutions
\bea \label{gendsmarr6D}
  \sum_{k=1}^{q}\mathfrak{t}_k \mathfrak{s}_k
  = \frac{3-n}{4} \mu \ .
\eea
The dimensionless temperature $\mathfrak{t}_k$ and entropy $\mathfrak{s}_k$
are defined in \eqref{tsdef}. The first law of thermodynamics takes
the form given in \eqref{Gen1stLaw}.

\subsection{Map between five- and six-dimensional physical quantities \label{secmap56}}

We construct in this section a map between the five-dimensional
$(p,q)$ solutions and the six-dimensional $(p,q)$ solutions. The
map can take any $(p,q)$ solution in five dimensions and map it to a
$(p,q)$ solution in six dimensions, and vice versa.

To write down the map we need to be careful regarding the subtlety
mentioned at the end of Section \ref{s:6dsol} that the parameters
$a_1,...,a_N$ have dimension
length in five dimensions and dimension length squared in six
dimensions. This means that we cannot in general identify the
$a_i$ parameters in five and six dimensions. Thus, in general we
should introduce a set of parameters $a^{(5D)}_i$, $i=1,...,N$,
for five dimensions and another set $a_i$, $i=1,...,N$, for six
dimensions, and then fix a relation between them to make the map.
However, the simplest way of dealing with this is to fix
$a^{(5D)}_N - a^{(5D)}_1 = 1$ and $a_N - a_1 = 1$. This can be
regarded as a choice of units. With this choice the map between
five- and six-dimensional solutions is simply given by%
\footnote{Note that to obtain the physical quantities from
the dimensionless quantities discussed below one should
use of course the appropriate five or six-dimensional Newton's constant.}
\begin{equation}
\label{amap} a_i = a^{(5D)}_i \ , i=1,...,N \, .
\end{equation}
That the map from five to six dimensions works is basically a
consequence of Eq.~\eqref{delphik}, which in turn is a consequence of
the similarity between the rod configurations in the five- and six-dimensional
case. Under the map \eqref{amap} we see
that a regular five-dimensional $(p,q)$ solution is mapped to a
regular six-dimensional $(p,q)$ solution, and vice versa, since
$(\Delta \phi)_k / (\Delta \phi^{(5D)})_k$ does not depend on $k$.

We now write down how the physical parameters of the solutions
transform under the map \eqref{amap}. From \eqref{delphik} we see
that
\begin{equation}
\label{circs5to6} L^{(6D)} = \frac{1}{\sqrt{2}} L^{(5D)} \, ,
\end{equation}
where $L^{(5D)}$ ($L^{(6D)}$) is the period of the
circle parameterized by $\phi$ for the five (six) dimensional $(p,q)$
solution. Using that $c^{(6D)}_t = 2 c^{(5D)}_t$ and
$c^{(6D)}_\phi = 2 c^{(5D)}_\phi$ we get
\begin{equation}
\label{mun5to6} \mu^{(6D)} = \frac{2\pi}{3L^{(5D)}} \mu^{(5D)}
\left( 5 - n^{(5D)} \right) \spa n^{(6D)} =
\frac{5n^{(5D)}-1}{5-n^{(5D)}} \, .
\end{equation}
For the inverse temperatures we see from \eqref{invtemp6D} that
$\beta^{(6D)}_k = \beta_k^{(5D)} / \sqrt{2}$. Using this, we get for the
dimensionless temperature and entropy (defined in \refeq{tsdef})
\begin{equation}
\label{ts5to6} \mathfrak{t}^{(6D)}_k = \mathfrak{t}^{(5D)}_k \spa
\mathfrak{s}^{(6D)}_k = \frac{4\pi}{L^{(5D)}}
\mathfrak{s}^{(5D)}_k \, .
\end{equation}
In summary, we see that Eqs.~\eqref{circs5to6}-\eqref{ts5to6} map
the physical parameters of a five-dimensional $(p,q)$ solution to
those of a six-dimensional $(p,q)$ solution. One can easily check
that the Smarr formulas \eqref{gendsmarr} and \eqref{gendsmarr6D}
for five and six dimensions are consistent with this map.

Note that we can also write down
Eqs.~\eqref{circs5to6}-\eqref{ts5to6} without choosing $a^{(5D)}_N
- a^{(5D)}_1 = 1$ and $a_N - a_1 = 1$. One simply substitutes
$L^{(5D)}$ with $L^{(5D)}/(a^{(5D)}_N - a^{(5D)}_1)$ and
$L^{(6D)}$ with $L^{(6D)}/\sqrt{a_N - a_1}$.

We can also formulate the map \eqref{circs5to6}-\eqref{ts5to6}
on the level of curves, in the sense that
given the curves $\mu^{(5D)} (n^{(5D)})$ and $\mathfrak{s}^{(5D)} (n^{(5D)})$
and $L^{(5D)} (n^{(5D)})$ for
a five-dimensional $(p,q)$ solution, the
curves for the corresponding six-dimensional $(p,q)$
solution are given by
\bea
\label{5to6mu}
\mu^{(6D)}(n^{(6D)}) = \frac{16 \pi}{5 +n^{(6D)} }
\frac{\mu^{(5D)} \big( (5n^{(6D)} +1)/(5 + n^{(6D)})\big)}{
L^{(5D)}\big( (5n^{(6D)} +1)/(5 + n^{(6D)})\big) } \ ,
\eea
\bea
\label{5to6s}
\mathfrak{s} ^{(6D)}(n^{(6D)}) = 4 \pi
\frac{\mathfrak{s}^{(5D)} \big( (5n^{(6D)} +1)/(5 + n^{(6D)})\big)}{
L^{(5D)} \big( (5n^{(6D)} +1)/(5 + n^{(6D)})\big)} \ .
\eea
The curve for $L^{(6D)}(n^{(6D)})$ can easily be found from
\eqref{circs5to6}.

\section{Solutions with equal temperatures \label{s:eqt} }

In this section we consider $(p,q)$ solutions for
which the temperatures for the $q$ event horizons all
are equal.
Apart from being interesting in their own right, they have
two important properties, as we show in the following.
Firstly, we explain that the equal temperature solutions
precisely are the class of regular $(p,q)$ solutions
that transform into regular $(q,p)$ solutions after
a double Wick rotation of the $t$ and $\phi$ directions.
Secondly, we show that for fixed mass the entropy
is extremized for the equal temperature solution.

\subsection{Double Wick rotation of $(p,q)$ solutions as a duality map
\label{secWick} }

Consider a $(p,q)$ solution in five or six dimensions.
Let $\beta_1,...,\beta_q$ be the inverse temperatures
of the $q$ event horizons and let
$(\Delta \phi)_1,...,(\Delta \phi)_p$ correspond to the
periods of $\phi$ associated with each of the $p$ Kaluza-Klein bubbles
(see Sections \ref{s:regu5d} and \ref{s:regu6d}).
Regularity of the solution requires that for given size $L$ of the
Kaluza-Klein circle at infinity $(\Delta \phi)_k = L$ for all
$k$. For now, assume that these $p$ constraints are not necessarily
obeyed. Then, if we make the double Wick rotation
$t' = i\phi$ and $\phi' = it$, we see that since the metric
is independent of $t$ and $\phi$, this correspond to the transformation
\begin{equation}
\label{doubWick}
g_{t't'} = - g_{\phi \phi} \spa g_{\phi'\phi'} = - g_{tt} \ ,
\end{equation}
with all other components of the metric unchanged.
Comparing for example Eq.~\eqref{nearr0} with Eq.~\eqref{timerod}
for $D=5$ (or in terms of rod configurations, compare for example
Figure \ref{fig5D1} and Figure \ref{fig5D2}), we see
that the double Wick rotation exchanges Kaluza-Klein bubbles and
event horizons.
Thus, the $(p,q)$ solution transforms into a $(q,p)$ solution.
Moreover, the temperatures and the periods are interchanged:
\begin{equation}
\label{betaphitr}
\beta'_k = (\Delta \phi)_k,\ k=1,...,p \ , ~~~~
(\Delta \phi')_l = \beta_l ,\ l=1,...,q \ .
\end{equation}
Now, if we want to consider a regular $(p,q)$ solution
transforming to a regular $(q,p)$ solution, we see from
Eq.~\eqref{betaphitr} that we need to have
$(\Delta \phi)_1=\cdots =(\Delta \phi)_p = L$
and moreover, $\beta_1 = \cdots = \beta_q = \beta$,
i.e. in addition to the conditions on the $\phi$
periods, we need to impose that all the (inverse) temperatures
of the event horizons are equal. It is always possible  to choose the
$q$ free parameters of the $(p,q)$ solution to give a one-parameter family of
regular equal temperature solutions, which we shall denote by $(p,q)_{\mathfrak{t}}$.
Note that $\beta$ is defined here as the inverse temperature of
all of the event horizons in the given configuration.
Obviously, the transformed solution, which is a regular $(q,p)$ solution,
also has equal temperatures, and we get the transformation law:
\begin{equation}
\label{dW1}
\beta' = L \spa L' = \beta \ .
\end{equation}
{}From Eq.~\eqref{doubWick} and Eq.~\eqref{gttzz} we see that
under the double Wick rotation
\begin{equation}
\label{mapc}
c_t' = - c_\phi \, , ~~\mbox{and}~~ c_\phi' = - c_t \, .
\end{equation}
Using this, we get the following map between regular equal temperature
$(p,q)_{\mathfrak{t}}$ and $(q,p)_{\mathfrak{t}}$ solutions:
\begin{equation}
\label{dW2}
\mu' = \frac{L^{d-3}}{\beta^{d-3}} n \mu \spa
n' = \frac{1}{n} \spa
\mathfrak{t}' = \frac{1}{\mathfrak{t}} \spa
\mathfrak{s}' = \frac{(d-2)n-1}{d-2-n} \frac{L^{d-1}}{\beta^{d-1}}
\mathfrak{s} \, ,
\end{equation}
where the transformation rule for the total rescaled entropy
$\mathfrak{s} = \sum_k \mathfrak{s}_k$
can be found from the Smarr formula Eq.~\eqref{smarr1}.
The simplicity of the map for $n$ follows directly using \eqref{mapc}
in the expression \eqref{mun} for $n$, and that for $\mathfrak{t}$
follows from $\mathfrak{t} = L/\beta$ using \eqref{dW1}.

We can also formulate the duality map \eqref{dW1}-\eqref{dW2}
on the level of curves $\mu(n)$ and $\mathfrak{s}(n)$,
so that given an equal temperature $(p,q)_{\mathfrak{t}}$
 solution with curves
$\mu(n)$ and $\mathfrak{s}(n)$ the corresponding
curves of the dual equal temperature $(q,p)_{\mathfrak{t}}$ solution are
\bea
\label{dualmu}
\mu'(n') = \frac{1}{(d-1)^{d-3}}
\left[ \frac{ n'(d-2) -1}{\mathfrak{s} (1/n')} \right]^{d-3}
\left[ \frac{ \mu (1/n') }{n'} \right]^{d-2} \ ,
\eea
\bea
\label{duals}
\mathfrak{s}'(n') = \frac{d-2-n'}{(d-1)^{d-1}}
\left[ \frac{ n'(d-2) -1}{\mathfrak{s} (1/n')} \right]^{d-2}
\left[ \frac{ \mu (1/n') }{n'} \right]^{d-1} \ .
\eea
As a check, we note that when applying now the Smarr formula to compute
the temperature of the dual solution, one indeed finds that
$\mathfrak{t}'(n') = [\mathfrak{t}(1/n')]^{-1}$.
Note also the interesting fact that curves of equal temperature
$(p,p)_{\mathfrak{t}}$ solutions have the property that they are self-dual, i.e.
invariant under the transformations \eqref{dualmu}-\eqref{duals}.

In conclusion, we have thus found that a regular $(p,q)_{\mathfrak{t}}$
solution with equal temperatures for the event horizons
transforms into a regular $(q,p)_{\mathfrak{t}}$ solution, also with equal temperatures,
with the physical parameters transforming according
to Eqs.~\eqref{dW1} and \eqref{dW2}, or, equivalently,
Eqs.~\eqref{dualmu}-\eqref{duals}.

\subsection{Solutions with equal temperatures extremize the entropy
\label{secext}}

In this section we consider the entropy $\mathfrak{s}$ for a given
mass $\mu$. We find that for fixed mass, the solutions with equal
temperatures extremize the entropy.

Consider a regular $(p,q)$ solution in five or six dimensions
with $q$ temperatures $\mathfrak{t}_k$
and entropies $\mathfrak{s}_k$, $k=1,...,q$.
Define the total entropy
\begin{equation}
\mathfrak{s} = \sum_{k=1}^q \mathfrak{s}_k \, .
\end{equation}
As explained in Sections \ref{s:regu5d} and \ref{s:regu6d},
a solution is specified by $q$ dimensionless parameters and
therefore $\mathfrak{s}$ is a function of $q$ dimensionless parameters.
Taking $\mu$ to be one of the parameters we can find
$q-1$ dimensionless parameters $x_1,x_2,...,x_{q-1}$
so that we can write
\begin{equation}
\mathfrak{s} = \mathfrak{s}(\mu,x_1,...,x_{q-1}) \, .
\end{equation}
Let now a $\mu$ be given.
Choose for this $\mu$ the parameters $x_1,...,x_{q-1}$
so that all the temperatures are equal, i.e.
$\mathfrak{t}_1 = \mathfrak{t}_2 = \cdots = \mathfrak{t}_q =
\mathfrak{t}$.
This specifies uniquely a solution.
If we then consider another solution
$(\mu',x_1',...,x_{q-1}')$ in the neighborhood of this solution
we have from the first law of thermodynamics \eqref{Gen1stLaw} that
\begin{equation}
\delta \mu = \sum_{i=1}^q \mathfrak{t}_i \delta \mathfrak{s}_i =
\mathfrak{t} \delta \mathfrak{s}
\end{equation}
with $\mu' = \mu + \delta \mu$. If we furthermore consider a solution
with $\mu' = \mu$, we see from this that
$\delta \mathfrak{s} = 0$. This means that
\begin{equation}
\frac{\partial \mathfrak{s}}{\partial x^i} = 0
\end{equation}
for $i=1,...,q-1$,
with $\mathfrak{s}$ being a function of $(\mu,x_1,...,x_{q-1})$.
Therefore, for a given $\mu$, $\mathfrak{s}(\mu,x_1,...,x_{q-1})$
is at an extremum for a solution with equal temperatures.
For all the cases we have examined explicitly (see Section
\ref{s:prop}) we find that this extremum is a minimum.

\section{Properties of specific solutions \label{s:prop}}

Having obtained the solutions for sequences with $p$ bubbles and $q$
black holes in five and six dimensions, we now turn to examine in more
detail the properties of specific solutions. In particular, we
consider the first few $(p,q)$ solutions for low values of $p$ and $q$
 and focus on how these solutions appear in the
$(\mu,n)$ phase diagram. We also discuss the thermodynamics of
these solutions in some detail.

In the analysis of $(p,q)$ solutions below the general results of
Sections \ref{s:5dKK}, \ref{s:6dKK} and \ref{s:eqt} are used throughout.
In particular we note here that:
\begin{enumerate}
\item In specifying the rod structure for a given $(p,q)$ solution
by the positions $a_i$, $i=1\ldots N=p+q-1$, we use translational
invariance
 to set $a_1=0$, while the scale is set by choosing $a_N=1$.
\item The rod structure and corresponding exact expressions of
the solutions we discuss can be found in Sections \ref{s:5dsol}
and  \ref{s:6dsol} for the five and six-dimensional case
respectively.
\item The topology of the bubbles and event horizons in the configurations
follows from the general discussion in Sections \ref{s:regu5d},
\ref{s:eh5D} and \ref{s:regu6d}, \ref{s:eh6D}.
\item The physical quantities we obtain are the dimensionless mass
$\mu$ and relative tension $n$, which follow
from the results of Sections \ref{s:asym5d} and \ref{s:asym6d}, while
the entropy $\mathfrak{s}$ and temperature $\mathfrak{t}$
follow from the analysis in Sections
\ref{s:eh5D} and \ref{s:eh6D} respectively.
 \item For each choice of $(p,q)$, we first treat the five-dimensional
case, and then apply to it the mapping of Section \ref{secmap56} to
obtain the results for the corresponding six-dimensional solutions.
\item For equal temperature solutions, the duality map of
Section \ref{secWick} is used to find the properties of the
$(p+1,p)_\mathfrak{t}$ solution
 from those of the $(p,p+1)_\mathfrak{t}$ case.
\end{enumerate}

\subsection{Configuration with $(p,q)=(1,1)$ \label{sec11} }

Since the $(0,1)$ and $(1,0)$ solution correspond to the uniform black
string and the static bubble, respectively, the simplest
non-trivial case is the $(1,1)$ solution.

\subsubsection*{$(1,1)$ configuration in $D=5$: black hole on a bubble}

\begin{figure}[htb]
   \centerline{\epsfig{file=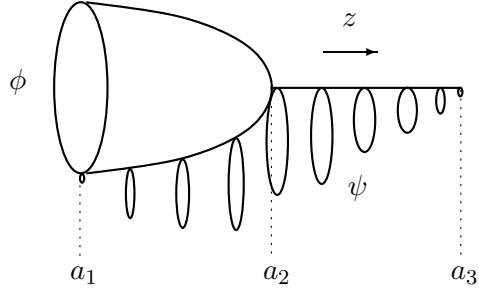,width=5.5cm}}
   \caption{Sketch of the $(1,1)$ configuration describing
a single black hole attached to a Kaluza-Klein bubble.}
   \label{sketch11}
\vspace{2mm}
\begin{picture}(0,0)(0,0)
\put(119,130){$\phi$}
\put(247,90){$\psi$}
\put(245,154){$z$}
\put(238,144){\vector(1,0){20}}
\put(142,58){$a_1$}
\put(215,58){$a_2$}
\put(286,58){$a_3$}
\end{picture}
\end{figure}

The $(1,1)$ solution in five dimensions describes a single black hole on
 a Kaluza-Klein bubble \cite{Emparan:2001wk}. The exact metric is
 given in \eqref{pqcaseU}, \eqref{pqcasenu} with $N=3$.
  As discussed
 in Section \ref{s:regu5d}, the topology of the black hole is a
 three-sphere $S^3$ and the bubble is a disk:%
\footnote{The coordinates of \eqref{pqcaseU} and \eqref{pqcasenu} do not cover the full space-time.
In analogy
to the maximally extended Schwarzschild geometry, one can extend the
geometry to include a second copy of the space-time on the other side
of the black hole throat. This is naturally done by extending the
coordinate $r$ to negative values, and noting that the metric is
symmetric under $r \to -r$. In the extended space-time, there is
another copy of the bubble disk for $r<0$, and the two disks join up
smoothly at $r=0$ to form a two-sphere. We can then say that
the extended geometry of the $(1,1)$ solution describes the configuration
\begin{eqnarray*}
  \begin{array}{ccc}
  \rom{black~hole}&-&\rom{bubble} \\
  S^3 &&  S^2
  \end{array}
\end{eqnarray*}
}
\begin{equation}
\label{11config}
  \begin{array}{ccc}
  \rom{black~hole}&-&\rom{bubble} \\
  S^3 &&  D
  \end{array}
\end{equation}
It is important to note that
the black hole is not localized on the KK circle,
and for a given size of the circle at
infinity the black hole can be  arbitrarily big.
In Figure \ref{sketch11} we have sketched the geometry of the
$(1,1)$ configuration.

\begin{figure}[t]
\centerline{\epsfig{file=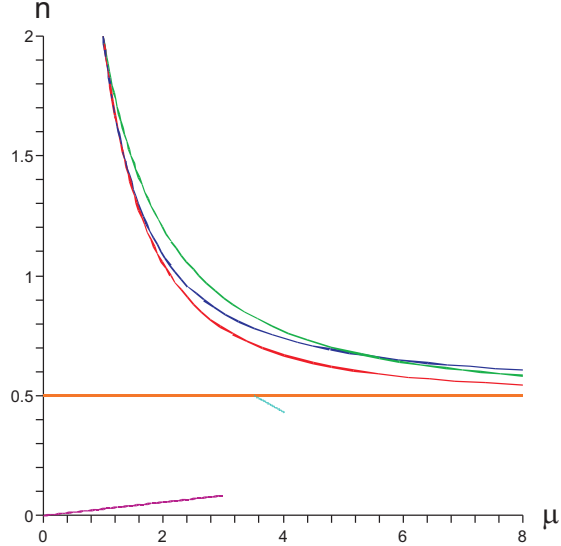,width=8cm,height=8cm}}
\caption{Curves in the $(\mu,n)$ phase diagram for the
$(p,q)=(1,1)$, $(1,2){}_{\mathfrak{t}}$ and $(2,1)$ solutions in
five dimensions. These curves lie in the region $1/2 < n \leq 2$.
The lowest (red) curve corresponds to the $(1,1)$ solution. The
(blue) curve that has highest $n$ for high values of $\mu$ is the
equal temperature $(1,2){}_{\mathfrak{t}}$ solution. The (green)
curve that has highest $n$ for small values of $\mu$ is the
$(2,1)$ solution. The entire phase space of the $(1,2)$
configuration is the wedge bounded by the equal temperature
$(1,2){}_{\mathfrak{t}}$ curve and the $(1,1)$ curve. For
completeness we have also included the uniform black string phase
and sketched the non-uniform black string phase in the region $0 \leq n \leq 1/2$.
In this region we also plot the small black hole branch using the
results of \cite{Harmark:2003yz}.}
 \label{fig5D}
\end{figure}

We now derive the curve for this solution in the $(\mu,n)$ phase diagram.
Using the conventions described in the beginning of this section, the
configuration has rod structure specified
by
\begin{equation}
(a_1,a_2,a_3 ) = (0,x,1) \spa 0 < x < 1 \, .
\end{equation}
Inserting this in \eqref{ct1_5D} we find  $c_t = x $, $c_\phi = x -1$, and hence
we obtain from \eqref{mu5}, \eqref{genL3}  the dimensionless mass,
relative tension and (rescaled) length of the compact circle
\begin{equation}
\label{mun11} \mu = \frac{4 \pi}{L} (1+x) \spa n = \frac{2 -x}{x
+1} \spa L = 4 \pi \sqrt{1-x} \, .
\end{equation}
We substitute the expression for $L$ into $\mu$ and use the
expression for $n$ to
 eliminate the dimensionless variable $x$. This yields the curve
\begin{equation}
\label{mu11} \mu_{(1,1)} (n) =\frac{3}{\sqrt{(n+1)(2n-1)}} \spa
\frac{1}{2} < n < 2 \, ,
\end{equation}
which we have plotted in the $(\mu,n)$ phase diagram of Figure \ref{fig5D}.

Inverting \eqref{mu11} we find
\begin{equation}
\label{nmu11} n_{(1,1)}(\mu )
= \frac{1}{4} \left[ -1 + 3 \sqrt{1 + \frac{8}{\mu^2}} \right]
\end{equation}
We see  that the lowest value for $\mu$ is obtained
for the static bubble solution ($\mu = \mu_{\rm b}=1$ and $n=2$,
see Table \ref{tabmuc}), while for large $\mu$ the value
$n = 1/2$ of the uniform black string in five dimensions is approached.
We conclude that the branch of $(1,1)$ solutions consists of a curve
in the $(\mu,n)$ phase diagram that starts in the point corresponding to
the static bubble and approaches for infinite mass the horizontal line
$n=1/2$ of the uniform black string.
This will be the case for all curves that we explicitly obtain in
this section.

To obtain the other thermodynamic variables of the solution  we
use \eqref{betak3} (and the formula below it) to compute the
dimensionless entropy (defined in \eqref{tsdef}) in terms of
$x$. This result is then easily written as a function of $n$
using \eqref{mun11}, yielding
\begin{equation}
\label{s11}
 \mathfrak{s}_{(1,1)}(n) =\frac{(2-n)^{3/2}}{(n+1)^{1/2}(2n-1)} \, .
\end{equation}
{}From this we can find the temperature $\mathfrak{t}(n)$
for example, by using the
Smarr formula \eqref{gendsmarr} and the form of $\mu$ in
\eqref{mu11}. Alternatively, the temperature can
be computed directly from \eqref{betak3}, or one may use the first law of
thermodynamics \eqref{first1} to obtain $\mathfrak{t} (n) =
\frac{\partial \mu}{\partial n} \left[\frac{\partial s}{\partial
n} \right]^{-1}$. We also note that the curves
$\mu_{(1,1)}(n)$ and $\mathfrak{s}_{(1,1)}(n)$ are correctly
invariant under the map \eqref{dualmu}, \eqref{duals} respectively.

\begin{figure}[t]
\centerline{\epsfig{file=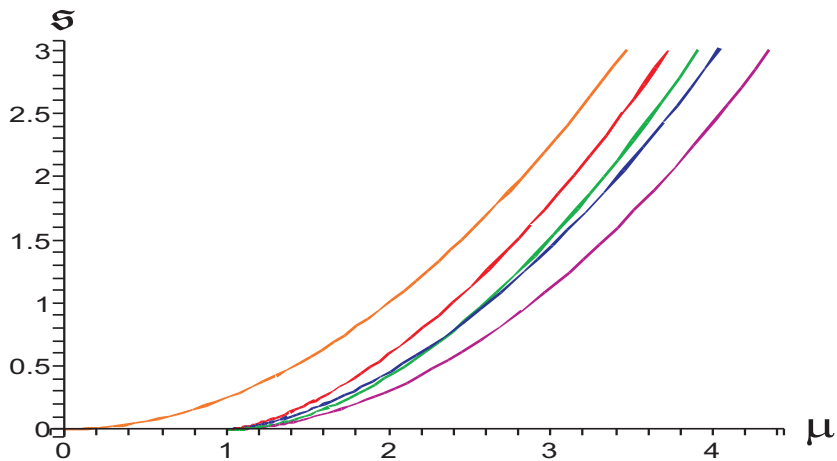,width=11cm,height=6cm}}
\caption{The dimensionless entropy $\mathfrak{s}$ as a function of
the dimensionless mass $\mu$ for the $(0,1)$, $(1,1)$, $(2,1)$,
$(1,2)_{\rm \mathfrak{t}}$ and $(2,3)_{\rm \mathfrak{t}}$
solutions in five dimensions. Here we have ordered the solutions
as they appear from left to right in the figure for high mass (in
terms of the colors used: orange, red, green, blue, purple). All
$(p,q)$ curves approach for large $\mu$ the entropy curve of the
$(0,1)$ solution, i.e. the uniform black string which has
$\mathfrak{s}(\mu) = \mu^2/4$. The entropy curve of the $(2,1)$
solution (green) intersects the $(1,2)_{\rm \mathfrak{t}}$
solution (blue) in the point $\mu = \mu_\star \approx 2.42$. For
comparison, the Gregory-Laflamme mass is $\mu_\rom{GL} \approx
3.52$. } \label{figs5d}
\end{figure}

The entropy $\mathfrak{s}$ as a function of $\mu$ can now
be obtained by combining \eqref{s11} with \eqref{mu11}, yielding
\begin{equation}
\label{lims11}
\mathfrak{s}_{(1,1)} (\mu) =
\frac{1}{2} \frac{\left(3 -\sqrt{1 + \frac{8}{\mu^2}}\right)^{3/2}}{
\left(1  +\sqrt{1 + \frac{8}{\mu^2}}\right)^{1/2}
\left(-1  +\sqrt{1 + \frac{8}{\mu^2}}\right)} \ .
\end{equation}
We have plotted the resulting curve in Figure \ref{figs5d}.
We see from \eqref{lims11} that the entropy goes to zero, as expected,
when the pure
bubble configuration is approached at $\mu=1$. Moreover,  we observe that
leading term for large $\mu$ precisely corresponds to the
(dimensionless) entropy of the uniform black string
\begin{equation}
\label{sbs5d}
\mathfrak{s}_{\rm bs} (\mu) = \frac{\mu^2}{4} \spa d =4 \, .
\end{equation}
The latter result easily follows from the definitions \eqref{tsdef}
and the standard thermodynamics of the uniform black string in
five dimensions. More genereally, the result in \eqref{lims11} explicitly
shows
\begin{equation}
\label{s11suni}
\mathfrak{s}_{(1,1)}(\mu) < \mathfrak{s}_{\rm bs}(\mu) \spa \forall
\;\, \mu \geq 1 \, ,
\end{equation}
so that the $(1,1)$ configuration has lower entropy than that of the
uniform black string.
This property can also be argued from the Intersection Rule of
Ref.~\cite{Harmark:2003dg} using the fact that in the $(\mu,n)$ phase
diagram the $(1,1)$ solution approaches
the uniform string from above for $\mu \rightarrow \infty$ (see Figure
\ref{fig5D}).
As we will see in all our examples, this is a general
feature of all the bubble-black hole sequences; some of the mass is
spent in the bubbles, and hence the bubble-black hole
sequences are going to have smaller horizon areas compared to the
uniform black string of the same mass $\mu$.

Finally, for the mapping from the five-dimensional $(1,1)$ solution to
the six-dimensional $(1,1)$ solution we need the length
of the compact circle for the $(1,1)$ solution
\begin{equation}
\label{L11}
L_{(1,1)} (n) = 4 \pi \sqrt{\frac{2n-1}{n+1}} \, ,
\end{equation}
as follows easily from \eqref{mun11}.

\subsubsection*{$(1,1)$ configuration in $D=6$: black ring on bubble}

The $(1,1)$ solution in six dimensions describes the configuration%
\footnote{Note that just as for the
five-dimensional $(1,1)$ solution, we can extend the
geometry to include another copy of the space-time on the other side
the black hole throat. That gives a second copy of the bubble disk,
and joining the two smoothly gives the geometry
\begin{eqnarray*}
  \begin{array}{ccc}
  \rom{black~ring}&-&\rom{bubble} \\
  S^3 \times S^1 &&  S^2 \times S^1
  \end{array}
\end{eqnarray*}
}
\begin{equation}
  \begin{array}{ccc}
  \rom{black~ring}&-&\rom{bubble} \\
  S^3 \times S^1 &&  D \times S^1
  \end{array}
\end{equation}
consisting of an $S^3 \times S^1$ black ring on a bubble of topology
a disk times a circle. The exact form of the metric
can be computed from \eqref{potD6},
\eqref{nu5nu6} with $N=3$. This solution was first obtained
in Ref.~\cite{Emparan:2001wk}.

The $S^1$ of the black ring does not wrap
the Kaluza-Klein circle, but it is supported by the bubble. Without
the bubble, a static black ring would collapse under its gravitational
self-attraction, but just as a bubble can hold apart two black holes,
we see that a bubble can also balance a black ring.

In order to obtain the thermodynamic results for the
six-dimensional $(1,1)$ configuration, we use the map in \eqref{5to6mu},
\eqref{5to6s} and the curves \eqref{mu11}, \eqref{s11}, \eqref{L11}
of the five-dimensional $(1,1)$ solution. Dropping
the labels ``6D'', the results are \bea
\label{mus116d}
  \mu_\rom{(1,1)}(n) = \frac{4}{3n-1} \, , \hspace{1cm}
  \mathfrak{s}_\rom{(1,1)}(n)   = \left(\frac{3-n}{3n-1}\right)^{3/2}
  \spa  1/3 < n < 3 \, ,
\eea
and one can verify again that these are correctly invariant under \eqref{dualmu},
\eqref{duals}.

\begin{figure}[t]
\centerline{\epsfig{file=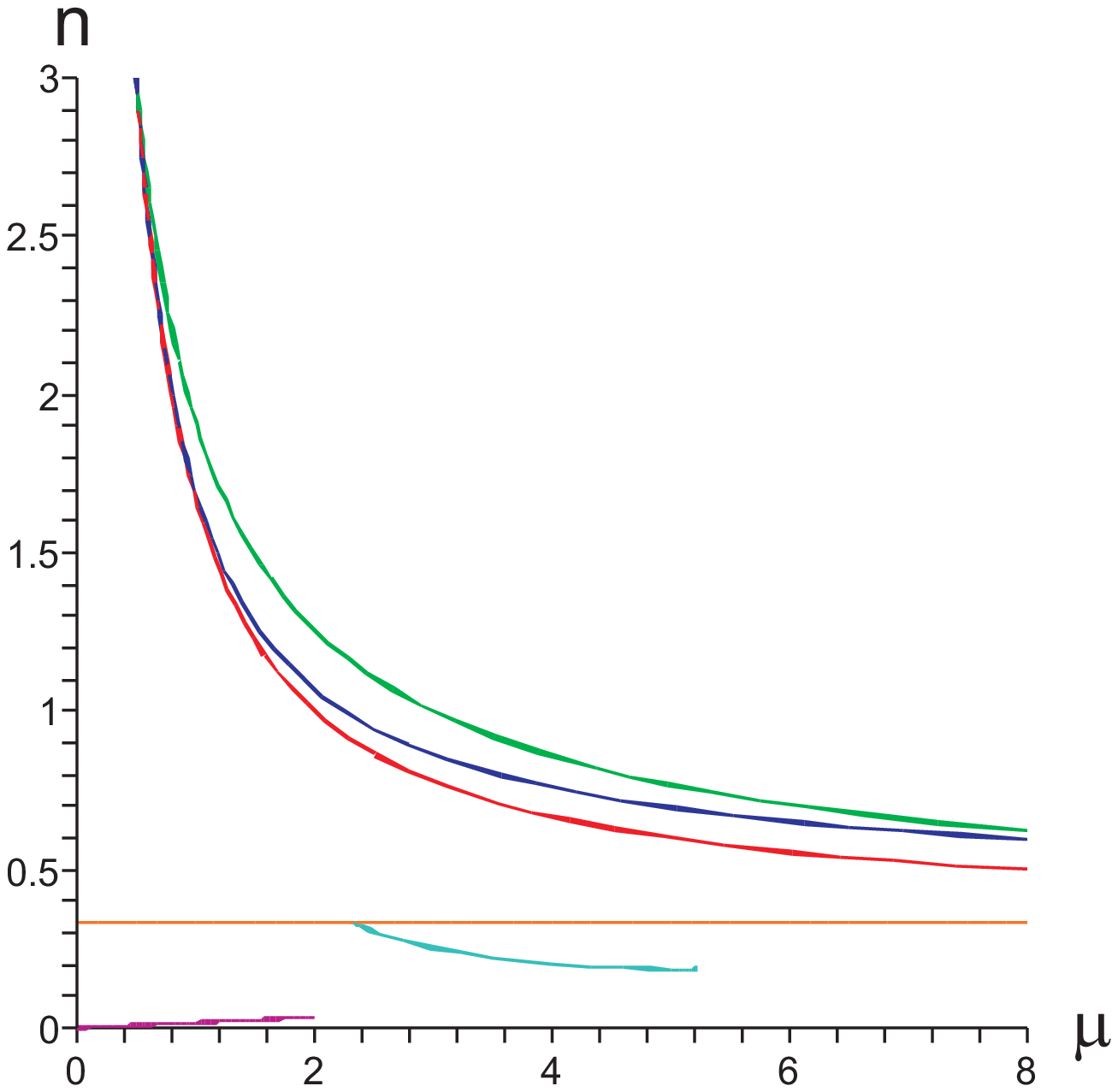,width=8cm,height=8cm}}
\caption{Curves in the $(\mu,n)$ phase diagram for the
$(p,q)=(1,1)$, $(1,2)_{\mathfrak{t}}$ and $(2,1)$ solutions in six dimensions.
These curves lie in the region $1/3 < n \leq 3$. The lowest (red)
curve corresponds to the $(1,1)$ solution. The (blue) curve that
has highest $n$ for high values of $\mu$ is the equal temperature
$(1,2)_{\mathfrak{t}}$ solution. The (green) curve that has
highest $n$ for small values of $\mu$ is the $(2,1)$ solution. The
entire phase space of the $(1,2)$ configuration is the wedge
bounded by the equal temperature $(1,2)_{\mathfrak{t}}$ curve and
the $(1,1)$ curve. For completeness we have also included the
uniform and non-uniform black string phase (numerical data
courtesy T.~Wiseman), as well as the small black hole branch,
lying in the region $0 \leq n \leq 1/3$ (see Figure
\ref{fig_neut}).}
 \label{fig6D}
\end{figure}

The  $ \mu_\rom{(1,1)}(n)$ curve is drawn in the $(\mu,n)$ phase diagram
of Figure \ref{fig6D}. In this case, the curve is easily inverted to
give
\begin{equation}
n_{(1,1)} (\mu ) =\frac{1}{3}  + \frac{4}{3\mu} \, .
\end{equation}
This shows that the lowest value of $\mu$ is obtained for the
static bubble solution ($\mu=\mu_{\rm b}=1/2$ and $n =3$, see Table \ref{tabmuc}), while for large $\mu$
the value $n =1/3$ of the uniform black string is approached.

The two expressions in \eqref{mus116d} can be combined to give
the entropy function
\begin{equation}
\mathfrak{s}_{(1,1)} (\mu)=
\bigg[ \frac{2}{3} \Big( \mu -\frac{1}{2} \Big) \bigg]^{3/2} \, .
\end{equation}
We see again that the entropy goes to zero as the pure bubble
configuration is approached. Furthermore, the leading term for
large $\mu$ corresponds to the (dimensionless) entropy of
the uniform black string
\begin{equation}
\label{sbs6d}
\mathfrak{s}_{\rm bs} (\mu) =
\left(\frac{2}{3} \, \mu \right)^{3/2}  \spa d =5
\end{equation}
in six dimensions. Again it is not difficult to verify that the entropy
of the entire $(1,1)$ branch is lower than that of the uniform black string
for all $\mu \geq 1/2$.

\subsection{Configurations with $(p,q)=(1,2)$ and $(2,1)$ \label{sec12}}

The next two cases to be considered are the $(1,2)$ and $(2,1)$
 configurations.  As already discussed in Section \ref{s:5dKK}, in
 five dimensions the $(1,2)$ solution describes two black holes
held apart by a static bubble, and the $(2,1)$ solution is a black ring
 between two bubbles; we describe these five-dimensional
 configurations first.
After that the corresponding six dimensional configurations are given.

\subsubsection*{$(1,2)$ configuration in $D=5$: Two black holes held apart by a bubble}

The $(1,2)$ solution in five dimensions describes the configuration%
\footnote{As in the previous cases, we can also consider the maximal extension of the geometry of the $(1,2)$ solution. In this case we have two horizons, and as discussed in \cite{Elvang:2002br} it is then natural to periodically identify the extended space-time. The global bubble topology is then a torus $T^2 = S^1 \times S^1$.}
\begin{equation}
\label{12config}
  \begin{array}{ccccc}
  \rom{black~hole}&-&\rom{bubble} &-& \rom{black~hole} \\
  S^3 &&  S^1 \times I   & & S^3
  \end{array}
\end{equation}
i.e. two $S^3$ black holes held apart by a bubble of cylinder
topology $S^1 \times I$. The metric is obtained
by  setting $N=4$ in \eqref{metric5d}, \eqref{gennu5D}. This
solution was studied in detail in \cite{Elvang:2002br}.

As for the black hole in the $(1,1)$ solution,
the black holes in the $(1,2)$ solution are not
localized on the KK circle, and for a given size of the circle at
infinity the black holes can be  arbitrarily big. The sizes of the
two black holes can be varied independently, and the bubble will
vary its size and shape to balance the black holes and keep them
in static equilibrium. In particular, one of the black holes can
be removed, and the resulting configuration is an $S^3$ black hole
on a bubble which is now topologically a disk $D$, i.e. the
$(1,1)$ solution in \eqref{11config}.

To further analyze the solution quantitatively, we choose
\begin{equation}
(a_1,a_2,a_3,a_4) = (0,x,y,1) \spa 0 < x < y < 1 \, .
\end{equation}
Using this in \eqref{ct1_5D}, we find from \eqref{mu5},
\eqref{genL} the dimensionless mass, relative tension and
(rescaled) length of the compact circle
\bea
  \mu= \frac{4\pi}{L} (2+x-y) \spa
   n=\frac{1-x+y}{2+x-y} \spa
   L
    = 4 \pi \frac{y-x}{\sqrt{y(1-x)}} \, .
  \label{mun12}
\eea
We can substitute $L$ into $\mu$, and also use the expression for $n$
to eliminate $y$. This then determines the following family of
curves in the $(\mu,n)$ phase diagram
\bea \label{mu12}
  \mu_{(1,2)}(n;\al)  = \frac{3}{(n+1) (2n-1)}
    \sqrt{[ n+1 + (n-2)\al] [2n-1 + (2-n) \al] }
\eea
for
\begin{equation}
\frac{1}{2} < n < 2 \spa 0 \leq \al \leq 1 \, ,
\end{equation}
where  $\al$ parameterizes the branches in the family.
Here, $\al = x/x_{\rm max}$, $x_\rom{max}\equiv\frac{2-n}{n+1}$ and
the upper bound on $x$ comes from the condition $y<1$.
The limits $\al =0$ and $\al=1$ correspond to removing either of the black
holes, so we recover the $(1,1)$ configuration with one black hole
on a bubble (see Section \ref{sec11}). These are in fact
the solutions where for given $n$, the mass $\mu$ is minimized.

Another special case is where the two black holes have
equal temperature. It is easily seen from \eqref{betak1}
that this condition requires  the two black holes to be of equal size.
In order to arrange this we thus
need to take the two black hole rods to have the same length
($a_2-a_1 = a_4-a_3$) so that $y=1-x$ in the general expressions
\eqref{mun12}, and then $x$ runs in the range $0<x<1/2$.
It is then not difficult to see that this corresponds to setting
$\al = 1/2$ in  \eqref{mu12}, from which it also follows that
these are the solutions that maximize $\mu$ for fixed $n$.
 The resulting curve in the $(\mu,n)$
phase diagram is given by
\bea
\label{mu12t}
\mu_\rom{(1,2)_{\mathfrak{t}}} (n) \equiv \mu_\rom{(1,2)}(n; \frac{1}{2}) = \frac{9n}{2(n+1)(2n-1)} \spa
 ~~\mbox{for}~~~
 \frac{1}{2} < n < 2 \, ,
\eea
and is shown in Figure \ref{fig5D}.
We  denote this branch as the $(1,2){}_{\mathfrak{t}}$
solution, where the $\mathfrak{t}$ subscript signifies the equal
temperature property.

At this point some remarks are in order: \newline
$\bullet$ Qualitatively, the
parameter $\delta = | \al -1/2|$ can be regarded as a measure of the
difference in size of the two black holes. \newline
$\bullet$ To obtain
physically distinct solutions it suffices to consider $0 \leq \al \leq
1/2$ as the other half  of the configurations are
related by the  $\Z_2$ symmetry $\al \rightarrow 1-\al$.
\newline
$\bullet$ The family of curves \eqref{mu12} fills a wedge in the
phase diagram bounded by the $(1,1)$ curve \eqref{mu11} (minimal
$\mu$ curve) and the equal temperature $(1,2)_{\mathfrak{t}}$
curve \eqref{mu12t} (maximal $\mu$  curve), both of which
are shown in Figure \ref{fig5D}. \newline $\bullet$ As seen above in
Section \ref{11config} for the $(1,1)$ curve, each branch labelled by
$\al$ has  $ 1
\leq \mu \leq \infty $, and starts in the point corresponding to the
static bubble and approaches for infinite mass the line $n=1/2$ of the
 uniform black string.
 \newline

Similarly, we can compute the total dimensionless entropy
$\mathfrak{s}= \frac{16\pi G_N}{L^3}(S_1+S_2) $
using the entropy
 $S_k$ (given below \refeq{betak1}) for each black hole.
This gives the total dimensionless entropy for the general $(1,2)$ solution as
\bea
\label{s12}
 \mathfrak{s}_{(1,2)}(n;\al) =
\frac{(2-n)^{3/2}}{(n+1)(2n-1)^2}
\left[ \sqrt{\al^{3}[ n+1 + \al(n-2)]^2[2n-1 +(2-n)\al] } \right.
\eea
$$
 + \left.
  \sqrt{(1-\al)^{3}[ n+1 + \al(n-2)] [2n-1 +(2-n)\al]^2}  \right]
$$
The temperatures $\mathfrak{t}_1$, $\mathfrak{t}_2$ of each of
the black holes can be computed from \eqref{betak1}.

For the special case of two
equal-size black holes we obtain from \eqref{s12}
\bea
\label{s12t}
\mathfrak{s}_\rom{(1,2)_{\mathfrak{t}}} (n) \equiv \mathfrak{s}_{(1,2)}(n;\frac{1}{2})
  =
\frac{3\sqrt{3}}{4}\frac{n^{3/2} (2-n)^{3/2}}{(n+1)(2n-1)^2} \, .
\eea
In this case, since the temperatures are the same,
one can use for example the Smarr formula \eqref{gendsmarr}
to obtain $\mathfrak{t}(n)$ from \eqref{mu12t} and \eqref{s12t}.

The expression \eqref{s12}  can be combined with \eqref{mu12},
which determines a mass $\mu$ for given values of $(n;\al)$,
to give the entropy as a function of mass.
The entropy of all curves
in the wedge satisfies the same property \eqref{s11suni}  as for the $(1,1)$ solution,
and in particular for large $\mu$ the leading behavior is that of the
uniform black string \eqref{sbs5d}.

Within the family of solution branches parameterized by $\al$, one
finds that for a given mass $\mu$, the configuration with highest
entropy is the one with a single black hole on a bubble, while the
configuration with lowest entropy is the one with two equal-size
black holes. The latter fact is in accordance with the general
result of Section \ref{secext} that solutions with equal temperatures
extremize (in this case minimize) the entropy. The other
configurations $ 0 < \al < 1/2$  interpolate between these
two entropy extremes. This is illustrated in Figure
\ref{figs5d}, in which we plot $\mathfrak{s}$ versus $\mu$ for
various configurations.

Finally, for use below  we give the (rescaled) length
of the compact circle for the $(1,2)$ solution
\begin{equation}
\label{L12}
L_{(1,2)} (n;\al) = 4 \pi \frac{(2n-1)}{\sqrt{[ n+1 + (n-2)\al]
[2n-1 + (2-n) \al] }} \ ,
\end{equation}
which follows from \eqref{mun12}.

\subsubsection*{$(2,1)$ configuration in $D=5$: Black ring between two
equal size bubbles}

The $(2,1)$ solution  in five dimensions is described by the configuration%
\footnote{Extending the metric as in the previous cases, we find on the other side of the black hole throat another copy of the geometry. In particular, the bubble disks meet up smoothly and the global topology of each bubble is therefore an $S^2$.}
\bea
\label{21config}
 \begin{array}{ccccc}
  \rom{bubble}&-&\rom{black~ring}&-&\rom{bubble}  \\
  D  &&   S^2 \times S^1 && D
  \end{array}
\eea
i.e. a black ring of topology $S^2 \times S^1$ supported by
two bubbles of disk topology. The exact form of the metric
 is obtained by setting $N=4$ in \eqref{met5D2}.
This solution was studied in detail in \cite{Elvang:2002br}.

The $S^1$ of the black ring is not the same $S^1$ as
the KK circle at infinity. The black hole is therefore not
topologically supported, which is the reason for calling it
a black ring. The bubbles are topologically disks and they
support the black ring from collapsing under its gravitational
self-attraction.
Note that this regular bubble-black ring solution
is different from the asymptotically flat static black ring in
\cite{Emparan:2001wk} which was supported by a conically singular
disk.

The physical parameters of this configuration can of course
be obtained directly following similar steps as
above, while taking into account the regularity constraint that the two
bubbles should have equal size. It is easier, however, to obtain
the physical parameters of the solution using the duality map of
Section \ref{secWick}. This uses the fact that the $(2,1)$ solution
 can be
obtained by a double Wick rotation $(t,\phi)\to
(i\phi,it)$ from the configuration \eqref{12config} with two
equal-size black holes, i.e. the $(1,2){}_{\mathfrak{t}}$ solution.

Then, using the duality map \eqref{dualmu}, \eqref{duals}  we find
from the mass and entropy curves \eqref{mu12t}, \eqref{s12t} of
the $(1,2){}_{\mathfrak{t}}$ solution the dual curves \bea
\label{mus21}
  \mu_{(2,1)}(n) = \frac{3\sqrt{3}}{(n+1)\sqrt{2n-1}}
  \spa
  \mathfrak{s}_{(2,1)}(n) =\frac{2(2-n)^2}{(n+1)(2n-1)}
\eea of the $(2,1)$ solution. We have plotted the curve
$\mu_{(2,1)}(n) $  in the $(\mu,n)$ phase diagram of Figure
\ref{fig5D}.  Note that for small $\mu$ the black ring branch
starts outside the wedge of solutions with two black holes on a
bubble, but at a certain mass it cuts into the wedge and stays
there. The intersection point in the $(\mu,n)$ phase diagram of
the $(2,1)$ branch \eqref{mus21} with the
 $(1,2){}_{\mathfrak{t}}$ branch \eqref{mus126} (equal size black holes)
  is at $n=\frac{2}{3}$
and $\mu=\frac{27}{5}$. This is an explicit example of non-uniqueness
in the $(\mu,n)$ phase diagram, as the two solutions are physically distinct.
The corresponding curve for $\mathfrak{s}_{(2,1)}(\mu)$
is drawn in Figure \ref{figs5d}, which asymptotes for large $\mu$ again
to the expression in \eqref{sbs5d}.

We can now also compare the entropies of the curves obtained so far.
For small $\mu < \mu_\star \approx 2.42$ the $(2,1)$ solution
(of a black ring between two bubbles)
 is the lowest entropy configuration. For $\mu >
\mu_\star$, the $(2,1)$ solution has higher entropy than
the $(1,2)_{\mathfrak{t}}$ solution (of two equal-size black holes
on a bubble), but always lower
entropy than the $(1,1)$ solution of a single black hole on the bubble.
We observe
(see Figure \ref{figs5d}) that the uniform black string ($n=1/2$
for all $\mu$) has higher entropy than any of the bubble-black
hole solutions, but we also note that for $\mu<\mu_\rom{GL}
\approx 3.52$ the uniform black string is classically unstable.
Based on the entropy consideration, the bubble-black hole
sequences are globally unstable, however, this does not
necessarily provide information about the local, or classical,
stability of the configurations.

\subsubsection*{$(1,2)$ and $(2,1)$ configurations in $D =6$}

The $(1,2)$ solution in six dimensions is described by the configuration%
\footnote{As in the five-dimensional case, the extended space-time can be periodically identified, and the global topology of the bubble is then a three-torus, $T^3 = S^1 \times S^1 \times S^1$.}
\beastar
  \begin{array}{ccccc}
  \rom{black~ring}&-&\rom{bubble}&-&\rom{black~ring}  \\
  S^3 \times S^1 &&   T^2 \times I && S^3 \times S^1
  \end{array}
\eeastar
consisting of two $S^3\times S^1$ black rings held apart by a bubble
of topology cylinder times a circle.
 The bubble plays two roles for keeping the black rings in
static equilibrium: it prevents each black ring from collapsing
under its gravitational self-attraction (this is necessary since
the rings are not topologically supported), and it also keeps the
two black rings apart.

In order to obtain the physical parameters for the six-dimensional
$(1,2)$ configuration we use again the mapping in \eqref{5to6mu},
\eqref{5to6s} and the five-dimensional results \eqref{mu12},
\eqref{s12}, \eqref{L12}.
After some algebra we find
\bea
\label{mu126}
  \mu_{(1,2)}(n;\al) =
\frac{2}{(n+1) (3n-1)^2}
    [ 2n+2 + (n-3)\al] [3n-1 + (3-n) \al] \, ,
\eea
\bea
\label{s126}
 \mathfrak{s}_{(1,2)}(n;\al) &=&
\frac{(3-n)^{3/2}}{2(n+1)(3n-1)^3}
\left[ \sqrt{\al^{3}[ 2n+2 + \al(n-3)]^3[3n-1 +(3-n)\al]^2 } \right. \\ \nonumber
&& \hspace{1.6cm}
 + \left.
  \sqrt{(1-\al)^{3}[ 2n+2 + \al(n-3)]^2 [3n-1 +(3-n)\al]^3}  \right] \spa
\eea
for
\begin{equation}
\frac{1}{3} < n < 3 \spa 0 \leq \al \leq 1 \, .
\end{equation}
The solution is qualitatively similar to the five-dimensional
$(1,2)$ solution, and all remarks above apply here as well.
In particular, we recover for  $\al =0,1$
the $(1,1)$ solution of one black ring on a bubble given in Section
\ref{sec11}.
For $\al=1/2$ we find the configuration of two equal-size (and equal temperature)
black rings on a bubble. The mass and entropy of this solution are
 \begin{equation}
 \label{mus126}
 \mu_{(1,2)_\mathfrak{t}}(n ) =
  \frac{(5n+1)^2}{2(n+1)(3n-1)^2} \spa
  \mathfrak{s}_{(1,2)_\mathfrak{t}}(n ) =
  \frac{ (5n+1)^{5/2} (3-n)^{3/2} }{16(n+1)(3n-1)^3} \ .
 \end{equation}
  The $ \mu_{(1,2)_\mathfrak{t}}(n )$ curve is shown in Figure
  \ref{fig6D}. {}From the two curves in \eqref{mus126} one may then
  obtain the entropy function $ \mathfrak{s}_{(1,2)_\mathfrak{t}}(\mu )$.
  This has again the property that for large $\mu$ the entropy \eqref{sbs6d}
  of the uniform black string is approached.

Finally, we turn to the six dimensional $(2,1)$ configuration%
\footnote{In the extended space-time, there is another copy of each bubble disk on the other side of the black hole throat, and the two copies join up to form an $S^2$. The global topology of each bubble is hence $S^2 \times S^1$.}
 \beastar
  \begin{array}{ccccc}
  \rom{bubble}&-&\rom{black~tuboid}&-&\rom{bubble}  \\
  D \times S^1 &&   S^2 \times T^2 && D \times S^1
  \end{array}
\eeastar
consisting of a black tuboid of topology $S^2\times T^2$
supported by two equal size bubbles of topology $D \times S^1$.
The $S^1$'s of the tuboid are
prevented from collapse by the two equal size bubbles.
In Section \ref{s:eh6D} we analyzed the horizon topology of the
black tuboid. One interesting property is that the torus $T^2 =
S^1 \times S^1$ of the horizon is non-trivially fibred over the
$S^2$.

The physical parameters of this solution can be either obtained
directly, or from the five dimensional $(2,1)$ solution using
the map \eqref{5to6mu}, \eqref{5to6s},
or from the six dimensional $(1,2){}_{\mathfrak{t}}$  solution
above using the duality map of \eqref{dualmu}, \eqref{duals}.
The results are
\begin{equation}
\label{mus216}
\mu_{(2,1)}(n) = \frac{2(n+5)}{(n+1)(3n-1)} \spa
s_{(2,1)}(n) =  \frac{ (n+5)^{1/2} (3-n)^2}{(n+1) (3n-1)^{3/2}} \, ,
\end{equation}
and the $\mu_{(2,1)}(n)$ curve is plotted in the phase diagram
of Figure \ref{fig6D}. Comparing to the five-dimensional
phase diagram Figure \ref{fig5D}, we see that the qualitative
features are the same. Also in six dimensions,
we see the feature that the $(2,1)$ branch of the tuboid intersects the
$(1,2)_\mathfrak{t}$ branch of two equal-size black rings on bubble, this time in the point  $n=7/13$ and $\mu=117/10$.

>From the two curves in \eqref{mus216} one can
  obtain the entropy function $ \mathfrak{s}_{(2,1)}(\mu )$,
  which again approaches the entropy \eqref{sbs6d} of the uniform black string
 for large  $\mu$. In all, the $(\mu,\mathfrak{s})$ diagram is  qualitatively the
 same as in five dimensions (see Figure \ref{figs5d}).
 Again we find that among the bubble-black hole
 solutions, the entropically favored one is the $(1,1)$ solution with a single black ring on a bubble.
 The $(2,1)$ solution has higher entropy than the $(1,2){}_{\mathfrak{t}}$
  solution for $\mu > \mu_\star \approx 2.61$. For
comparison, $\mu_\rom{GL} \approx 2.31$. The six-dimensional uniform black
string with entropy \eqref{sbs6d}   is entropically favored over any of the bubble-black hole configurations that we have studied here.

\subsection{Configurations with $(p,q)=(2,3)$, $(3,2)$ and $(2,2)$
\label{sec23}}

Finally, we discuss some more involved, but still tractable, configurations, starting with the case $(p,q) = (2,3)$.

\subsubsection*{$(2,3)$ configuration in $D=5$}

\begin{figure}[htb]
\centerline{\epsfig{file=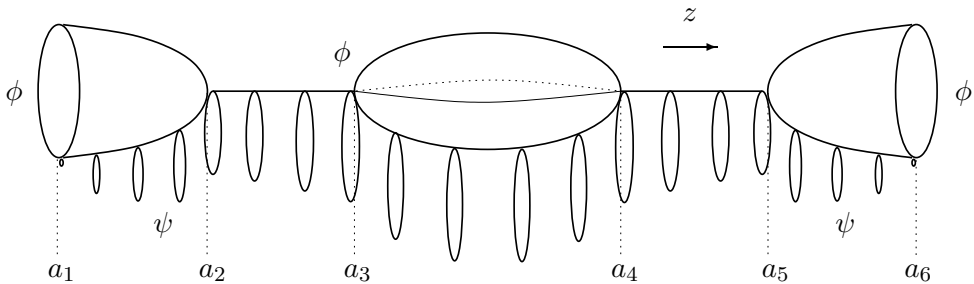,width=12cm}}
\vspace{2mm}
   \caption{Sketch of the $(2,3)$ configuration of black holes and
bubbles.}
   \label{sketch1}
\begin{picture}(0,0)(0,0)
\put(31,110){$\phi$}
\put(155,125){$\phi$}
\put(390,110){$\phi$}
\put(87,60){$\psi$}
\put(345,60){$\psi$}
\put(287,140){$z$}
\put(280,130){\vector(1,0){20}}
\put(47,43){$a_1$}
\put(104,43){$a_2$}
\put(159,43){$a_3$}
\put(260,43){$a_4$}
\put(317,43){$a_5$}
\put(371,43){$a_6$}
\end{picture}
\end{figure}

The $(2,3)$ solution in five dimensions is described by the configuration
 \bea
 \label{config23}
  \begin{array}{ccccccccc}
\rom{black~hole}&-&\rom{bubble}&-&\rom{black~ring}&-&\rom{bubble}&-&\rom{black~hole} \\
  S^3 && S^1 \times I && S^2 \times S^1 && S^1 \times I && S^3
  \end{array}
\eea
In Figure \ref{sketch1} we sketch the geometry of the
configuration. The general $(p,p+1)$ configuration has extra
``$-\rom{bubble}-\rom{black~ring}-$'' insertions.
The explicit metric follows by setting $N=6$ in \eqref{metric5d}.

According to the discussion of Section \ref{s:regu5d}, as soon as we have
more than one bubble in the configuration, the regularity condition
\refeq{Deltaphi} imposes constraints on the parameters $a_i$ specifying
the solution. In particular, for the $(2,3)$ solution the
constraint is $L =(\Delta\phi)_1 =(\Delta\phi)_2$. Using the general
expressions \eqref{genL} for $(\Delta \phi)_k$, this implies that
\begin{equation}
  \frac{a_3-a_2}{a_4-a_2} \sqrt{\frac{a_4-a_1}{a_3-a_1}}
  =
  \frac{a_5-a_4}{a_5-a_3} \sqrt{\frac{a_6-a_3}{a_6-a_4}} \, .
  \label{N6constr}
\end{equation}
Taken together with our conventions $a_1=0$, $a_6=1$, this means that
there are three independent dimensionless parameters characterizing
the general $(2,3)$ solution.

To simplify matters, we choose here to focus only on
 the subset of $(2,3)$ solutions for which the rod configurations have a $\Z_2$-symmetry of reflection in the line through its center. Such a configuration can generally be specified by choosing the $a_i$'s as follows
\begin{equation}
\label{a23}
(a_1,a_2,a_3,a_4,a_5,a_6) =(0,x,x+y,1-2(x+y),1-(x+y),1-x,1) \spa
\end{equation}
with two parameters $x,y$ such that
\begin{equation}
 0 < x < \frac{1}{2}-y \spa 0 < y < \frac{1}{2} \, .
 \end{equation}
This is a two-parameter solution of the constraint \eqref{N6constr}. Though we do not consider the entire phase space of the $(2,3)$
solution, this subset serves already as a useful illustration of the
features of the solution. It is clear that the $\Z_2$ symmetry in \eqref{a23} implies that the two black holes at the ends of the bubble-black hole sequence \eqref{config23} are equal in size and that they have equal temperatures.
Since we expect that the
maximal $\mu$ curve in the $(\mu,n)$ phase diagram for the
general $(2,3)$ solution corresponds
to a configuration with  this $\Z_2$ symmetry, this means that
the subset under consideration here suffices to determine this curve.

As we have already provided details of
the procedure of extracting the physical quantities for the simpler
$(1,1)$, $(1,2)$ and $(2,1)$ solutions discussed above, we will be less extensive in the discussion below.

Using the rod structure  \eqref{a23} we find again $\mu$, $n$, and $L$
from \eqref{ct1_5D}, \eqref{mu5}, and \eqref{genL}. The equation for $n$ can be solved for $y$ to
give $y = (2n-1)/(2(n+1))$. Setting $x= x_{\rm max} \alpha$,
$x_{\rm max} = 1/2-y = (n-2)/(2(n+1))$, we obtain the following one-parameter family of curves
\begin{equation}
\label{mu23}
\mu_{(2,3)} (n;\alpha) = \frac{3}{2}
\frac{ [3 + 2 \al (n -2)][2 (n+1)  + \al (n-2)] }
{(n+1)(2n-1)[n+1 + \al (n-2)]}
\sqrt{\frac{2n-1 -\al (n-2)}{3 + \al (n-2)}}
\end{equation}
for
\bea
 \frac{1}{2} < n < 2 \spa 0 \leq \al \leq 1 \, .
\eea
The curve for $\al =0$ reduces to the $(2,1)$ case (see \eqref{mus21} ) since
the two $S^3$ black holes on the outside shrink to zero size.
On the other hand,  $\al =1$  reduces to the $(1,2){}_{\mathfrak{t}}$
case (see \eqref{mu12t})
since the bubble in the middle is eliminated, and the black holes on the outside have equal temperature.

\begin{figure}[t]
\centerline{\epsfig{file=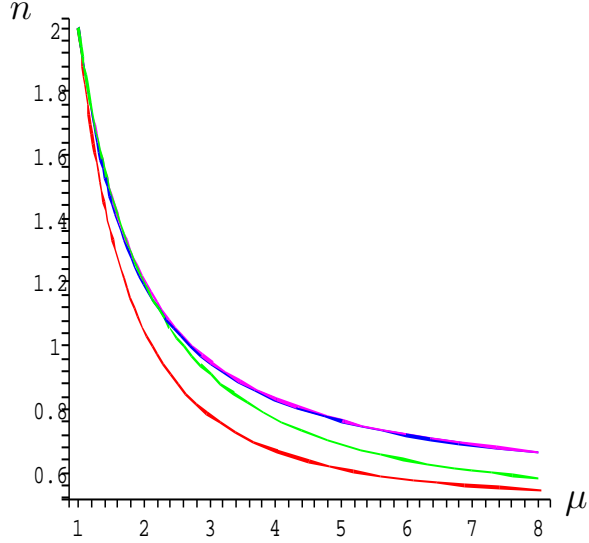,width=8cm,height=8cm}}
\caption{Curves in the $(\mu,n)$ phase diagram relevant for the
$(p,q)=(2,3)$ solution in five dimensions. The highest (maximal
$\mu$) curve (purple) is obtained in Appendix \ref{23details}. On
this curve, the two black holes at the endpoint are at equal
temperature, but the middle one has a lower temperature. The
lowest (minimal $\mu$) curve (red) corresponds to the $(1,1)$
solution of Section \ref{sec11}. The $(2,3){}_\mathfrak{t}$
solution with all temperatures equal (and which also is the curve
of minimal entropy) is the blue curve, which lies almost on top of
the maximal $\mu$ curve. The green curve corresponds to the
$(2,1)$ solution.}
\label{fig235D}
\begin{picture}(0,0)(0,0)
\put(310,160){\Large $\mu$}
\put(100,345){\Large $n$}
\end{picture}
\end{figure}

The family of curves \eqref{mu23} describes again a wedge of solutions in
the $(\mu,n)$ phase diagram. The curves that form the bounds of the wedge
are the maximal and minimal $\mu$ curves, found by extremizing the function
$\mu_{(2,3)}$ as a function of $\al$. Details of this are given in Appendix
\ref{23details}. The minimal $\mu$ curve for the symmetric solution
is given in \eqref{mu23min} and consists of a combination of the
$(1,2){}_{\mathfrak{t}}$ and $(2,1)$  solution (see  also Figure \ref{fig5D}).
However, the general
$(2,3)$ configuration (without the symmetry) includes the $(1,1)$ solution,
which we have seen gives the overall minimal $\mu$ curve. More interesting
is therefore to consider the maximal $\mu$ curve of \eqref{mu23}, as we expect
this to be also the maximal $\mu$ curve of the general $(2,3)$ configuration.
This curve is given in \eqref{mu23max} and shown in Figure \ref{fig235D}.
It involves the solution  $\al = \al_{\mu ({\rm max})}(n)$ of
 a quartic equation on $\al$, with $n$-dependent coefficients in the range
 $0 \leq \al \leq 1$.

We also consider the total entropy of the one-parameter
family,
\begin{equation}
\label{s23} \mathfrak{s}_{(2,3)}(n;\al) = \frac{(n-2)^{2} [2 (n+1)
+ \al (n-2)][2n-1 -\al (n-2)] (1-\al)^2}{ (n+1)(2n-1)^2 [n+1 + \al
(n-2)]^2  }
\end{equation}
$$
+ \frac{ (n-2)^{3/2}  [3 + 2 \al (n -2)]^2 [2 (n+1)  + \al
(n-2)]^{3/2} [2n-1 -\al (n-2)]^{1/2} \al^{3/2}}{ 4 (n+1)(2n-1)^2
[n+1 + \al (n-2)]^2 [3 + \al (n-2)]^{1/2} }
$$
which can be combined with \eqref{mu12} to give the entropy function
$\mathfrak{s}_{(2,3)} (\mu)$.
 It is interesting to find out what
are the curves  that minimize
and maximize the entropy for a give $\mu$. The details of this
computation can be found in Appendix \ref{23details}, and we present
here the main results.

Within the family \eqref{mu12}, the curve of maximum entropy is
(just as the minimal $\mu$ curve)
described by a combination \eqref{s23max} of the $(1,2){}_{\mathfrak{t}}$ and $(2,1)$
solutions (see  also Figure \ref{figs5d}). However, for the general $(2,3)$ solution, we expect
that the overall maximal
entropy curve will be given by the $(1,1)$ solution.

More interesting is to determine the minimal entropy curve.
For the
symmetric $(2,3)$ solution this curve follows by solving a quintic
equation on $\al$ with $n$-dependent coefficients and we denote its solution
as $\al =  \al_{\mathfrak{s}(\rm min)}(n)$. The curve in the
$(\mu,n)$ phase diagram is given by \eqref{s23minmu} and is shown
in Figure \ref{fig235D}, while the curve in the $(\mu,\mathfrak{s})$ diagram
is given in \eqref{s23min} and plotted in Figure \ref{figs5d}.

We can now explicitly check the general result of Section \ref{secext}
that the equal temperature case should correspond to an extremum of
the entropy. As remarked above, for any $\al$ we have that
$\mathfrak{t}_1=\mathfrak{t}_3$
(because the solution is symmetric) but not necessarily equal to
$\mathfrak{t}_2$. Equating also $\mathfrak{t}_1=\mathfrak{t}_2$,
we find from \eqref{betak1} a quintic equation on $\al$, whose solutions
we denote by $\al = \al_{ \mathfrak{t}} (n)$.
We then find, in accordance with Section \ref{secext} that
\begin{equation}
\al_{ \mathfrak{t}} (n) =\al_{\mathfrak{s}(\rm min)}(n)
\end{equation}
i.e. the curve minimizing the total entropy coincides
with the curve describing the special case where all  temperatures
of the three black holes in the $(2,3)$ solution are equal. We denote this
configuration therefore as the $(2,3){}_{\mathfrak{t}}$ solution.
The corresponding curve (coinciding with the minimum entropy case) is
shown in Figures \ref{fig235D} and \ref{figs5d}.

Finally, we comment on the issue of non-uniqueness in the $(\mu,n)$
phase diagram. First, we note that the $(2,3){}_{\mathfrak{t}}$ solution
intersects the $(2,1)$ solution in the point $(\mu,n)\approx (2.149,1.139)$.
Just as the earlier observed intersection point of the
$(1,2){}_{\mathfrak{t}}$ and $(2,1)$ solution, this is another explicit
example of non-uniqueness.

More generally, we can argue for infinite non-uniqueness in the
$(2,3)$ solution as follows. We have seen that the two-parameter family of
$\Z_2$ symmetric $(2,3)$ solutions describe a two-dimensional wedge in the
phase diagram bounded from above by the maximal $\mu$ curve of the $(2,3)$
solution and from below by a combination of the
$(1,2){}_{\mathfrak{t}}$ solution and the $(2,1)$ solution.
More generally, we argued that the full $(2,3)$ solution
should cover the wedge bounded  from above by this maximal $\mu$ curve and
from below by the $(1,1)$ solution.
If one now imagines relaxing the $\Z_2$ condition,
we know that: i) A larger region of the phase diagram starts to be covered as
we are now able to go all the way down to the $(1,1)$ curve. ii)
Since the $(2,3)$ solution has $q=3$ free parameters we can
write $n=n (\mu,\al,\beta)$ where $\beta$ parameterizes the asymmetry.
Since we have one more additional parameter that we can vary this
implies that in the entire two-dimensional wedge accessible to the
$(2,3)$ solution there is a subregion for which there should be infinite
non-uniqueness specified by the extra parameter.

\subsubsection*{$(3,2)$ configuration in $D=5$}

The $(3,2)$ solution in five dimensions is described by the configuration
\bea
\label{config32}
\begin{array}{ccccccccc}
  \rom{bubble}&-&\rom{black~ring}&-&\rom{bubble}&-&\rom{black~ring}&-&\rom{bubble} \\
  D && S^2 \times S^1 &&   S^1 \times I && S^2 \times S^1 && D
  \end{array}
\eea
We restrict our quantitative analysis to the case where the two
black holes have equal temperatures. We call  this the
$(3,2){}_{\mathfrak{t}}$ solution.

>From Section \ref{secWick} we know that this case can be obtained
by Wick rotation from the equal temperature $(2,3){}_{\mathfrak{t}}$
solution, which we just discussed. Substituting the solution
 $\al =\al_{ \mathfrak{t}} (n)$ in the curves
\eqref{mu23}, \eqref{s23} and subsequently using the mapping
\eqref{dualmu}, \eqref{duals} it is relatively easy to find the
corresponding curve for the $(3,2){}_{\mathfrak{t}}$ solution in the
$(\mu,n)$ phase diagram. It  starts for low $\mu$
above the wedge containing the $(2,3)$ solution, but then for some value
of $\mu$ cuts into it and stays within the wedge. This feature is
thus similar as that seen before for the $(2,1)$ solution, which
cuts into the wedge containing the $(1,2)$ solutions (cf. Figure \ref{fig5D}).

\subsubsection*{$(2,2)$ configuration in $D=5$}

The $(2,2)$ solution in five dimensions is described by the configuration
\bea
 \label{config22}
  \begin{array}{ccccccc}
  \rom{black~hole}&-&\rom{bubble}&-&\rom{black~ring}&-&\rom{bubble}
   \\
  S^3 && S^1 \times I && S^2 \times S^1 && D
  \end{array}
\eea
Note that this is the simplest solution that has objects of all possible
topologies encountered in the five dimensional case in it.
However, due to the lack of symmetry, this case is more difficult to
treat in practice than the cases considered above,
 even though there are only two independent dimensionless quantities
 specifying the configuration.

\subsubsection*{$(2,3)$, $(3,2)$ and $(2,2)$ configurations in $D=6$}

The preceding analysis in five dimensions can easily be repeated
for six dimensions, and the topology of the objects will change
accordingly. Moreover, the results above can be used to find
the corresponding six dimensional ones using the map in
\eqref{5to6mu}, \eqref{5to6s}. In particular, we can use these
transformations directly on the curves of Figures \ref{fig235D},
\ref{figs5d}, and it is found that all
qualitative features are the same as in five dimensions.

\section{Conclusions}
\label{s:concl}

We conclude the paper with the following general remarks
and open problems:

\noindent {\sl Phase diagram: }
We have seen in this paper that the five- and six-dimensional
bubble-black hole sequences that we constructed fall into
 the upper region, $ 1/(d-2) < n \leq d-2$ of the $(\mu,n)$
phase diagram of static Kaluza-Klein black holes. We also
have the lower bound $\mu \geq \mu_{\rm b}$ for the
(dimensionless) mass of each sequence, where $\mu_{\rm b}$ is the
mass of the static Kaluza-Klein bubble.

Based on the general construction as
well as the examples, it appears that the set of all bubble-black
hole sequences covers a very large part of this  region of the
 phase diagram and that there is a high degree of non-uniqueness.
 In particular, we note the following features:
 \begin{itemize}
\item Among all the examples, the $(1,1)$ solution
gives the curve in the $(\mu,n)$ phase
 diagram  that has the lowest mass for given relative tension $n$.
In other words for all other examples the curves lie above the
curve of the $(1,1)$ solution. We believe this to hold in
general for all bubble-black hole sequences.
\item As we increase
the number of bubbles and/or black holes in the sequence the
curves of maximal mass (for given $n$) move upward.
Physically this means that as we increase the
number of objects in the sequence, one can achieve a higher
relative tension for a given mass.
It would be interesting to examine whether the entire area of the phase
diagram above the $(1,1)$ solution
is filled up all the way to (but
not including) the horizontal line $n=d-2$.
\item For a given $(p,q)$
solution with $p$ bubbles and $q$ black holes there are $q$ free
parameters, corresponding to the relative sizes of the black
holes. At the level of the phase diagram this implies the
following important facts. For $q=1$ the solution is a curve in
the phase diagram and for $q=2$ the solution is a two-dimensional
wedge in the phase diagram. For $q \geq 3$ we continue to have
two-dimensional wedges in the phase diagram, but there must be
infinite non-uniqueness since there are still $q-2$ further parameters
characterizing the solution.

To see this more clearly, note that
we can write $n= n(\mu,x_1,\ldots, x_{q-1})$, since
we can choose $\mu$ as one of the $q$ parameters of the $(p,q)$
solution.
If $q=1$ we find a one-dimensional curve in the $(\mu,n)$ phase diagram.
If $q=2$ there
is one additional parameter so that we find a two-dimensional wedge
in the $(\mu,n)$ phase diagram.
For higher $q$ we have in addition $q-2$ parameters
that can vary independently for given $\mu$ and $n$.
We thus conclude that for a given $(p,q)$  solution
with $q \geq 3$  there
is an infinite non-uniqueness in the accessible region.%
\footnote{Infinite non-uniqueness has also been found in
\cite{Emparan:2004wy}
for black rings with dipole charges in asymptotically flat
space.}

\item The non-uniqueness becomes even more plentiful once we consider the
entire set of  bubble-black hole sequences in the phase
diagram. To see this note that any $(p,p)$ or $(p,p+1)$ solution includes
as a special case the $(1,1)$ solution. In other words the parameters
specifying these solutions can be chosen arbitrarily close to the
$(1,1)$ solution. So as we increase the number of objects in the sequence
we have two effects. On the one hand, we find more solutions in the (``low'')
region where we already had solutions. On the other hand, we find that we
start to cover a ``higher'' region that was not covered before.
It would be interesting to examine this further and prove these assertions
mathematically rigorous.

\item We also recall here that we have observed explicit cases of non-uniqueness
in the intersection points of the $(1,2){}_{\mathfrak{t}}$ solution
and the $(2,1)$ solution, as well as the intersection of $(2,3){}_{\mathfrak{t}}$ and the $(2,1)$ solution.

\end{itemize}

It is also possible to arrive at general statements regarding the entropy
as a function of the mass.
\begin{itemize}
\item In the examples, the curve with highest entropy for a given
mass $\mu$, is the one obtained for the $(1,1)$ solution. We believe that this
curve represents the upper bound for
the entropy curves of all $(p,q)$ solutions.
Moreover, as the number of objects in the
sequence increases, the corresponding entropy curves
will tend to lie lower, but there can be non-trivial intersections.
\item We have proven that, within a family of $(p,q)$ solutions
the curve for the equal temperature
$(p,q)_{\mathfrak{t}}$ solution corresponds to an extremum of the
entropy, and we have observed in the examples that this is always the
minimum entropy curve.
This feature can be physically understood by considering
two widely separated Schwarzschild black holes.
For fixed total mass, the entropy of this configuration  is
minimized when the black holes have the same radius
(hence same temperature), while the maximal entropy configuration is the
one where all the mass is located in a single black hole.
\item From the examples it appears that for any branch of $(p,q)$
solutions the entropy function approaches
that of the uniform black string $\propto  \mu^{\frac{d-2}{d-3}}$ in the
limit of large mass, and
the uniform black string has higher entropy than any of the
$(p,q)$ solutions.
The physical reason to expect that all bubble-black hole sequences
have lower entropy than a uniform string of same mass, is that
some of the mass has gone into the bubble rather than the black
holes, giving a smaller horizon area for the same mass.
\end{itemize}

\noindent {\sl Topology: } The bubble-black hole sequences that we
have constructed involve event horizons of various topologies. In particular,
in five dimensions we have $S^3$ (black hole) and $S^2\times S^1$
(black ring) topology. Moreover, in six dimensions we observed
$S^3 \times S^1$ (black ring) and $S^2 \times S^1 \times S^1$
(black tuboid) topology. In all cases, the $S^1$'s are
not in the Kaluza-Klein direction (and hence not topologically
supported), but supported by the Kaluza-Klein bubbles.
It is interesting to note that the
six-dimensional sequences we constructed do not involve
event horizons with black hole topology $S^4$.
It is not difficult to see that such sequences, if they exist,
would not have four but only three commuting
Killing isometries, and are hence not contained in the generalized
Weyl ansatz used here. It would be interesting to examine
whether more general classes of bubble-black hole sequences exist,
in less symmetric ans\"atze, that would include the topology $S^4$.
\newline

\noindent {\sl General $D$: } We have found a map between five- and
six-dimensional bubble-black hole solutions. The general structure
of the solutions was very similar in five and six dimensions, but
the topologies of bubbles and black holes changed. It is
interesting to ask if solutions for bubble-black hole sequences
exist for $D \geq 7$, and if so, whether they can be related to the
solutions of this paper by a map similar to the one mapping five-
to six-dimensional solutions. Along these lines also follows the
question about the $1/(d-2)<n \le d-2$ region of the phase
diagram. For $D \geq 7$, there are (apart from the static bubble
at $(\mu_{\rm b},d-2)$) no known solutions in this region. It is
tempting to speculate that for $D \geq 7$ this region is occupied by
bubble-black hole sequences with features similar to the ones
found in this paper for the five- and six-dimensional cases. It
would be interesting if one could find bubble-black hole sequences
in $\CM^d \times S^1$ for $D= d+1 \geq 7$. However, this may be
difficult since one cannot use the generalized Weyl ansatz for
such solutions.

\begin{figure}[t]
   \centerline{\epsfig{file=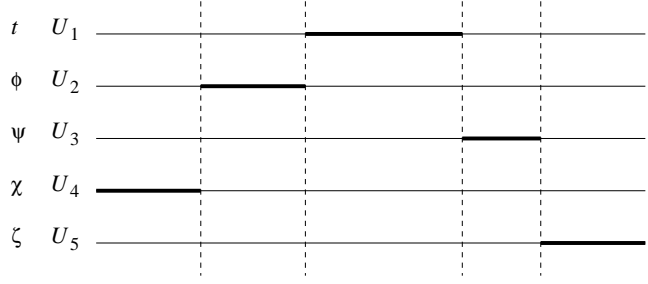,width=8.5cm}}
   \caption{Rod configuration for black tuboid with two bubbles in
   seven dimensions.}
   \label{fig7d}
\end{figure}

On the other hand,
one can easily construct rod configurations for Weyl solutions
describing regular bubble-black hole sequences in dimensions greater
than six. However, these solutions will have more than one
Kaluza-Klein circle, i.e.~asymptotics $\CM^{d}\times T^k$, and the
bubbles will be associated with different Kaluza-Klein directions. One
potentially interesting example arises from the rod configuration
depicted in Figure \ref{fig7d}.
This configuration describes a seven-dimensional black tuboid with
horizon topology $S^3 \times S^1 \times S^1$ supported by two
bubbles of topology $D \times S^1 \times S^1$. Asymptotically the
solution is $\CM^5 \times T^2$. It could be interesting to
study the
phase diagram for that class of solutions.
\newline

\noindent {\sl Stability:}
We have found that the entropy of the $(1,1)$ solution is less than that
of the uniform black string with the same mass $\mu$. Also the examples
showed that the $(p,q)$ solutions with more bubbles and black holes all
had entropy lower than the $(1,1)$ solution. This indicates that the
$(p,q)$ solutions are globally unstable. It is interesting to ask if the
bubble-black hole sequences are also classically unstable.

The bubbles hold the black holes in a fine-tuned equilibrium. For the
$(1,2)$ solution with two small black holes on a bubble, one can imagine
perturbing the position of one black hole, causing it to "slide" around
the bubble to merge with the other black hole. This could be viewed as a
decay of the $(1,2)$ solution to (a perturbed version of) the $(1,1)$
solution.

As discussed in Section \ref{bubble}, the static Kaluza-Klein bubble is
classically unstable to perturbations that make it collapse or expand.
There are $(p,q)$ solutions arbitrarily close to the pure bubble
solutions, so we may expect these solutions to be clasically unstable too.

Another classical instability mode is that of the black rings supported by
the bubbles. One can choose the parameters such that the black ring is
long and thin, in which case a Gregory-Laflamme type instability mode
could appear. It would be interesting to examine if the bubbles help
stabilizing the black ring, or cause it to be even more unstable.
\newline

\noindent {\sl Periodic array:}
Imagine extending a rod configuration to include infinitely many
finite rods arranged periodically. This effectively describes the
covering space of a rod configuration with a compact
$z$-direction. In five and six dimensions the asymptotic
space-time has (at least) two circles, one parameterized by $\phi$
and the other by $z$. Potentially this makes the six-dimensional
solution more interesting since it has four uncompactified
dimensions. The solutions describe periodic arrays of black holes
and bubbles. This is an uncharged analogue of arrays of
supersymmetric black holes, but for the neutral bubble-black hole
arrays we expect the bubbles to play a role in balancing the
configurations. Generalizing the phase diagram to include
configurations with non-trivial tensions in two compact
directions, it could be interesting to study the phases of black
hole arrays in Kaluza-Klein theory.
\newline

\noindent {\sl More solutions: }
Finally, we note that one can argue for the existence of new
classes of solutions similar to the bubble-black hole sequences
presented in this paper.
Imagine taking the five-dimensional
$(1,1)$ solution and throwing some matter into the black hole
in such a way that the solution acquires an angular momentum. This
suggest the existence of new $(1,1)$ solutions with angular
momentum. However, since these solutions are stationary rather
than static they are not part of the class of solutions studied in
this paper. But, if we make a double Wick rotation in the $t$ and
$\phi$-direction, as in Section \ref{secWick}, it seems likely
that we get a static solution with one black hole and one
Kaluza-Klein bubble that asymptote $\CM^4 \times S^1$. Therefore,
the existence of the rotating $(1,1)$ solution imply the existence
of a whole new continuous class of static solutions that should be
part of the $(\mu,n)$ phase diagram.

Another way to argue for this is to take the four-dimensional
Kerr solution \cite{Kerr:1963ud},
adding a trivial $\phi$-direction, and making
a double Wick rotation in the $t$ and $\phi$-direction.
This gives a continuous class of static Kaluza-Klein
bubble solutions \cite{Dowker:1995gb}%
\footnote{See \cite{Aharony:2002cx,Horowitz:2002cx} for
applications of this class of bubble solutions.}
which clearly all have $n=d-2$. This
class of bubble solutions precisely correspond to
the $n \rightarrow d-2$ limit of the new continuous class
of static solutions with
one bubble and one black hole that we argued above should exists.
Similarly, one could argue for new classes of solutions in
six dimensions.
\newline

\noindent {\sl Bubble-black hole sequences with charge: }
Using the methods of \cite{Harmark:2004ws}, one can
put charge on the bubble-black hole sequences considered
in this paper. This we explore in \cite{Harmark:2004bb}.

\section*{Acknowledgments}

We thank O.~Aharony, M.~Berkooz, J.~de~Boer, R.~Emparan, D.~Gorbonos, 
G.~Horowitz, B.~Kol, A.~Maharana, G.~Neergaard, H.~Reall,
K.~Skenderis and S.~Vandoren for illuminating discussions. HE is
grateful to the Niels Bohr Institute for hospitality during
various stages of this project. HE was supported by the Danish
Research Agency and NSF grant PHY-0070895.


\begin{appendix}

\section{Details on the $(p,q)=(2,3)$ solution \label{23details}}

Given the family of curves $\mu_{(2,3)}(n;\alpha)$ (see
\eqref{mu23}) of the $\Z_2$ symmetric
$(2,3)$ configuration, we give here some details on the various special
curves discussed in Section \ref{sec23}.

\subsubsection*{Extremal mass curves}

To find the curves that extremize the mass in the range $\frac{1}{2} < n < 2$,
we set $\partial \mu_{(2,3)}(n;\alpha)/\partial \al =0$ which results
in the following fourth order equation on the parameter $\al$
\begin{equation}
\label{quartic}
2 (n-2)^4 \al^4 + 2 (n +7)(n-2)^3 \al^3 + 2 (n+4)^2(n-2)^2 \al^2
\end{equation}
$$
- (n+1)(4n^2 - 7n -29) (n-2) \al - 9 (n+1) (2n^2-2n -1) =0
$$
but at the same time we should remember that $0 \leq \al \leq 1$,
so that the extrema can also occur on the boundaries of the parameter
space. Some analysis then shows that the minimal $\mu$ curve always
lies on the boundary of parameter space
\begin{equation}
\label{mu23min}
\mu_{(2,3)}^{\rm max} (n) = \left\{
\begin{array}{cc}
\mu_{(2,3)} (n ; 0) =\mu_{(1,2)_{\mathfrak{t}}} (n) & \spa 1 \leq \mu \leq \frac{27}{5}
\\
\mu_{(2,3)} (n ; 1) = \mu_{(2,1)} (n)  &  \spa \mu \geq \frac{27}{5}
\end{array} \right.
\end{equation}
where we recall that the $(1,2)_{\mathfrak{t}}$ and $(2,1)$ curves
are given in \eqref{mu12}, \eqref{mus21} respectively. This result should
also be clear from Figure \ref{fig5D}.

\begin{figure}[ht]
 \centerline{\epsfig{file=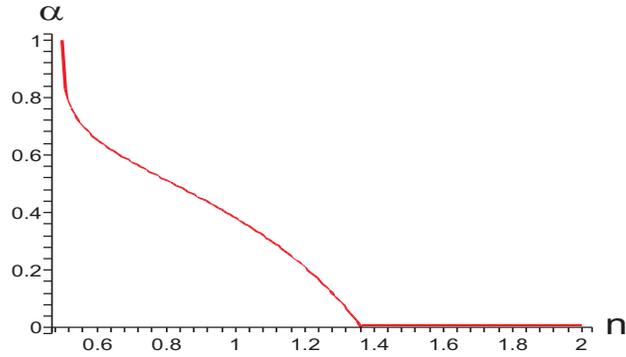,height=5cm,width=9cm}}
\caption{The solution $\alpha_{\mu ({\rm max})}(n)$
 that
determines the maximum $\mu$ curve for the symmetric $(2,3)$ solution.}
 \label{plot_alphanmu}
\end{figure}

On the other hand, the corresponding value $\alpha_{\mu ({\rm max})}(n)$
that maximizes $\mu$ is more complicated and is plotted
as a function of $n$ in Figure \ref{plot_alphanmu}.
The behavior is slightly non-trivial. For $ n \geq
\frac{1}{2} (1 + \sqrt{3}) \simeq 1.336 $ the maximum value occurs for
$\al =0$, i.e. the $(1,2)_{\mathfrak{t}}$ solution. For $n <
\frac{1}{2} (1 + \sqrt{3})$ we find a non-trivial solution
for $\al$ by solving \eqref{quartic}. This has
 the property that for $n \rightarrow \frac{1}{2}$ we approach
$\al =1$, i.e. the $(2,1)$ solution.

In conclusion, the curve that
maximizes $\mu$ for a given $n$ is given by
\begin{equation}
\label{mu23max}
\mu_{(2,3)}^{\rm max} (n) = \mu_{(2,3)} (n;\alpha_{\mu ({\rm max})}(n))
\end{equation}
where $\mu_{(2,3)}(n,\al)$ is given in \eqref{mu23} and
$\alpha_{\mu ({\rm max})}(n)$ the numerically obtained result in
Figure \ref{plot_alphanmu}.

\begin{figure}[ht]
 \centerline{\epsfig{file=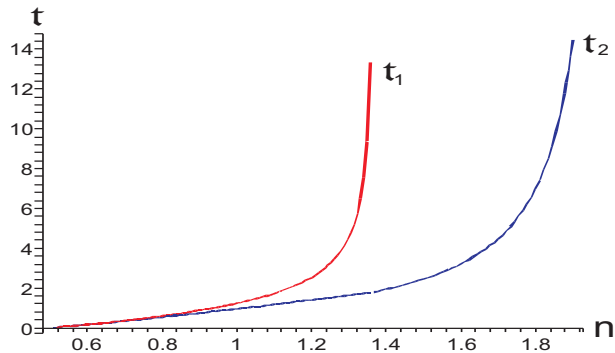,height=5cm,width=9cm}}
\caption{The temperatures $\mathfrak{t}_1 = \mathfrak{t_3}$ and $\mathfrak{t}_2$
of the black holes for the maximal mass  symmetric $(2,3)$ solution.}
 \label{tempfig}
\end{figure}

Using \eqref{betak1},
we have also computed the temperatures $\mathfrak{t}_1 = \mathfrak{t_3}$
and $\mathfrak{t}_2$ of the black holes for the symmetric $(2,3)$ solution
with maximal mass. The result is plotted in Figure \ref{tempfig}.
This shows that the temperature of the black holes at the two ends of the sequence
is higher than that of the black hole in the middle. We also see, for each
of these, that as the black hole becomes smaller its temperature approaches
infinity. For the black holes at the ends this happens at
$n=\frac{1}{2}(1 +\sqrt{3})$, after which the black holes have disappeared.
For the black hole in the middle the temperature diverges at $n=2$,
when only the bubble is left. We note, however, that the
product $\mathfrak{t}\mathfrak{s}$ in this point is correctly zero, as
required by the Smarr formula \eqref{gendsmarr}.

\subsubsection*{Extremal entropy curves}

To find the curves that extremize the total entropy $\mathfrak{s}(\mu)$
(which is obtained  from \eqref{mu23} and \eqref{s23}), we proceed as
follows.

After some standard variational calculus, the condition for
an extremum of the entropy at fixed $\mu$ is
\begin{equation}
\label{exent}
\frac{\partial s (n;\al)}{\partial n}
\frac{\partial \mu (n;\al)}{\partial \al} =
\frac{\partial s (n;\al)}{\partial \al}
\frac{\partial \mu (n;\al)}{\partial n}
\end{equation}
in the continuous region $ 0 \leq \al \leq 1$, but we should again
remember that extrema can occur on the boundaries of our parameter
space. After some algebra the condition \eqref{exent} gives rise
to the quintic equation on the parameter $\al$
\begin{equation}
\label{quintic} - 32 (2-n)^4 \al^5 +  8 (2-n)^3 (2n + 29) \al^4 +
8 (2-n)^2 (4n^2 - 16n - 83) \al^3
\end{equation}
$$
 -(2-n) (8n^3 + 180 n^2 - 366 n - 943) \al^2
- 2 (n+1) (8n^3 -36n^2 - 114n + 335) \al + 48 (n+ 1)^2 (2-n)=0
$$
where we have used also that we are only interested in solutions
for which $0 \leq \al \leq 1$.

We then find that the curve of maximum entropy
lies on the boundary of parameter space and is given by
\begin{equation}
\label{s23max}
\mathfrak{s}_{(2,3)}^{\rm max} (\mu) = \left \{
\begin{array}{cc}
\mathfrak{s}_{(2,3)} (\mu) \vert_{\al =0} = \mathfrak{s}_{(1,2)_{\mathfrak{t}}} (\mu)  & \spa  \mu \leq 2.42\\
\mathfrak{s}_{(2,3)} (\mu) \vert_{\al =1}  = \mathfrak{s}_{(2,1)} (\mu) &  \spa \mu \geq 2.42
\end{array} \right.
\end{equation}
i.e. a particular combination of the $(1,2){}_{\mathfrak{t}}$ and
$(2,1)$ solutions (just as the minimal $\mu$ curve \eqref{mu23min}).
 Again this should also be clear from Figure \ref{figs5d}.
The corresponding crossover point in the $(\mu,n)$ diagram
has $n=0.95$ on the $(1,2){}_{\mathfrak{t}}$ curve and
$n=1.05$ on the $(2,1)$ curve, so there is a discontinuity in $n$.

\begin{figure}[ht]
 \centerline{\epsfig{file=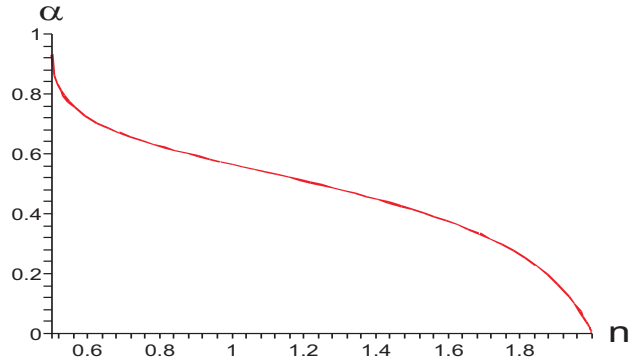,height=5cm,width=9cm}}
\caption{The solution $\al_{\mathfrak{s} ({\rm min})} (n)$ that
determines the minimum entropy curve for the symmetric $(2,3)$ solution.
This coincides with the solution $\al_{\mathfrak{t}}(n)$ that determines
the equal temperature $(2,3)$ solution.}
 \label{plot_alphan}
\end{figure}

On the other hand, for the minimum of the entropy, we find that $\al$
is given by the unique non-trivial solution
$\al_{\mathfrak{s} ({\rm min})} (n)$
of the quintic equation \eqref{quintic}
in the range $ 0 \leq \al \leq 1$. The numerically obtained solution
is shown in Figure  \ref{plot_alphan}. We conclude that the
minimum entropy curve is given by
\begin{equation}
\label{s23min}
\mathfrak{s}_{(2,3)}^{\rm min} (\mu) =
\mathfrak{s}_{(2,3)}(n;\al_{\mathfrak{s} ({\rm min})}) \vert_{n = n_\star(\mu)}
\end{equation}
where $n_\star(\mu)$ is found by inverting the curve
\begin{equation}
\label{s23minmu}
\mu_{(2,3)}^{\mathfrak{s}({\rm min})} (n) \equiv
\mu_{(2,3)} (n;\alpha_{\mathfrak{s} ({\rm min})}(n))
\end{equation}
where $\mathfrak{s}_{(2,3)}(\mu;\al)$ and  $\mu_{(2,3)}(\mu,\al)$ are given in
\eqref{s23}, \eqref{mu23} respectively, and
$\al_{\mathfrak{s} ({\rm min})} (n)$ the numerically obtained result in
 Figure  \ref{plot_alphan}.

\subsubsection*{Equal temperature curve}

To find the equal temperature curve we use the temperature
expressions in \eqref{betak1} along with the rod structure
\eqref{a23}. Requiring $\mathfrak{t}_1 = \mathfrak{t}_2$ and using
the definitions above \eqref{mu23}, one finds that this condition
becomes a quintic equation on $\al$ with $n$-dependent
coefficients. The solution of this equation in the range $0 \leq
\al  \leq 1$ which we denote by $\al_{\mathfrak{t}}(n)$ is then
seen to exactly agree with that of $\al_{\mathfrak{s} ({\rm min})}
(n)$ plotted
 in Figure  \ref{plot_alphan}. This explicitly shows that equal temperature
corresponds to an extremum (here a minimum) of the total entropy.

\end{appendix}

\addcontentsline{toc}{section}{References}



\providecommand{\href}[2]{#2}\begingroup\raggedright\endgroup

\end{document}